\documentclass[a4paper,11pt]{article}
\pdfoutput=1
\usepackage{jheppub}

\usepackage[T1]{fontenc}
\usepackage{amssymb}
\usepackage{amsmath}
\usepackage{graphicx}
\usepackage{subcaption}
\usepackage[colorlinks=true]{hyperref}
\usepackage{appendix}
\usepackage{amsbsy}

\usepackage{booktabs}
\usepackage{multirow}

\usepackage{tabularx}

\def\CA{C_A}
\def\CF{C_F}

\def\NF{N_F}
\def\Re{\mbox{Re}}
\def\e{\epsilon}
\def\d{\hbox{d}}
\newcommand{\code}[1]{\textsc{#1}}

\preprint{{\raggedleft%
ZU-TH 18/23
}}

\title{The parton-level structure of Higgs decays to hadrons at N$^3$LO}
\author[a,b]{Xuan Chen,}
\author[b]{Petr Jakub\v{c}\'{i}k,}
\author[b]{Matteo Marcoli}
\author[b]{and Giovanni Stagnitto}

\affiliation[a]{School of Physics, Shandong University,
            Jinan, Shandong 250100, China}
\affiliation[b]{Physik-Institut, Universit\"at Z\"urich,
  Winterthurerstrasse 190, CH-8057 Z\"urich, Switzerland}
\emailAdd{xuan.chen@uzh.ch}
\emailAdd{petr.jakubcik@physik.uzh.ch}
\emailAdd{matteo.marcoli@physik.uzh.ch}
\emailAdd{giovanni.stagnitto@physik.uzh.ch}

\abstract{%
  We present the quantum chromodynamics (QCD) corrections for Higgs boson decays
  to hadronic final states at next-to-next-to-next-to-leading order (N$^3$LO) in
  the strong coupling constant $\alpha_s$.
  In particular, we consider the Higgs boson decay to massless bottom quarks and
  the Higgs boson decay to a pair of gluons in the limit of a heavy top quark.
  The tree-level five-parton, the one-loop four-parton, the two-loop
  three-parton, and the three-loop two-parton matrix elements are integrated
  separately over the inclusive phase space and classified by partons appearing
  in the final state and by colour structure.
  As a check, we reproduce known results for the inclusive hadronic decay rates
  at N$^3$LO.  We study patterns of infrared singularity cancellation within the
  colour layers of the integrated expressions and we comment on the similarities
  between $H \to b\bar{b}$ and $\gamma^* \to q\bar{q}$.
  We anticipate that our result will be an essential ingredient for the
  formulation of N$^3$LO subtraction schemes.
}

\begin{document} 
\maketitle
\flushbottom

\section{Introduction}

The Higgs boson plays a special role in the Standard Model. Its discovery in
2012~\cite{ATLAS:2012yve,CMS:2012qbp} through the clean decay modes $H \to
\gamma\gamma$, $H \to 4l$ and $H \to 2l2\nu$ marked a major milestone in
particle physics. Amongst the remaining decay modes, the decay to bottom quarks
has the largest branching ratio and allows probing the Higgs boson coupling to
third generation fermions.
On the other hand, the loop-induced decays of a Higgs boson to gauge
bosons ($gg$, $\gamma\gamma$ or $\gamma Z$) are key to an indirect
determination of the couplings to $Z$, $W$ and $t$.
A summary of the current experimental evidence for the Higgs production and
decay modes can be found in reviews by ATLAS~\cite{ATLAS:2022vkf} and
CMS~\cite{CMS:2022dwd}.

The importance of Higgs decays to quarks and gluons extends beyond
phenomenology. Thanks to the simplicity of the $1 \to 2$ Born kinematics, the
computation of radiative QCD corrections to these processes has been achieved to
very high orders in perturbation theory.
The $H \to gg$ decay rate in an effective theory with the top quark integrated
out~\cite{Wilczek:1977zn,Shifman:1978zn,Inami:1982xt} is known up to
N$^4$LO~\cite{Herzog:2017dtz,Djouadi:1991tka,Chetyrkin:1997iv,Baikov:2006ch,
  Moch:2007tx,Davies:2017rle}.
The $H \to b\bar{b}$ decay rate in a massless approximation where only the
bottom Yukawa coupling is kept different from zero is also known up to
N$^4$LO~\cite{Herzog:2017dtz,Baikov:2005rw,Braaten:1980yq,Gorishnii:1990zu,Chetyrkin:1996sr}.
The correction to the total decay rate of the Higgs boson to hadrons, including
also contributions induced by the effective Higgs-bottom quark interaction where
only the dominant $m_b^2$ mass terms are retained, has been calculated at order
N$^3$LO in~\cite{Chetyrkin:1997vj} and at order N$^4$LO
in~\cite{Davies:2017xsp}.
We refer the reader to a review~\cite{Spira:2016ztx} for further discussion on
terms suppressed by the top-quark and bottom-quark mass and electroweak effects.

All the above results follow a common approach based on the optical theorem,
which amounts to the computation of the imaginary part of the Higgs
self-energy. This technique is completely agnostic to the number and species of
particles in the final state. Hence it does not reveal the infrared structure of
the result and obscures the interplay between real radiation and virtual
corrections in the different final-state partonic channels.

In this paper, we extend this analysis by separately integrating all the
possible physical cuts of the four-loop QCD correction to the Higgs two-point
function.
In other words, we analytically integrate the tree-level five-parton, the
one-loop four-parton, the two-loop three-parton, and the three-loop two-parton
matrix elements over the respective phase space:
\begin{equation}\label{eqn:matelems}
  \sigma^{(3)} = \int \d\Phi_5\,M_{5}^0
  + \int \d\Phi_4\,M_{4}^1
  + \int \d\Phi_3\,M_{3}^2
  + \int \d\Phi_2\,M_{2}^3\,,
\end{equation}
where $M_{n}^l$ denotes the $l$-loop matrix element for the decay of the Higgs
into $n$ final state QCD particles. We refer to the four terms
in~\eqref{eqn:matelems} as the triple-real (RRR), double-real-virtual (VRR),
double-virtual-real (VVR), and triple-virtual (VVV) \textit{layers} of the calculation. For completeness we recompute also the
next-to-leading order (NLO) and next-to-next-to-leading order (NNLO)
corrections.
Our method was first described in~\cite{Jakubcik:2022zdi} in the context of the
decay of a virtual photon into hadrons and leveraged the reverse unitarity
relation~\cite{Cutkosky:1960sp,Anastasiou:2002yz,Anastasiou:2002wq,Anastasiou:2003yy,Anastasiou:2003ds}
to gain access to modern multi-loop techniques.

The presented results are particularly relevant for the development of N$^3$LO
subtraction schemes. N$^3$LO precision is the state-of-the-art for simple
processes~\cite{Anastasiou:2015vya,Dreyer:2016oyx,Mistlberger:2018etf,Currie:2018fgr,Cieri:2018oms,Dulat:2018bfe,Dreyer:2018qbw,Mondini:2019gid,Duhr:2019kwi,Chen:2019lzz,Duhr:2020seh,Duhr:2020sdp,Chen:2022lwc,Duhr:2021vwj,Chen:2021vtu,Chen:2021isd,Billis:2021ecs,Chen:2022cgv,Baglio:2022wzu,Chen:2022vzo}
but a general local subtraction scheme at this order is still missing.
The matrix elements of $H \to gg$ at NNLO were used for the derivation of
gluon-gluon antenna functions in the context of the antenna subtraction
scheme~\cite{Gehrmann-DeRidder:2005btv} in order to encapsulate all the
unresolved radiation between a pair of hard
gluons~\cite{Gehrmann-DeRidder:2005alt}. It follows that the matrix elements for
$H \to gg$ at N$^3$LO are candidates for gluon-gluon antenna functions one order
higher.
Moreover, the integrated version of such matrix elements can shed light on how
the universal behaviour of N$^3$LO matrix elements in unresolved
configurations~\cite{Badger:2004uk,Duhr:2014nda,Duhr:2013msa,
  Li:2013lsa,Dixon:2019lnw,Catani:2003vu,Czakon:2022fqi,
  Catani:2021kcy,Zhu:2020ftr,Czakon:2022dwk,
  DelDuca:2019ggv,DelDuca:2020vst,Catani:2019nqv,DelDuca:2022noh,Catani:2022hkb}
translates to the integrated level and how it relates to the divergences of
virtual corrections.

The paper is organised as follows. In Section~\ref{sec:notation}, we introduce the
notation and we briefly describe our method.
In Section~\ref{sec:results}, we analyse our results which are explicitly reported
in Appendices~\ref{app:exprN3LO} and~\ref{app:exprlower}.
We conclude in Section~\ref{sec:conclusions} with an outlook of future applications.

\section{Method}
\label{sec:notation}

We consider the $H \to gg$ and $H \to b\bar{b}$ decays.
In the first case, we work in the heavy-top effective theory, with the QCD
Lagrangian supplemented with an effective Lagrangian given by
\begin{equation}
  \mathcal{L}_{gg} = - \frac{\lambda_0}{4} H G_a^{\mu\nu} G_{a,\mu\nu}\,,
\end{equation}
with $G_a^{\mu\nu}$ the renormalised gluon field-strength, $H$ the Higgs field
and $\lambda_0$ the bare effective coupling, obtained by matching the effective
theory to the full Standard Model~\cite{Chetyrkin:1997iv,Kniehl:1995tn,Chetyrkin:1997un}.
In the second case, we implement the standard vertex between the Higgs field and
a fermion line
\begin{equation}\label{eq:vHbb}
  \mathcal{L}_{q\bar{q}} = y^b_0 H \bar{\psi} \psi\,,
\end{equation}
with $\psi$ the bottom quark field and $y^b_0$ the bare Yukawa coupling. In our
calculation, the bottom quark is treated as massless but the Yukawa coupling is
non-vanishing.

We present results for the integration of renormalised squared amplitudes in the
$\overline{\text{MS}}$ scheme. We replace the bare coupling $\alpha_0$ with the
renormalised coupling $\alpha_s$ according to~\cite{Gehrmann:2010ue}
\begin{eqnarray}\label{eq:alfaren}
\alpha_0\,\mu_0^{2\e}\,S_\e &=& \alpha_s\,\mu^{2\e}\Bigg[
1- \frac{\beta_0}{\e}\left(\frac{\alpha_s}{2\pi}\right) 
+\left(\dfrac{\beta_0^2}{\e^2}-\dfrac{\beta_1}{2\e}\right)\left(\frac{\alpha_s}{2\pi}\right)^2 
\nonumber \\ && \phantom{\alpha_s\,\mu^{2\e}\Bigg[} 
- \left(\dfrac{\beta_0^3}{\e^3}-\dfrac{7}{6}\dfrac{\beta_1\beta_0}{\e^2}
+ \dfrac{1}{3}\dfrac{\beta_2}{\e}\right)   \left(\frac{\alpha_s}{2\pi}\right)^3
+{\cal O}(\alpha_s^4) \Bigg]\,,
\end{eqnarray}
with
\begin{eqnarray}
\beta_0 &=& \frac{11 \CA - 2 \NF}{6}\,,\label{eq:beta0}\\
\beta_1 &=& \frac{17 \CA^2 - 5 \CA \NF -3 \CF \NF}{6}\,,\label{eq:beta1}\\
\beta_2 &=& \frac{2857 \CA^3}{432} +  \frac{\CF^2 \NF}{8}
- \frac{205 \CF \CA \NF}{144}
- \frac{1415 \CA^2 \NF}{432} +  \frac{11 \CF \NF^2}{72}
+ \frac{79 \CA \NF^2}{432}\,,\label{eq:beta2}\\
S_\e &=& (4\pi)^\e e^{-\e\gamma}\,,\qquad \mbox{with Euler constant }
\gamma = 0.5772\ldots
\end{eqnarray}
where $\alpha_0$ is the bare coupling, $\mu_0^2$ is the mass parameter
introduced in dimensional regularisation to maintain a dimensionless coupling in
the bare QCD Lagrangian density. We fix the renormalisation scale $\mu^2$ to be the
invariant mass of the decaying particle $q^2$.

In the calculation of $H \to b\bar{b}$, the renormalisation of the bare Yukawa
coupling $y^b_0$ is needed, which is done through the replacement $y^b_0 =
Z_y\,y^b$, with $Z_y$ as in~\cite{Gehrmann:2014vha}
\begingroup
\allowdisplaybreaks
\begin{eqnarray}
Z_y &=& 
1 - \frac{3 \CF}{2\e}\left(\frac{\alpha_s}{2\pi}\right) 
\nonumber \\
&& + \Bigg[\CF^2 \left(\frac{9}{8 \e^2}-\frac{3}{16 \e} \right)
  + \CF \CA \left(\frac{11}{8 \e^2}-\frac{97}{48 \e}\right)
  + \CF \NF \left(-\frac{1}{4\e^2}+\frac{5}{24 \e}\right)\Bigg]
\left(\frac{\alpha_s}{2\pi}\right)^2
\nonumber \\
&& + \Bigg[
  \CF^3 \left(-\frac{9}{16 \e^3}+\frac{9}{32 \e^2}-\frac{43}{16 \e} \right)
  + \CF^2 \CA \left(-\frac{33}{16 \e^3}+\frac{313}{96 \e^2}+\frac{43}{32 \e} \right)
\nonumber \\
&& \phantom{+ \Bigg[} + \CF \CA^2 \left(-\frac{121}{72 \e^3}+\frac{1679}{432 \e^2}
-\frac{11413}{2592 \e} \right)
+ \CF^2 \NF \left(\frac{3}{8\e^3}-\frac{29}{48 \e^2}+\frac{1}{\e} \left(\frac{23}{24}-\zeta_3\right) \right)
\nonumber \\
&& \phantom{+ \Bigg[} + \CF \CA \NF \left(\frac{11}{18 \e^3}-\frac{121}{108 \e^2}+\frac{1}{\e} \left(\frac{139}{324}+\zeta_3\right) \right)
\nonumber \\
&& \phantom{+ \Bigg[} + \CF \NF^2 \left(-\frac{1}{18 \e^3}+\frac{5}{108 \e^2}+\frac{35}{648 \e} \right)
\Bigg]
\left(\frac{\alpha_s}{2\pi}\right)^3
+{\cal O}(\alpha_s^4) \; .
\end{eqnarray}
\endgroup

In the calculation of $H \to gg$, we additionally need to renormalise the
effective coupling $\lambda_0=Z_{\lambda}\lambda$~\cite{Spiridonov:1988md}
according to~\cite{Gehrmann:2010ue}
\begin{equation}
 Z_\lambda = 1 - \frac{\beta_0}{\e}\left(\frac{\alpha_s}{2\pi}\right)
 + \left(\dfrac{\beta_0^2}{\e^2}-\dfrac{\beta_1}{\e}\right)
 \left(\frac{\alpha_s}{2\pi}\right)^2
 - \left(\dfrac{\beta_0^3}{\e^3}-\dfrac{2\beta_1\beta_0}{\e^2}
 + \dfrac{\beta_2}{\e}\right)   \left(\frac{\alpha_s}{2\pi}\right)^3
 +{\cal O}(\alpha_s^4)\,.
\end{equation}

We follow the strategy outlined in~\cite{Jakubcik:2022zdi}, where the relevant
decay diagrams are generated with QGRAF~\cite{Nogueira:1991ex} as self-energies
of the Higgs boson with cut internal propagators. They are matched onto the
integral families reported in~\cite{Jakubcik:2022zdi} using
\code{Reduze2}~\cite{vonManteuffel:2012np} and the Feynman rules are inserted
and evaluated in \code{FORM}~\cite{Vermaseren:2000nd}.
The integrals appearing in the matrix elements have up to eleven propagators in
the denominator and a maximum of five scalar products in the numerator, compared
to four scalar products in the photon decay.
The integrals are reduced with the help of
\code{Reduze2}~\cite{vonManteuffel:2012np} to a set of 22, 27, 35 and 31 master
integrals for the four terms of~\eqref{eqn:matelems}, respectively. The master
integrals required for the NNLO calculation can be found
in~\cite{Gehrmann-DeRidder:2003pne} and have been extended up to weight 6
in~\cite{Jakubcik:2022zdi,Gituliar:2018bcr,Gehrmann:2010ue}.
The integrals required for the N$^3$LO  calculation were computed
in~\cite{Gituliar:2018bcr,Magerya:2019cvz}.

\section{Results}
\label{sec:results}

We illustrate the general structure of the different partonic contributions by
adopting a notation similar to~\cite{Jakubcik:2022zdi} for ease of reference.
Given a set of $n$ final-state particles denoted by ${\cal I}$, we can
generically write the associated amplitude for the $H \to ij$ process as
\begin{equation}
|{\cal M}_{ij}\rangle_{{\cal I}} =
|{\cal M}_{ij}^{(0)}\rangle_{{\cal I}} 
+ \left(\frac{\alpha_s}{2\pi}\right) |{\cal M}_{ij}^{(1)}\rangle_{{\cal I}} 
+ \left(\frac{\alpha_s}{2\pi}\right)^2 |{\cal M}_{ij}^{(2)}\rangle_{{\cal I}}
+ \left(\frac{\alpha_s}{2\pi}\right)^3 |{\cal M}_{ij}^{(3)}\rangle_{{\cal I}}
+ \ldots \,.
\end{equation}
We denote the integration over the respective phase space of the matrix element
$\langle{\cal M}_{ij}|{\cal M}_{ij}\rangle _{{\cal I}}$ summed over spins,
colours and quark flavours as
\begin{equation}\label{eq:Tll}
  {\cal T}^{ij,(k,\left[\ell \times \ell\right])}_{{\cal I}} = \int \d \Phi_n \,
  \langle{\cal M}_{ij}^{(\ell)}|{\cal M}_{ij}^{(\ell)}\rangle_{{\cal I}}
\end{equation}
and for $\ell_1 \neq \ell_2$
\begin{equation}\label{eq:Tlm}
  {\cal T}^{ij,(k,\left[\ell_1 \times \ell_2\right])}_{{\cal I}} = \int \d
  \Phi_n \, 2\,\Re\big[\langle{\cal M}_{ij}^{(\ell_1)}|{\cal
      M}_{ij}^{(\ell_2)}\rangle_{{\cal I}}\big]\,,
\end{equation}
where $ij=gg,q\bar{q}$.
The label $k$ denotes the perturbative order: contributions with
the same $k$ sum to the N$^k$LO result for the total cross section.
 The long explicit expressions for ${\cal T}^{ij,(3,\left[\ell_1 \times
     \ell_2\right])}_{{\cal I}}$ are provided in Appendix \ref{app:exprN3LO},
 while in Appendix \ref{app:exprlower} we report the lower-order results
 expanded up to transcendental weight six. We denote the coefficient of each colour factor
 ${\cal C}$ as ${\cal T}^{ij,(k,\left[\ell_1 \times \ell_2\right])}_{{\cal
     I}}\big|_{\cal C}$, and we omit the superscript $\left[\ell_1 \times
   \ell_2\right]$ in case of no ambiguity.
All results are also provided in the ancillary files in computer-readable
format, with the notation
\begin{equation}
  {\cal T}^{ij,(k,\left[\ell_1 \times \ell_2\right])}_{{\cal I}}
  = \texttt{[H}ij\texttt{\_}\mathcal{I}\texttt{\_}k\texttt{\_}\ell_1\texttt{x}\ell_2\texttt{]}\,.
\end{equation}

All expressions are renormalised and in time-like kinematics.
Higher-order results are normalised to
\begin{equation}\label{eq:Hgg0}
{\cal T}^{gg,(0)}_{gg} = \frac{1}{4}\,\lambda\,(N^2 - 1)\,(q^2)^2\,(1-\e)\,P_2\,
\end{equation}
for the decay to gluons and 
\begin{equation}\label{eq:Hbb0}
2\,\CF {\cal T}^{q\bar{q},(0)}_{q\bar{q}} = 4\,y^b\,(N^2 - 1)\,q^2\,P_2
\end{equation}
for the decay to bottom quarks, with $N$ the number of colours and $\CF =
(N^2-1)/(2N)$. ${\cal T}^{gg,(0)}_{gg}$ and ${\cal T}^{q\bar{q},(0)}_{q\bar{q}}$
are the respective Born-level cross sections.
Note that the factor $2\CF$ in~\eqref{eq:Hbb0} is included in the normalisation
of $H \to b\bar{b}$ as it appears in all colour layers starting from NLO.
Finally, $P_2$ is the volume of the two-particle phase space,
\begin{equation}
P_2  = \int \d \Phi_2 =
2^{-3+2\e}\, \pi^{-1+\e}\, \frac{\Gamma(1-\e)}{\Gamma(2-2\e)}\,
(q^2)^{-\e} \,.
\end{equation}
The structure or (non-)appearance of certain colour factors is key in
understanding the cancellation patterns between real and virtual corrections. We
therefore summarize the colour factors appearing in the various final-state
configurations in Appendix~\ref{app:coltables}.
In the final state with two quark lines, we explicitly separate the
configurations with same- or different-flavour quark pairs.
Note that for all colour factors ${\cal C}$ in ${\cal
  T}^{ij,(k)}_{q\bar{q}q'\bar{q}'(g)}$, we have that
\begin{equation}
  {\cal T}^{ij,(k)}_{q\bar{q}q'\bar{q}'(g)}\Big|_{\cal C}
  = (\NF-1)\,{\cal T}^{ij,(k)}_{q\bar{q}q\bar{q}(g)}\Big|_{{\cal C}/(\NF-1)}\,.
\end{equation}
For $H\to gg$, in the two-particle final state, the terms proportional to
$\NF^k$ for $k=1,2,3$ in ${\cal T}_{gg}^{gg,(k)}$ are introduced as part of
renormalisation and as such are not present in the finite part or beyond.
Note that the only allowed two-particle final state in the $H\to gg$ corrections
is a pair of gluons, even at higher loops. This is explained by the fact that
for a scalar particle to decay into a quark-antiquark pair, a chirality flip
along the fermionic line is needed due to spin conservation. Therefore, all the
diagrams contributing to the Higgs decay to a massless quark-antiquark pair via
QCD interactions vanish.

We perform several checks on our results.
First, we can directly compare our expressions for the two-particle final states
to the calculations of the quark and gluon form factors up to three
loops~\cite{Gehrmann:2010ue,Gehrmann:2014vha}.
Second, we observe the complete cancellation of infrared (IR) poles in the sum of all the
partonic final states at each perturbative order, both in $H \to gg$ and $H \to
b\bar{b}$.
Third, the finite parts agree with the known results for the total cross
sections e.g.\ in~\cite{Herzog:2017dtz}.
Indeed, the sum of ~\eqref{eq:Hbbstart}--\eqref{eq:Hbbend} is the N$^3$LO
coefficient of the inclusive decay rate for $H \to b\bar{b}$,
\begingroup
\allowdisplaybreaks
\begin{flalign}
  R^{H \to b\bar{b}}\Big|_{\alpha_s^3} &= \left(\frac{\alpha_s}{2\pi}\right)^3
  \sum_{{\cal C},\,{\cal I}} {\cal T}^{q\bar{q},(3)}_{\cal I}\Big|_{\cal C}
  \nonumber &\\ &=
  \left(\frac{\alpha_s}{2\pi}\right)^3 \Bigg[
  N^2 \left( \frac{25999999}{62208} - \frac{3803}{216}\pi^2 - \frac{4321}{24}\zeta_3
  + \frac{155}{6}\zeta_5  \right)
  \nonumber &\\& \phantom{=\left(\frac{\alpha_s}{2\pi}\right)^3 \Bigg[}    
  - \frac{76055}{384} + \frac{545}{48}\pi^2 + \frac{1567}{16}\zeta_3
  - \frac{235}{8}\zeta_5 
  \nonumber &\\& \phantom{=\left(\frac{\alpha_s}{2\pi}\right)^3 \Bigg[}  
    +\frac{1}{N^2} \left( \frac{23443}{768} - \frac{27}{16}\pi^2
    -\frac{239}{16} \zeta_3 + \frac{45}{8}\zeta_5 \right)
  \nonumber &\\& \phantom{=\left(\frac{\alpha_s}{2\pi}\right)^3 \Bigg[}
    + N \NF \left( - \frac{47731}{486}  + \frac{1727}{432}\pi^2
    + \frac{371}{12}\zeta_3 - \frac{\pi^4}{120} - \frac{10}{3}\zeta_5  \right)
  \nonumber &\\& \phantom{=\left(\frac{\alpha_s}{2\pi}\right)^3 \Bigg[}    
    + \frac{\NF}{N} \left( \frac{88}{3} -\frac{65}{48}\pi^2
    -\frac{65}{4} \zeta_3 -\frac{\pi^4}{120} + 5\zeta_5 \right)
  \nonumber &\\& \phantom{=\left(\frac{\alpha_s}{2\pi}\right)^3 \Bigg[}  
  + \NF^2 \left( \frac{15511}{3888} - \frac{11}{54}\pi^2 - \zeta_3 \right) \Bigg]\,. &
\end{flalign}
\endgroup
Similarly, by summing~\eqref{eq:Hggstart}--\eqref{eq:Hggend} we obtain the N$^3$LO
coefficient of the inclusive decay rate for \mbox{$H \to gg$},
\begingroup
\allowdisplaybreaks
\begin{flalign}
  R^{H \to gg}\Big|_{\alpha_s^3} &= \left(\frac{\alpha_s}{2\pi}\right)^3
  \sum_{{\cal C},\,{\cal I}} {\cal T}^{gg,(3)}_{\cal I}\Big|_{\cal C}
  \nonumber &\\ &=
  \left(\frac{\alpha_s}{2\pi}\right)^3 \Bigg[
    N^3 \left(
          \frac{15420961}{5832}
          - \frac{2816}{27}\pi^2
          - \frac{44539}{54}\zeta_3
          + \frac{385}{3}\zeta_5          
    \right)
    \nonumber &\\& \phantom{=\left(\frac{\alpha_s}{2\pi}\right)^3 \Bigg[}
       + \NF N^2 \left(
          - \frac{11918065}{7776}
          + \frac{465}{8}\pi^2
          + \frac{8171}{36}\zeta_3          
          - \frac{10}{3}\zeta_5
          \right)
     \nonumber &\\& \phantom{=\left(\frac{\alpha_s}{2\pi}\right)^3 \Bigg[}
       + \NF \left(
          \frac{11279}{72}
          - \frac{143}{72}\pi^2
          - \frac{389}{4}\zeta_3          
          + 10 \zeta_5
          \right)
    \nonumber &\\& \phantom{=\left(\frac{\alpha_s}{2\pi}\right)^3 \Bigg[}
       + \frac{\NF}{N^2} \left(
          \frac{221}{96}
          + 6 \zeta_3       
          - 10 \zeta_5
          \right)
     \nonumber &\\& \phantom{=\left(\frac{\alpha_s}{2\pi}\right)^3 \Bigg[}            
        + \NF^2 N \left(
           \frac{58346}{243}
           - \frac{359}{36}\pi^2
           - \frac{128}{9}\zeta_3
           \right)
   \nonumber &\\& \phantom{=\left(\frac{\alpha_s}{2\pi}\right)^3 \Bigg[}
        + \frac{\NF^2}{N} \left(
           - \frac{55}{2}
           + \frac{13}{36}\pi^2
           + 15 \zeta_3
           \right)
   \nonumber &\\& \phantom{=\left(\frac{\alpha_s}{2\pi}\right)^3 \Bigg[}
        + \NF^3  \left(
           - \frac{7127}{729}
           + \frac{14}{27}\pi^2
           + \frac{8}{27}\zeta_3
           \right)            
     \Bigg]\,. &
\end{flalign}
\endgroup

\subsection{Infrared structure of the deepest poles}
It is interesting to compare the colour and singularity structure of $H \to
b\bar{b}$ to that of $\gamma^* \to q\bar{q}$ derived in our previous
work~\cite{Jakubcik:2022zdi}, both representing decays of a colour-singlet state
into a quark-antiquark pair.
Once we normalise with respect to the respective Born-level cross sections, we
observe the presence of identical colour factors at all perturbative orders,
except for the singlet contributions which are only present in $\gamma^* \to
q\bar{q}$.
For the Higgs case, the singlet contribution vanishes because the Yukawa
interaction in \eqref{eq:vHbb} introduces a chirality flip which would break
chirality conservation along a closed massless fermionic loop.
We notice that for all layers and orders, the two deepest poles appearing in any
given colour factor of the integrated renormalised matrix elements coincide
between the $H \to b\bar{b}$ and $\gamma^* \to q\bar{q}$ processes. To clarify,
we refer to the two deepest non-vanishing poles in each partonic configuration
and colour layer, not simply to the $\e^{-2k}$ and $\e^{-2k+1}$ poles at order
$k$.
For the two-particle final states, the infrared singularity structure is
predicted through universal IR factorisation
formulae~\cite{Catani:1998bh,Becher:2009cu}.
In particular, the deepest poles can be interpreted in a completely
process-independent way in terms of the $I^{(1)}_{q\bar{q}}$ insertion
operator~\cite{Catani:1996vz}.
For purely real corrections up to NNLO, we can build on a recent algorithmic
construction of idealised NNLO antenna functions~\cite{Braun-White:2023sgd}. In this work,
the deepest poles of $M_4^0$ in the $\gamma^{*}\to q\bar{q}$ decay are
identified with structures coming from the integrations of double-unresolved
limits.
Defining the operator ${\cal P}_{n}[\cdot]$ which extracts from an expression
the $n$ deepest non-vanishing poles in $\e$, the relations read
\begin{align}
  {\cal P}_{2}\left[{\cal T}^{q\bar{q},(1)}_{q\bar{q}g}\Big|_{N^0}\right] &=
  {\cal P}_{2}\left[{\cal S}soft_g+2{\cal S}col_{q^hg}\right]	\,,\nonumber \\
  {\cal P}_{2}\left[{\cal T}^{q\bar{q},(2)}_{q\bar{q}gg}\Big|_{N}\right] &=
  {\cal P}_{2}\left[{\cal D}soft_{gg}
    + 2{\cal T}col_{q^hgg}\right]\,,\nonumber \\
    {\cal P}_{2}\left[{\cal T}^{q\bar{q},(2)}_{q\bar{q}gg}\Big|_{N^{-1}}\right] &=
    {\cal P}_{2}\left[-\frac{1}{2}\left({\cal D}soft_{\gamma\gamma}
    + 2{\cal T}col_{q^h\gamma\gamma}\right)\right]\,,\nonumber \\
  {\cal P}_{2}\left[{\cal T}^{q\bar{q},(2)}_{q\bar{q}q'\bar{q}'}\Big|_{(\NF-1)}\right] &=
  {\cal P}_{2}\left[{\cal D}soft_{q\bar{q}}
    +2{\cal T}col_{q^h\bar{q}'q'}\right]\,,\nonumber \\
  {\cal P}_{2}\left[{\cal T}^{q\bar{q},(2)}_{q\bar{q}q\bar{q}}\Big|_{N^0}\right] &=
       {\cal P}_{2}\left[{\cal D}soft_{q\bar{q}}
       +2{\cal T}col_{q^h\bar{q}'q'}\right]\,,\nonumber \\
  {\cal P}_{1}\left[{\cal T}^{q\bar{q},(2)}_{q\bar{q}q\bar{q}}\Big|_{N^{-1}}\right] &=
  {\cal P}_{1}\left[{\cal C}_4^0\right]\,,
            \label{eq:HP2}
\end{align}
where ${\cal S}soft$, ${\cal S}col$, ${\cal D}soft$ and ${\cal T}col$ are given
in Appendix A and B of~\cite{Braun-White:2023sgd}, with the superscript $h$ in 
${\cal S}col$ and ${\cal T}col$ indicating the hard radiator. The
quantities in the last line of~\eqref{eq:HP2} only exhibit a single pole and
${\cal C}_4^0$, which encapsulates the triple collinear limit of a $q \parallel
\bar{q} \parallel q$ configuration, is given in~(5.57)
of~\cite{Braun-White:2023sgd}. Our results confirm that~\eqref{eq:HP2} holds for
the $H \to b\bar{b}$ process as well.
Beyond the two deepest poles, the finite parts of lower-order integrated matrix
elements contribute to the singularities, resulting in differences between $H
\to b\bar{b}$ and $\gamma^* \to q\bar{q}$.
An analogous study for single-real emission at one loop will allow for a
physically motivated description of the singularities of the real-virtual
contribution~\cite{WGP:inprepare}.
The fact that the exact correspondence of the deepest non-vanishing poles (and
not just the $\e^{-2k}$ and $\e^{-2k+1}$ poles at order $k$) between the two
processes extends also to N$^3$LO in all the layers, colour factors and partonic
final states is notable.
Therefore, we expect that a universal and process-independent description exists for
unresolved radiation between a hard quark-antiquark pair also at this
perturbative order.

We observe that for every power of $\e$, the terms with the highest
trascendental weight always coincide in the $\gamma^* \to q\bar{q}$ and the $H
\to b\bar{b}$ process.
This is reflected in the N$^3$LO decay rates: the weight 4 contribution in the
$\NF^2$ colour layer is absent in both processes; the weight 5 contribution in
the $\NF/N$ colour layer is identically $5\,\zeta_5$; the weight 5 contribution
in the $\NF N$ colour layer is identically $-10/3\,\zeta_5$; the weight 6
contributions are absent in both processes.
In fact, the correspondence between the highest weight contribution in every
colour factor extends to the N$^4$LO decay rates of $\gamma^* \to q\bar{q}$ and $H
\to b\bar{b}$~\cite{Baikov:2010je,Herzog:2017dtz}.
Moreover, we observe that the weight $2k$ contribution vanishes in the decay rate
of $\gamma^* \to q\bar{q}$, $H \to b\bar{b}$ and $H \to gg$ all the way up to
N$^4$LO.
Since the decay rate at N$^k$LO is given by the absorptive part of a $(k+1)$-loop
propagator, this is directly related to the maximal trascendental weight
appering in the two-point function.
It would be interesting to analyse our observations in the framework
of~\cite{Henn:2021aco}.

For $H \to gg$, no analogous processes have been computed, so it is not possible
to perform a comparison as done for $H \to b\bar{b}$.
From~\cite{Braun-White:2023sgd}, it is possible to argue that for the real
radiation corrections up to NNLO in the purely gluonic final state, the following
relations hold:
\begin{align}
  {\cal P}_{2}\left[{\cal T}^{gg,(1)}_{ggg}\Big|_{N}\right] &=
  {\cal P}_{2}\left[{\cal S}soft_g+2{\cal S}col_{gg}\right]	\,,\nonumber \\
	{\cal P}_{2}\left[{\cal T}^{gg,(2)}_{gggg}\Big|_{N^2}\right] &=
	{\cal P}_{2}\left[4{\cal D}soft_{gg} +2{\cal D}soft_{\gamma\gamma}
	+ 8{\cal T}col_{g^hgg}+4{\cal T}col_{gg^hg}\right].
\end{align}
On the other hand, the dependence on lower orders is manifest already at the
second-deepest poles in the $q\bar{q}gg$ and $q\bar{q}q'\bar{q}'$ partonic
channels~\cite{Braun-White:2023sgd}.
Nonetheless, one can expect a weaker statement to hold also for $H\to gg$, namely
the universality of the $\e^{-2k}$ and $\e^{-2k+1}$ poles at order $k$
across all the partonic final states and colour factors.

\subsection{Cancellation of infrared singularities}
The pattern of cancellation of infrared singularities among the four layers
contributing at N$^3$LO is highly non-trivial due to the interplay of different
soft currents and splitting functions in the real-emission contributions.
Despite recently becoming available in their unintegrated
form~\cite{Badger:2004uk,Duhr:2014nda,Duhr:2013msa,
  Li:2013lsa,Dixon:2019lnw,Catani:2003vu,Czakon:2022fqi,
  Catani:2021kcy,Zhu:2020ftr,Czakon:2022dwk,
  DelDuca:2019ggv,DelDuca:2020vst,Catani:2019nqv,DelDuca:2022noh,Catani:2022hkb},
such structures are yet to be fully understood at the integrated level. One can
expect the cancellation pattern of the deepest poles to be easier to explain in
terms of universal factorization properties of QCD. Moreover, the deeper the
pole, the fewer infrared-divergent partonic configurations contribute to it,
simplifying the analysis. To facilitate the inspection of the deepest
singularities, in Tables~\ref{tab:polesHbb} and~\ref{tab:polesHgg} we display
the coefficients of the $\e^{-6}$ and $\e^{-5}$ poles for different colour
layers and partonic final states. Trivially, the sum of the coefficients in each
column vanishes. In the following, we make some basic observations, postponing a
thorough analysis, which would require detailed knowledge of the integrated
structures contributing at each layer.

For $H\to b\bar{b}$ in Table~\ref{tab:polesHbb}, the easiest colour factor to
inspect is $N^{-2}$, because it only receives contributions from the emission of
abelian gluons from the hard pair of quarks.
For both the virtual and real corrections, the coefficients of the $\e^{-6}$ and
$\e^{-5}$ poles are proportional to $(I^{(1)}_{q\bar{q}})^3$. Moreover, we
notice a pair-wise cancellation between the poles of the triple-virtual and
triple-real, and the double-real-virtual and double-virtual-real contributions,
which can be explained by the following argument.
Due to a complete factorization in the deepest poles, each virtual (or real)
abelian emission from a hard quark-antiquark pair contributes to the poles
independently and with a factor $I^{(1)}_{q\bar{q}}$ (or $-I^{(1)}_{q\bar{q}}$).
Hence at N$^k$LO, the structure in every layer is proportional to
$(I^{(1)}_{q\bar{q}})^k$ and the relative value of the coefficients between the
layers is entirely of combinatorial origin. Namely if the deepest poles of the
virtual layer are captured by ${\cal N}(I^{(1)}_{q\bar{q}})^k$, then the poles
of the $r$-fold real, $(k-r)$-fold virtual correction are given by
\begin{eqnarray}
 {\cal T}^{q\bar{q},(k)}_{{\cal I}}\Big|_{\rm abelian}
  &=& {\cal N}\binom{k}{r}\left(-I^{(1)}_{q\bar{q}}\right)^{r}\left(I^{(1)}_{q\bar{q}}\right)^{k-r}
  + \mathcal{O}(\e^{-2k+2})
  \nonumber \\ 
  &=& {\cal N}\binom{k}{r}(-1)^{r} \left(I^{(1)}_{q\bar{q}}\right)^k
  + \mathcal{O}(\e^{-2k+2})\,,
  \label{eq:binomial}
\end{eqnarray}
since there are $\binom{k}{r}$ ways to cut open $r$ of the $k$ loops
and substitute a real emission for a virtual correction.
In~\eqref{eq:binomial}, ${\cal N}$ is an overall normalisation factor common to
all layers at order $k$.

At NLO, \eqref{eq:binomial} trivially yields ${\cal N}$ and $-{\cal N}$ for the
virtual and the real corrections, while at NNLO, as one can see in
\eqref{eq:HbbX22_1}, \eqref{eq:HbbX22_2}, \eqref{eq:HbbX31} and
\eqref{eq:HbbX40}, the coefficients for the double-virtual, real-virtual and
double-real contributions are ${\cal N}$, $-2{\cal N}$ and ${\cal N}$.  Finally,
at N$^3$LO, the coefficients multiplying $(I^{(1)}_{q\bar{q}})^3$ are ${\cal
  N}$, $-3{\cal N}$, $3{\cal N}$ and $-{\cal N}$ for the VVV, VVR, VRR and RRR
contributions, as shown in Table~\ref{tab:polesHbb}, justifying the pair-wise
cancellation.
In~general, the cancellation of the deepest abelian poles at any order $k$ is
guaranteed by
\begin{equation}\label{eq:newton}
  \sum_{r=0}^{k} {\cal N} \binom{k}{r} \left(-I^{(1)}_{q\bar{q}}\right)^r\left(I^{(1)}_{q\bar{q}}\right)^{k-r}
  = {\cal N}\left(I^{(1)}_{q\bar{q}}-I^{(1)}_{q\bar{q}}\right)^k = 0\,.
\end{equation}
The second term in the previous equation reflects the exponentiation of multiple
abelian emissions. In other words, the cancellation of infrared singularities
proceeds independently for each emission.

With a similar combinatorial logic, one can predict the highest poles of the
individual $\ell_1 \times \ell_2$ terms within each layer with $\ell = \ell_1 +
\ell_2$ loops. One can show that in the purely abelian case the coefficient is
split among the loop configurations according to
\begin{equation}\label{eq:newtonloop}
  \frac{{\cal T}^{q\bar{q},(k,[\ell_1 \times \ell_2])}_{{\cal I}}\Big|_{\rm abelian}}%
       {{\cal T}^{q\bar{q},(k)}_{{\cal I}}\Big|_{\rm abelian}}
  = \frac{1}{2^{\ell-1}} \binom{\ell}{\ell_1} \left(\frac{1}{2}\right)^{\delta_{\ell_1,\ell_2}}
  + \mathcal{O}(\e^2) \,.
\end{equation}
This can be observed at NNLO in~\eqref{eq:HbbX22_1} and in~\eqref{eq:HbbX22_3},
and at N$^3$LO in~\eqref{eq:HbbT321} and in~\eqref{eq:Hbb330}, and
in~\eqref{eq:Hbb311} and in~\eqref{eq:Hbb320}.

The coefficients of the remaining colour factors for purely gluonic emissions, $N^2$ and $N^0$, are more
complicated due to the appearance of non-abelian effects. For the triple-virtual
contribution, the only colour structure present in the $\e^{-6}$ coefficients is
$\CF^2$, as indicated in the first line of Table~\ref{tab:polesHbb}, in
accordance with~\cite{Gehrmann:2014vha} once we account for the extra colour
factor of $2\CF$ due to our normalisation in~\eqref{eq:Hbb0}.
For the other layers, the colour factors $\CF\CA$ and $\CA^2$ also contribute to
the deepest singularities, indicating that the description of the poles
associated with real emission contains ingredients different from
$I^{(1)}_{q\bar{q}}$ already at $\e^{-6}$. An equivalent observation at NNLO was
the basis of the analysis of the infrared structure performed
in~\cite{Gehrmann-DeRidder:2004ttg}.
Additionally, the values of the coefficients in the $N^2$ contributions to the
three- and four-particle final state suggest the presence of an integrated
structure cancelling between these two layers.
For the $\NF N^{-1}$ colour factor, we see a complete cancellation between the
double-real-virtual and triple-real layers, explained by a single soft
gluon emission on top of a four-quark final state.
 \begin{table}[t]
   \centering
   \begin{tabular}{c c c c c c c}
     \toprule
     & Final-state $\mathcal{I}$ & 
     $N^2$ & $N^0$ & $N^{-2}$ & $\NF N$ & $\NF N^{-1}$ \\
     \midrule
     \midrule
     \multirow{1}{*}{{\footnotesize VVV}} & {\footnotesize $q\bar{q}$}
     & $-\frac{1}{6}\frac{1}{\e^6} -\frac{17}{8}\frac{1}{\e^5}$
     & $+\frac{1}{3}\frac{1}{\e^6}+\frac{23}{8}\frac{1}{\e^5}$
     & $-\frac{1}{6}\frac{1}{\e^6} -\frac{3}{4}\frac{1}{\e^5}$
     & $+\frac{1}{4}\frac{1}{\e^5}$ 
     & $-\frac{1}{4}\frac{1}{\e^5}$ 
     \\
     \midrule
     \multirow{1}{*}{{\footnotesize VVR}} & {\footnotesize $q\bar{q}g$}
     & $+\frac{29}{36}\frac{1}{\e^6} +\frac{1663}{216}\frac{1}{\e^5}$
     & $-\frac{5}{4}\frac{1}{\e^6} -\frac{53}{6}\frac{1}{\e^5}$
     & $+\frac{1}{2}\frac{1}{\e^6} +\frac{9}{4}\frac{1}{\e^5}$
     & $-\frac{20}{27}\frac{1}{\e^5}$
     & $+\frac{7}{12}\frac{1}{\e^5}$
     \\
     \midrule
     \multirow{2}{*}[-0.2em]{{\footnotesize VRR}} & {\footnotesize $q\bar{q}gg$}
     & $-\frac{41}{36}\frac{1}{\e^6} -\frac{311}{36}\frac{1}{\e^5}$
     & $+\frac{3}{2}\frac{1}{\e^6} +\frac{217}{24}\frac{1}{\e^5}$
     & $-\frac{1}{2}\frac{1}{\e^6} -\frac{9}{4}\frac{1}{\e^5}$
     & $+\frac{1}{2}\frac{1}{\e^5}$
     & $-\frac{1}{3}\frac{1}{\e^5}$
     \\
     \cmidrule{2-7}
     & {\footnotesize $q\bar{q}q'\bar{q}' + q\bar{q}q\bar{q}$}
     & 
     & 
     & 
     & $+\frac{13}{108}\frac{1}{\e^5}$
     & $-\frac{11}{108}\frac{1}{\e^5}$
     \\    
     \midrule
     \multirow{2}{*}[-0.2em]{{\footnotesize RRR}} & {\footnotesize $q\bar{q}ggg$}
     & $+\frac{1}{2}\frac{1}{\e^6} +\frac{311}{108}\frac{1}{\e^5}$
     & $-\frac{7}{12}\frac{1}{\e^6} -\frac{37}{12}\frac{1}{\e^5}$
     & $+\frac{1}{6}\frac{1}{\e^6} +\frac{3}{4}\frac{1}{\e^5}$
     & 
     & 
     \\
     \cmidrule{2-7}     
     & {\footnotesize $q\bar{q}q'\bar{q}'g + q\bar{q}q\bar{q}g$}
     & 
     & 
     & 
     & $-\frac{7}{54}\frac{1}{\e^5}$
     & $+\frac{11}{108}\frac{1}{\e^5}$
     \\  
     \bottomrule
     \end{tabular}
   \caption{Coefficients of $\e^{-6}$ and $\e^{-5}$ poles for different colour
     factors of $H \to b\bar{b}$ at N$^3$LO. Blank cells indicate vanishing
     coefficients for these poles. Note that an overall factor of $2\CF$ is
     factored out as indicated in~\eqref{eq:Hbb0}.}
   \label{tab:polesHbb}
 \end{table}

For $H\to gg$ in Table \ref{tab:polesHgg}, the rows representing final states
containing four quarks are left blank since they only exhibit poles starting
from $\e^{-4}$. This happens because the most infrared-divergent behaviour at
the triple-real level with four quarks in the final state is given by the
emission of two collinear quark-antiquark pairs with one of the pairs also soft,
or alternatively by the emission of two collinear quark-antiquark pairs in
association with a single soft gluon emission. Both of these configurations
yield at most $\e^{-4}$ poles. The triple-virtual correction does not have deep
poles in the $\NF$ and $\NF N^{-2}$ colour factors, so the poles can be
interpreted as NNLO-type corrections to the $q\bar{q}g$ final state on top of a
collinear quark-antiquark configuration. Including the $\e^{-4}$ poles for this
colour layer, given
in~\eqref{eq:HggX32_1},~\eqref{eq:HggX32_2},~\eqref{eq:HggX41}
and~\eqref{eq:HggX50}, the two deepest poles also follow the pattern ${\cal K}$,
$-2{\cal K}$ and ${\cal K}$ for the VVR, VRR and RRR respectively, where
\begin{equation}
	{\cal K} = -\frac{1}{9}\frac{1}{\e^5}-\frac{61}{54}\frac{1}{\e^4}\,.
\end{equation}
One can expect a structure similar to the abelian case above because this colour
factor receives contributions only from abelian gluon emission from the
quark-antiquark pair. However, the poles are not directly proportional to
$(I^{(1)}_{q\bar{q}})^2$ due to the additional integration over the
unresolved quark-antiquark pair.
For the VVR contribution, \eqref{eq:newtonloop} holds, as can be seen
in~\eqref{eq:HggX32_1} and~\eqref{eq:HggX32_2}.
In the $\NF$ colour factor we notice substantial cancellations between the four-
and five-particle final states. On the other hand, the $\NF N^2$ coefficients
exhibit a highly non-trivial interplay between real quark-antiquark pair
emissions and fermionic loops.

\begin{table}[t]
  \centering
  \begin{tabular}{c c c c c c}
    \toprule
    & Final-state $\mathcal{I}$ & 
    $N^3$ & $\NF N^2$ & $\NF$ & $\NF N^{-2}$ \\
    \midrule
    \midrule
    \multirow{1}{*}{{\footnotesize VVV}} & {\footnotesize $gg$}
    & $-\frac{4}{3}\frac{1}{\e^6} -\frac{77}{6}\frac{1}{\e^5}$
    & $+\frac{7}{3}\frac{1}{\e^5}$
    & 
    & 
    \\
    \midrule
    \multirow{2}{*}[-0.2em]{{\footnotesize VVR}} & {\footnotesize $ggg$}
    & $+\frac{46}{9}\frac{1}{\e^6} +\frac{4609}{108}\frac{1}{\e^5}$
    & $-\frac{305}{54}\frac{1}{\e^5}$
    & 
    & 
    \\ 
    \cmidrule{2-6}    
    & {\footnotesize $q\bar{q}g$}
    & 
    & $-\frac{4}{3}\frac{1}{\e^5}$
    & $+\frac{2}{3}\frac{1}{\e^5}$
    & $-\frac{1}{9}\frac{1}{\e^5}$
    \\
    \midrule
    \multirow{3}{*}[-0.4em]{{\footnotesize VRR}} & {\footnotesize $gggg$}
    & $-\frac{113}{18}\frac{1}{\e^6} -\frac{1661}{36}\frac{1}{\e^5}$
    & $+\frac{10}{3}\frac{1}{\e^5}$
    & 
    & 
    \\
    \cmidrule{2-6}        
    & {\footnotesize $q\bar{q}gg$}
    & 
    & $+\frac{92}{27}\frac{1}{\e^5}$
    & $-\frac{77}{54}\frac{1}{\e^5}$
    & $+\frac{2}{9}\frac{1}{\e^5}$
    \\
    \cmidrule{2-6}        
    & {\footnotesize $q\bar{q}q'\bar{q}' + q\bar{q}q\bar{q}$}
    & 
    & 
    & 
    & 
    \\    
    \midrule
    \multirow{3}{*}[-0.4em]{{\footnotesize RRR}} & {\footnotesize $ggggg$}
    & $+\frac{5}{2}\frac{1}{\e^6} +\frac{440}{27}\frac{1}{\e^5}$
    & 
    & 
    & 
    \\
    \cmidrule{2-6}        
    & {\footnotesize $q\bar{q}ggg$}
    & 
    & $-\frac{113}{54}\frac{1}{\e^5}$
    & $+\frac{41}{54}\frac{1}{\e^5}$
    & $-\frac{1}{9}\frac{1}{\e^5}$
    \\
    \cmidrule{2-6}        
    & {\footnotesize $q\bar{q}q'\bar{q}'g + q\bar{q}q\bar{q}g$}
    & 
    & 
    & 
    & 
    \\        
    \bottomrule
  \end{tabular}
  \caption{Coefficients of $\e^{-6}$ and $\e^{-5}$ poles for different colour
    factors of $H \to gg$ at N$^3$LO. Blank cells indicate vanishing
    coefficients for these poles.}
  \label{tab:polesHgg}
\end{table}

\section{Conclusions and outlook}
\label{sec:conclusions}

In this paper, we carried out the analytical integration over the inclusive
phase space for all the possible partonic channels for the decay of a Higgs
boson into a pair of gluons and a bottom quark-antiquark pair up to N$^3$LO.

This work is a natural continuation of~\cite{Jakubcik:2022zdi}, where an
analogous calculation was performed for the decay of a virtual photon into
hadrons. Remarkably, the deepest IR poles of photon and Higgs decay to quarks
coincide completely for any partonic final state. We observe a similar agreement
also in the highest transcendental weight numbers in the coefficients of all
powers of $\e$.

These results constitute an important contribution to the study of unresolved
QCD radiation at the integrated level. In the context of the antenna subtraction
method, they are necessary for the extraction of N$^3$LO gluon-gluon antenna
functions in final-final kinematics.

The set of integrated antenna functions at N$^3$LO ought to be completed with
analogous results for a quark-gluon pair of hard radiators.
At NNLO, the expressions were extracted from QCD corrections to a neutralino
decay~\cite{GehrmannDeRidder:2005hi}.
We foresee extending this analysis to N$^3$LO in a forthcoming publication.

\section*{Acknowledgements}
We are indebted to Thomas Gehrmann and Nigel Glover for their feedback and encouragement
to pursue this work.
We thank Oscar Braun-White, Christian Preuss, Kay Sch\"onwald, Vasily
Sotnikov and Tong-Zhi Yang for elucidating discussions and suggestions on the
manuscript.
This work was supported by the Swiss National Science Foundation (SNF) under
contract 200020-204200 and by the European Research Council (ERC) under the
European Union's Horizon 2020 research and innovation programme grant agreement
101019620 (ERC Advanced Grant TOPUP).

\appendices

\section{N$^3$LO results}\label{app:exprN3LO}

\begingroup
\allowdisplaybreaks

\subsection{Higgs to bottom quarks}

\subsubsection{Two-particle final states}


\begin{flalign}\label{eq:Hbbstart}
{\cal T}^{q\bar{q},(3,\left[2\times 1\right])}_{q\bar{q}}\Big|_{N^{2}} = &
+\frac{1}{\e^6}\left(-\frac{1}{8}\right)
+\frac{1}{\e^5}\left(-\frac{5}{4}\right)
+\frac{1}{\e^4}\left(-\frac{89}{36}+\frac{7}{96}\pi^2\right)
\nonumber &\\&
+\frac{1}{\e^3}\left(-\frac{1865}{864}+\frac{83}{192}\pi^2\right)
\nonumber &\\&
+\frac{1}{\e^2}\left(-\frac{45169}{10368}+\frac{3155}{3456}\pi^2+\frac{97}{36}\zeta_{3}-\frac{11}{768}\pi^4\right)
\nonumber &\\&
+\frac{1}{\e}\left(-\frac{174337}{15552}-\frac{23893}{20736}\pi^2-\frac{685}{288}\zeta_{3}-\frac{1619}{23040}\pi^4+\frac{11}{48}\pi^2\zeta_{3}-\frac{33}{10}\zeta_{5}\right)
\nonumber &\\&
-\frac{548719}{23328}+\frac{8561}{31104}\pi^2-\frac{22783}{1728}\zeta_{3}+\frac{3179}{138240}\pi^4
\nonumber &\\&
+\frac{3859}{1728}\pi^2\zeta_{3}-\frac{56}{15}\zeta_{5}-\frac{389}{193536}\pi^6+\frac{95}{16}\zeta_{3}^2 + \mathcal{O}(\e), &
\end{flalign}

\begin{flalign}
{\cal T}^{q\bar{q},(3,\left[2\times 1\right])}_{q\bar{q}}\Big|_{N^{0}} = &
+\frac{1}{\e^6}\left(\frac{1}{4}\right)
+\frac{1}{\e^5}\left(\frac{29}{16}\right)
+\frac{1}{\e^4}\left(\frac{1063}{288}-\frac{\pi^2}{6}\right)
\nonumber &\\&
+\frac{1}{\e^3}\left(\frac{3809}{864}-\frac{161}{192}\pi^2-\frac{13}{8}\zeta_{3}\right)
\nonumber &\\&
+\frac{1}{\e^2}\left(\frac{24637}{2592}-\frac{77}{54}\pi^2-\frac{599}{72}\zeta_{3}+\frac{59}{5760}\pi^4\right)
\nonumber &\\&
+\frac{1}{\e}\left(\frac{5902}{243}+\frac{2335}{5184}\pi^2-\frac{469}{36}\zeta_{3}+\frac{3047}{23040}\pi^4+\frac{37}{32}\pi^2\zeta_{3}+\frac{9}{40}\zeta_{5}\right)
\nonumber &\\&
+\frac{336379}{5832}-\frac{29917}{15552}\pi^2-\frac{19637}{432}\zeta_{3}-\frac{29719}{69120}\pi^4
\nonumber &\\&
+\frac{6239}{1728}\pi^2\zeta_{3}-\frac{1469}{120}\zeta_{5}+\frac{6647}{241920}\pi^6+\frac{125}{8}\zeta_{3}^2 + \mathcal{O}(\e), &
\end{flalign}

\begin{flalign}\label{eq:HbbT321}
{\cal T}^{q\bar{q},(3,\left[2\times 1\right])}_{q\bar{q}}\Big|_{N^{-2}} = &
+\frac{1}{\e^6}\left(-\frac{1}{8}\right)
+\frac{1}{\e^5}\left(-\frac{9}{16}\right)
+\frac{1}{\e^4}\left(-\frac{39}{32}+\frac{3}{32}\pi^2\right)
\nonumber &\\&
+\frac{1}{\e^3}\left(-\frac{9}{4}+\frac{13}{32}\pi^2+\frac{13}{8}\zeta_{3}\right)
\nonumber &\\&
+\frac{1}{\e^2}\left(-\frac{659}{128}+\frac{197}{384}\pi^2+\frac{45}{8}\zeta_{3}+\frac{47}{11520}\pi^4\right)
\nonumber &\\&
+\frac{1}{\e}\left(-\frac{837}{64}+\frac{539}{768}\pi^2+\frac{493}{32}\zeta_{3}-\frac{119}{1920}\pi^4-\frac{133}{96}\pi^2\zeta_{3}+\frac{123}{40}\zeta_{5}\right)
\nonumber &\\&
-\frac{1093}{32}+\frac{211}{128}\pi^2+\frac{3753}{64}\zeta_{3}+\frac{6251}{15360}\pi^4
\nonumber &\\&
-\frac{187}{32}\pi^2\zeta_{3}+\frac{639}{40}\zeta_{5}-\frac{24643}{967680}\pi^6-\frac{345}{16}\zeta_{3}^2 + \mathcal{O}(\e), &
\end{flalign}

\begin{flalign}
{\cal T}^{q\bar{q},(3,\left[2\times 1\right])}_{q\bar{q}}\Big|_{\NF N} = &
+\frac{1}{\e^5}\left(\frac{1}{8}\right)
+\frac{1}{\e^4}\left(\frac{35}{144}\right)
+\frac{1}{\e^3}\left(\frac{25}{432}-\frac{\pi^2}{96}\right)
\nonumber &\\&
+\frac{1}{\e^2}\left(-\frac{593}{2592}-\frac{65}{864}\pi^2-\frac{23}{72}\zeta_{3}\right)
\nonumber &\\&
+\frac{1}{\e}\left(-\frac{4913}{3888}+\frac{1567}{5184}\pi^2+\frac{31}{72}\zeta_{3}-\frac{7}{2304}\pi^4\right)
\nonumber &\\&
-\frac{33425}{5832}+\frac{1141}{1944}\pi^2+\frac{1033}{432}\zeta_{3}+\frac{71}{4320}\pi^4
\nonumber &\\&
-\frac{53}{864}\pi^2\zeta_{3}+\frac{5}{24}\zeta_{5} + \mathcal{O}(\e), &
\end{flalign}

\begin{flalign}
{\cal T}^{q\bar{q},(3,\left[2\times 1\right])}_{q\bar{q}}\Big|_{\NF N^{-1}} = &
+\frac{1}{\e^5}\left(-\frac{1}{8}\right)
+\frac{1}{\e^4}\left(-\frac{35}{144}\right)
+\frac{1}{\e^3}\left(-\frac{25}{432}+\frac{\pi^2}{96}\right)
\nonumber &\\&
+\frac{1}{\e^2}\left(\frac{593}{2592}+\frac{65}{864}\pi^2+\frac{23}{72}\zeta_{3}\right)
\nonumber &\\&
+\frac{1}{\e}\left(\frac{4913}{3888}-\frac{1567}{5184}\pi^2-\frac{31}{72}\zeta_{3}+\frac{7}{2304}\pi^4\right)
\nonumber &\\&
+\frac{33425}{5832}-\frac{1141}{1944}\pi^2-\frac{1033}{432}\zeta_{3}-\frac{71}{4320}\pi^4
\nonumber &\\&
+\frac{53}{864}\pi^2\zeta_{3}-\frac{5}{24}\zeta_{5} + \mathcal{O}(\e), &
\end{flalign}



\begin{flalign}
{\cal T}^{q\bar{q},(3,\left[3\times 0\right])}_{q\bar{q}}\Big|_{N^{2}} = &
+\frac{1}{\e^6}\left(-\frac{1}{24}\right)
+\frac{1}{\e^5}\left(-\frac{7}{8}\right)
+\frac{1}{\e^4}\left(-\frac{4813}{1296}+\frac{17}{96}\pi^2\right) \nonumber &\\&
+\frac{1}{\e^3}\left(\frac{3293}{3888}+\frac{6229}{5184}\pi^2-\frac{7}{12}\zeta_{3}\right)
\nonumber &\\&
+\frac{1}{\e^2}\left(\frac{64441}{31104}-\frac{21731}{31104}\pi^2-\frac{151}{108}\zeta_{3}-\frac{2299}{20736}\pi^4\right)
\nonumber &\\&
+\frac{1}{\e}\left(-\frac{929183}{139968}-\frac{400709}{186624}\pi^2+\frac{3727}{864}\zeta_{3}-\frac{18307}{207360}\pi^4+\frac{151}{54}\pi^2\zeta_{3}-\frac{631}{60}\zeta_{5}\right)
\nonumber &\\&
+\frac{4215335}{209952}+\frac{1307401}{279936}\pi^2+\frac{59365}{15552}\zeta_{3}+\frac{1134673}{1244160}\pi^4
\nonumber &\\&
-\frac{175}{576}\pi^2\zeta_{3}-\frac{53}{10}\zeta_{5}+\frac{208037}{3732480}\pi^6-\frac{163}{144}\zeta_{3}^2 + \mathcal{O}(\e), &
\end{flalign}

\begin{flalign}
{\cal T}^{q\bar{q},(3,\left[3\times 0\right])}_{q\bar{q}}\Big|_{N^{0}} = &
+\frac{1}{\e^6}\left(\frac{1}{12}\right)
+\frac{1}{\e^5}\left(\frac{17}{16}\right)
+\frac{1}{\e^4}\left(\frac{595}{288}-\frac{3}{8}\pi^2\right)
\nonumber &\\&
+\frac{1}{\e^3}\left(\frac{1163}{864}-\frac{313}{192}\pi^2-\frac{11}{24}\zeta_{3}\right)
\nonumber &\\&
+\frac{1}{\e^2}\left(\frac{224}{81}+\frac{535}{864}\pi^2-\frac{479}{72}\zeta_{3}+\frac{1573}{5760}\pi^4\right)
\nonumber &\\&
+\frac{1}{\e}\Bigg(\frac{161377}{15552}+\frac{1879}{5184}\pi^2+\frac{3}{8}\zeta_{3}-\frac{20657}{207360}\pi^4
\nonumber &\\&
\phantom{+\frac{1}{\e}\Bigg(}+\frac{593}{288}\pi^2\zeta_{3}+\frac{33}{40}\zeta_{5}\Bigg)
\nonumber &\\&
+\frac{112981}{5832}-\frac{37219}{15552}\pi^2+\frac{1067}{48}\zeta_{3}-\frac{920941}{622080}\pi^4
\nonumber &\\&
+\frac{679}{192}\pi^2\zeta_{3}-\frac{6409}{360}\zeta_{5}-\frac{35857}{1088640}\pi^6+\frac{35}{8}\zeta_{3}^2 + \mathcal{O}(\e), &
\end{flalign}

\begin{flalign}\label{eq:Hbb330}
{\cal T}^{q\bar{q},(3,\left[3\times 0\right])}_{q\bar{q}}\Big|_{N^{-2}} = &
+\frac{1}{\e^6}\left(-\frac{1}{24}\right)
+\frac{1}{\e^5}\left(-\frac{3}{16}\right)
+\frac{1}{\e^4}\left(-\frac{13}{32}+\frac{19}{96}\pi^2\right)
\nonumber &\\&
+\frac{1}{\e^3}\left(-\frac{23}{32}+\frac{11}{32}\pi^2+\frac{25}{24}\zeta_{3}\right)
\nonumber &\\&
+\frac{1}{\e^2}\left(-\frac{235}{128}+\frac{191}{384}\pi^2+\frac{31}{8}\zeta_{3}-\frac{649}{3840}\pi^4\right)
\nonumber &\\&
+\frac{1}{\e}\left(-\frac{275}{48}+\frac{899}{768}\pi^2+\frac{1117}{96}\zeta_{3}+\frac{287}{1920}\pi^4-\frac{1457}{288}\pi^2\zeta_{3}+\frac{161}{40}\zeta_{5}\right)
\nonumber &\\&
-\frac{265}{24}+\frac{3635}{1152}\pi^2+\frac{1877}{64}\zeta_{3}+\frac{4495}{9216}\pi^4
\nonumber &\\&
-\frac{383}{32}\pi^2\zeta_{3}+\frac{983}{40}\zeta_{5}-\frac{19301}{1741824}\pi^6-\frac{913}{48}\zeta_{3}^2 + \mathcal{O}(\e), &
\end{flalign}

\begin{flalign}
{\cal T}^{q\bar{q},(3,\left[3\times 0\right])}_{q\bar{q}}\Big|_{\NF N} = &
+\frac{1}{\e^5}\left(\frac{1}{8}\right)
+\frac{1}{\e^4}\left(\frac{1283}{1296}\right)
+\frac{1}{\e^3}\left(-\frac{1711}{3888}-\frac{419}{2592}\pi^2\right)
\nonumber &\\&
+\frac{1}{\e^2}\left(-\frac{10195}{7776}+\frac{839}{7776}\pi^2+\frac{47}{216}\zeta_{3}\right)
\nonumber &\\&
+\frac{1}{\e}\left(-\frac{43339}{69984}+\frac{39491}{46656}\pi^2-\frac{301}{648}\zeta_{3}+\frac{3941}{103680}\pi^4\right)
\nonumber &\\&
-\frac{3396275}{419904}-\frac{36857}{139968}\pi^2-\frac{2029}{1296}\zeta_{3}-\frac{583}{4320}\pi^4
\nonumber &\\&
+\frac{41}{288}\pi^2\zeta_{3}-\frac{205}{72}\zeta_{5} + \mathcal{O}(\e), &
\end{flalign}

\begin{flalign}
{\cal T}^{q\bar{q},(3,\left[3\times 0\right])}_{q\bar{q}}\Big|_{\NF N^{-1}} = &
+\frac{1}{\e^5}\left(-\frac{1}{8}\right)
+\frac{1}{\e^4}\left(-\frac{35}{144}\right)
+\frac{1}{\e^3}\left(\frac{23}{432}+\frac{17}{96}\pi^2\right)
\nonumber &\\&
+\frac{1}{\e^2}\left(\frac{641}{2592}-\frac{199}{864}\pi^2+\frac{55}{72}\zeta_{3}\right)
\nonumber &\\&
+\frac{1}{\e}\left(\frac{967}{7776}-\frac{3235}{5184}\pi^2-\frac{221}{216}\zeta_{3}-\frac{577}{20736}\pi^4\right)
\nonumber &\\&
+\frac{145375}{46656}-\frac{22571}{15552}\pi^2-\frac{15131}{1296}\zeta_{3}+\frac{4411}{38880}\pi^4
\nonumber &\\&
+\frac{85}{96}\pi^2\zeta_{3}+\frac{193}{72}\zeta_{5} + \mathcal{O}(\e), &
\end{flalign}

\begin{flalign}
{\cal T}^{q\bar{q},(3,\left[3\times 0\right])}_{q\bar{q}}\Big|_{\NF^2} = &
+\frac{1}{\e^4}\left(-\frac{11}{162}\right)
+\frac{1}{\e^3}\left(-\frac{1}{243}\right)
+\frac{1}{\e^2}\left(\frac{23}{324}+\frac{\pi^2}{108}\right)
\nonumber &\\&
+\frac{1}{\e}\left(\frac{2417}{17496}-\frac{5}{324}\pi^2-\frac{\zeta_{3}}{81}\right)
\nonumber &\\&
+\frac{259}{6561}+\frac{97}{972}\pi^2-\frac{25}{243}\zeta_{3}+\frac{43}{9720}\pi^4 + \mathcal{O}(\e), &
\end{flalign}


\subsubsection{Three-particle final states}


\begin{flalign}
{\cal T}^{q\bar{q},(3,\left[1\times 1\right])}_{q\bar{q}g}\Big|_{N^{2}} = &
+\frac{1}{\e^6}\left(\frac{29}{72}\right)
+\frac{1}{\e^5}\left(\frac{71}{24}\right)
+\frac{1}{\e^4}\left(\frac{1681}{144}-\frac{373}{864}\pi^2\right)
\nonumber &\\&
+\frac{1}{\e^3}\left(\frac{2231}{48}-\frac{367}{96}\pi^2-\frac{685}{72}\zeta_{3}\right)
\nonumber &\\&
+\frac{1}{\e^2}\left(\frac{6565}{32}-\frac{23713}{1728}\pi^2-\frac{4135}{72}\zeta_{3}-\frac{2737}{34560}\pi^4\right)
\nonumber &\\&
+\frac{1}{\e}\Bigg(\frac{533087}{576}-\frac{36157}{576}\pi^2-\frac{111853}{432}\zeta_{3}+\frac{4807}{11520}\pi^4
\nonumber &\\&
\phantom{+\frac{1}{\e}\Bigg(}+\frac{9749}{864}\pi^2\zeta_{3}-\frac{12349}{120}\zeta_{5}\bigg)
\nonumber &\\&
+\frac{1644281}{384}-\frac{332527}{1152}\pi^2-\frac{180955}{144}\zeta_{3}-\frac{182003}{207360}\pi^4
\nonumber &\\&
+\frac{22273}{288}\pi^2\zeta_{3}-\frac{70603}{120}\zeta_{5}-\frac{94961}{967680}\pi^6+\frac{18773}{144}\zeta_{3}^2 + \mathcal{O}(\e), &
\end{flalign}

\begin{flalign}
{\cal T}^{q\bar{q},(3,\left[1\times 1\right])}_{q\bar{q}g}\Big|_{N^{0}} = &
+\frac{1}{\e^6}\left(-\frac{5}{8}\right)
+\frac{1}{\e^5}\left(-\frac{179}{48}\right)
+\frac{1}{\e^4}\left(-\frac{631}{48}+\frac{199}{288}\pi^2\right)
\nonumber &\\&
+\frac{1}{\e^3}\left(-\frac{4975}{96}+\frac{695}{144}\pi^2+\frac{137}{8}\zeta_{3}\right)
\nonumber &\\&
+\frac{1}{\e^2}\left(-\frac{5299}{24}+\frac{2227}{144}\pi^2+\frac{5729}{72}\zeta_{3}+\frac{6817}{34560}\pi^4\right)
\nonumber &\\&
+\frac{1}{\e}\left(-\frac{186863}{192}+\frac{1107}{16}\pi^2+\frac{7607}{24}\zeta_{3}-\frac{1981}{6912}\pi^4-\frac{7411}{288}\pi^2\zeta_{3}+\frac{1903}{8}\zeta_{5}\right)
\nonumber &\\&
-\frac{1701065}{384}+\frac{356963}{1152}\pi^2+\frac{54101}{36}\zeta_{3}+\frac{66763}{34560}\pi^4
\nonumber &\\&
-\frac{107867}{864}\pi^2\zeta_{3}+\frac{39227}{40}\zeta_{5}+\frac{607987}{2903040}\pi^6-\frac{5805}{16}\zeta_{3}^2 + \mathcal{O}(\e), &
\end{flalign}

\begin{flalign}\label{eq:Hbb311}
{\cal T}^{q\bar{q},(3,\left[1\times 1\right])}_{q\bar{q}g}\Big|_{N^{-2}} = &
+\frac{1}{\e^6}\left(\frac{1}{4}\right)
+\frac{1}{\e^5}\left(\frac{9}{8}\right)
+\frac{1}{\e^4}\left(\frac{61}{16}-\frac{13}{48}\pi^2\right)
\nonumber &\\&
+\frac{1}{\e^3}\left(\frac{469}{32}-\frac{45}{32}\pi^2-\frac{85}{12}\zeta_{3}\right)
\nonumber &\\&
+\frac{1}{\e^2}\left(\frac{481}{8}-\frac{817}{192}\pi^2-\frac{191}{8}\zeta_{3}-\frac{41}{384}\pi^4\right)
\nonumber &\\&
+\frac{1}{\e}\left(\frac{24727}{96}-\frac{439}{24}\pi^2-\frac{4279}{48}\zeta_{3}+\frac{29}{768}\pi^4+\frac{1505}{144}\pi^2\zeta_{3}-\frac{2113}{20}\zeta_{5}\right)
\nonumber &\\&
+\frac{440579}{384}-\frac{5059}{64}\pi^2-\frac{1211}{3}\zeta_{3}-\frac{16933}{23040}\pi^4
\nonumber &\\&
+\frac{1159}{32}\pi^2\zeta_{3}-\frac{11721}{40}\zeta_{5}-\frac{161467}{1451520}\pi^6+\frac{1265}{8}\zeta_{3}^2 + \mathcal{O}(\e), &
\end{flalign}

\begin{flalign}
{\cal T}^{q\bar{q},(3,\left[1\times 1\right])}_{q\bar{q}g}\Big|_{\NF N} = &
+\frac{1}{\e^5}\left(-\frac{5}{24}\right)
+\frac{1}{\e^4}\left(-\frac{67}{72}\right)
+\frac{1}{\e^3}\left(-\frac{125}{48}+\frac{13}{48}\pi^2\right)
\nonumber &\\&
+\frac{1}{\e^2}\left(-\frac{157}{16}+\frac{365}{432}\pi^2+\frac{55}{18}\zeta_{3}\right)
\nonumber &\\&
+\frac{1}{\e}\left(-\frac{5203}{144}+\frac{865}{288}\pi^2+\frac{601}{54}\zeta_{3}-\frac{41}{576}\pi^4\right)
\nonumber &\\&
-\frac{406}{3}+\frac{3389}{288}\pi^2+\frac{315}{8}\zeta_{3}-\frac{2611}{25920}\pi^4
\nonumber &\\&
-\frac{149}{36}\pi^2\zeta_{3}+\frac{143}{6}\zeta_{5} + \mathcal{O}(\e), &
\end{flalign}

\begin{flalign}
{\cal T}^{q\bar{q},(3,\left[1\times 1\right])}_{q\bar{q}g}\Big|_{\NF N^{-1}} = &
+\frac{1}{\e^5}\left(\frac{1}{6}\right)
+\frac{1}{\e^4}\left(\frac{1}{2}\right)
+\frac{1}{\e^3}\left(\frac{73}{48}-\frac{2}{9}\pi^2\right)
\nonumber &\\&
+\frac{1}{\e^2}\left(\frac{131}{24}-\frac{23}{48}\pi^2-\frac{53}{18}\zeta_{3}\right)
\nonumber &\\&
+\frac{1}{\e}\left(\frac{1879}{96}-\frac{85}{48}\pi^2-\frac{13}{2}\zeta_{3}+\frac{19}{432}\pi^4\right)
\nonumber &\\&
+\frac{13763}{192}-\frac{1261}{192}\pi^2-\frac{863}{36}\zeta_{3}+\frac{103}{1920}\pi^4
\nonumber &\\&
+\frac{493}{108}\pi^2\zeta_{3}-\frac{299}{10}\zeta_{5} + \mathcal{O}(\e), &
\end{flalign}

\begin{flalign}
{\cal T}^{q\bar{q},(3,\left[1\times 1\right])}_{q\bar{q}g}\Big|_{\NF^2 } = &
+\frac{1}{\e^4}\left(\frac{1}{36}\right)
+\frac{1}{\e^3}\left(\frac{1}{24}\right)
+\frac{1}{\e^2}\left(\frac{7}{48}-\frac{7}{432}\pi^2\right)
\nonumber &\\&
+\frac{1}{\e}\left(\frac{127}{288}-\frac{7}{288}\pi^2-\frac{25}{108}\zeta_{3}\right)
\nonumber &\\&
+\frac{85}{64}-\frac{49}{576}\pi^2-\frac{25}{72}\zeta_{3}-\frac{71}{51840}\pi^4 + \mathcal{O}(\e), &
\end{flalign}



\begin{flalign}
{\cal T}^{q\bar{q},(3,\left[2\times 0\right])}_{q\bar{q}g}\Big|_{N^{2}} = &
+\frac{1}{\e^6}\left(\frac{29}{72}\right)
+\frac{1}{\e^5}\left(\frac{128}{27}\right)
+\frac{1}{\e^4}\left(\frac{21907}{1296}-\frac{343}{288}\pi^2\right)
\nonumber &\\&
+\frac{1}{\e^3}\left(\frac{455953}{7776}-\frac{18173}{2592}\pi^2-\frac{557}{72}\zeta_{3}\right)
\nonumber &\\&
+\frac{1}{\e^2}\left(\frac{11841379}{46656}-\frac{393785}{15552}\pi^2-\frac{5489}{72}\zeta_{3}+\frac{99149}{103680}\pi^4\right)
\nonumber &\\&
+\frac{1}{\e}\Bigg(\frac{300578239}{279936}-\frac{10451453}{93312}\pi^2-\frac{123899}{432}\zeta_{3}+\frac{55093}{20736}\pi^4
\nonumber &\\&
\phantom{+\frac{1}{\e}\Bigg(}+\frac{20885}{864}\pi^2\zeta_{3}-\frac{31687}{360}\zeta_{5}\Bigg)
\nonumber &\\&
+\frac{7825448299}{1679616}-\frac{287499737}{559872}\pi^2-\frac{3288509}{2592}\zeta_{3}+\frac{7339967}{622080}\pi^4
\nonumber &\\&
+\frac{113437}{864}\pi^2\zeta_{3}-\frac{92413}{120}\zeta_{5}-\frac{5842331}{26127360}\pi^6+\frac{14893}{144}\zeta_{3}^2 + \mathcal{O}(\e), &
\end{flalign}

\begin{flalign}
{\cal T}^{q\bar{q},(3,\left[2\times 0\right])}_{q\bar{q}g}\Big|_{N^{0}} = &
+\frac{1}{\e^6}\left(-\frac{5}{8}\right)
+\frac{1}{\e^5}\left(-\frac{245}{48}\right)
+\frac{1}{\e^4}\left(-\frac{1127}{72}+\frac{547}{288}\pi^2\right)
\nonumber &\\&
+\frac{1}{\e^3}\left(-\frac{12541}{216}+\frac{6911}{864}\pi^2+\frac{135}{8}\zeta_{3}\right)
\nonumber &\\&
+\frac{1}{\e^2}\left(-\frac{1276453}{5184}+\frac{141881}{5184}\pi^2+\frac{7249}{72}\zeta_{3}-\frac{17101}{11520}\pi^4\right)
\nonumber &\\&
+\frac{1}{\e}\Bigg(-\frac{32867617}{31104}+\frac{3792287}{31104}\pi^2+\frac{37321}{108}\zeta_{3}-\frac{30331}{10368}\pi^4
\nonumber &\\&
\phantom{+\frac{1}{\e}\Bigg(}-\frac{16825}{288}\pi^2\zeta_{3}+\frac{5579}{24}\zeta_{5}\Bigg)
\nonumber &\\&
-\frac{875425903}{186624}+\frac{104433767}{186624}\pi^2+\frac{2151911}{1296}\zeta_{3}-\frac{1784969}{124416}\pi^4
\nonumber &\\&
-\frac{58523}{288}\pi^2\zeta_{3}+\frac{124961}{120}\zeta_{5}+\frac{2075989}{8709120}\pi^6-\frac{17695}{48}\zeta_{3}^2 + \mathcal{O}(\e), &
\end{flalign}

\begin{flalign}\label{eq:Hbb320}
{\cal T}^{q\bar{q},(3,\left[2\times 0\right])}_{q\bar{q}g}\Big|_{N^{-2}} = &
+\frac{1}{\e^6}\left(\frac{1}{4}\right)
+\frac{1}{\e^5}\left(\frac{9}{8}\right)
+\frac{1}{\e^4}\left(\frac{61}{16}-\frac{37}{48}\pi^2\right)
\nonumber &\\&
+\frac{1}{\e^3}\left(\frac{233}{16}-\frac{65}{32}\pi^2-\frac{103}{12}\zeta_{3}\right)
\nonumber &\\&
+\frac{1}{\e^2}\left(\frac{3871}{64}-\frac{511}{64}\pi^2-\frac{239}{8}\zeta_{3}+\frac{3289}{5760}\pi^4\right)
\nonumber &\\&
+\frac{1}{\e}\Bigg(\frac{100315}{384}-\frac{38461}{1152}\pi^2-\frac{1863}{16}\zeta_{3}+\frac{7363}{11520}\pi^4
\nonumber &\\&
\phantom{+\frac{1}{\e}\Bigg(}+\frac{4211}{144}\pi^2\zeta_{3}-\frac{6449}{60}\zeta_{5}\Bigg)
\nonumber &\\&
+\frac{902531}{768}-\frac{342539}{2304}\pi^2-\frac{50483}{96}\zeta_{3}+\frac{16631}{4608}\pi^4
\nonumber &\\&
+\frac{6949}{96}\pi^2\zeta_{3}-\frac{13321}{40}\zeta_{5}+\frac{14593}{622080}\pi^6+\frac{4943}{24}\zeta_{3}^2 + \mathcal{O}(\e), &
\end{flalign}

\begin{flalign}
{\cal T}^{q\bar{q},(3,\left[2\times 0\right])}_{q\bar{q}g}\Big|_{\NF N} = &
+\frac{1}{\e^5}\left(-\frac{115}{216}\right)
+\frac{1}{\e^4}\left(-\frac{1439}{648}\right)
+\frac{1}{\e^3}\left(-\frac{20165}{3888}+\frac{739}{1296}\pi^2\right)
\nonumber &\\&
+\frac{1}{\e^2}\left(-\frac{116501}{5832}+\frac{1457}{972}\pi^2+\frac{64}{9}\zeta_{3}\right)
\nonumber &\\&
+\frac{1}{\e}\left(-\frac{4655821}{69984}+\frac{11455}{2916}\pi^2+\frac{2435}{108}\zeta_{3}-\frac{193}{3240}\pi^4\right)
\nonumber &\\&
-\frac{89920699}{419904}+\frac{1668419}{139968}\pi^2+\frac{9539}{162}\zeta_{3}+\frac{43861}{155520}\pi^4
\nonumber &\\&
-\frac{1501}{216}\pi^2\zeta_{3}+\frac{190}{3}\zeta_{5} + \mathcal{O}(\e), &
\end{flalign}

\begin{flalign}
{\cal T}^{q\bar{q},(3,\left[2\times 0\right])}_{q\bar{q}g}\Big|_{\NF N^{-1}} = &
+\frac{1}{\e^5}\left(\frac{5}{12}\right)
+\frac{1}{\e^4}\left(\frac{71}{72}\right)
+\frac{1}{\e^3}\left(\frac{293}{108}-\frac{193}{432}\pi^2\right)
\nonumber &\\&
+\frac{1}{\e^2}\left(\frac{25271}{2592}-\frac{1543}{2592}\pi^2-\frac{227}{36}\zeta_{3}\right)
\nonumber &\\&
+\frac{1}{\e}\left(\frac{500171}{15552}-\frac{28375}{15552}\pi^2-\frac{2033}{216}\zeta_{3}+\frac{1661}{51840}\pi^4\right)
\nonumber &\\&
+\frac{9786161}{93312}-\frac{486667}{93312}\pi^2-\frac{37751}{1296}\zeta_{3}-\frac{44509}{311040}\pi^4
\nonumber &\\&
+\frac{775}{144}\pi^2\zeta_{3}-\frac{3019}{60}\zeta_{5} + \mathcal{O}(\e), &
\end{flalign}

\begin{flalign}
{\cal T}^{q\bar{q},(3,\left[2\times 0\right])}_{q\bar{q}g}\Big|_{\NF^2} = &
+\frac{1}{\e^4}\left(\frac{1}{12}\right)
+\frac{1}{\e^3}\left(\frac{1}{8}\right)
+\frac{1}{\e^2}\left(\frac{7}{16}-\frac{7}{144}\pi^2\right)
\nonumber &\\&
+\frac{1}{\e}\left(\frac{127}{96}-\frac{7}{96}\pi^2-\frac{25}{36}\zeta_{3}\right)
\nonumber &\\&
+\frac{255}{64}-\frac{49}{192}\pi^2-\frac{25}{24}\zeta_{3}-\frac{71}{17280}\pi^4 + \mathcal{O}(\e), &
\end{flalign}


\subsubsection{Four-particle final states}


\begin{flalign}
{\cal T}^{q\bar{q},(3)}_{q\bar{q}gg}\Big|_{N^{2}} = &
+\frac{1}{\e^6}\left(-\frac{41}{36}\right)
+\frac{1}{\e^5}\left(-\frac{311}{36}\right)
+\frac{1}{\e^4}\left(-\frac{54325}{1296}+\frac{1151}{432}\pi^2\right)
\nonumber &\\&
+\frac{1}{\e^3}\left(-\frac{1695641}{7776}+\frac{45337}{2592}\pi^2+\frac{380}{9}\zeta_{3}\right)
\nonumber &\\&
+\frac{1}{\e^2}\left(-\frac{8560249}{7776}+\frac{358123}{3888}\pi^2+\frac{33013}{108}\zeta_{3}-\frac{16537}{10368}\pi^4\right)
\nonumber &\\&
+\frac{1}{\e}\Bigg(-\frac{772552649}{139968}+\frac{45794413}{93312}\pi^2+\frac{543997}{324}\zeta_{3}-\frac{28033}{3840}\pi^4
\nonumber &\\&
\phantom{+\frac{1}{\e}\Bigg(}-\frac{7435}{72}\pi^2\zeta_{3}+\frac{40319}{90}\zeta_{5}\Bigg)
\nonumber &\\&
-\frac{5789936701}{209952}+\frac{1424319259}{559872}\pi^2+\frac{23462453}{2592}\zeta_{3}-\frac{12003187}{311040}\pi^4
\nonumber &\\&
-\frac{65477}{96}\pi^2\zeta_{3}+\frac{166594}{45}\zeta_{5}+\frac{4433837}{13063680}\pi^6-\frac{64345}{72}\zeta_{3}^2 + \mathcal{O}(\e), &
\end{flalign}

\begin{flalign}
{\cal T}^{q\bar{q},(3)}_{q\bar{q}gg}\Big|_{N^{0}} = &
+\frac{1}{\e^6}\left(\frac{3}{2}\right)
+\frac{1}{\e^5}\left(\frac{217}{24}\right)
+\frac{1}{\e^4}\left(\frac{2995}{72}-\frac{65}{18}\pi^2\right)
\nonumber &\\&
+\frac{1}{\e^3}\left(\frac{180305}{864}-\frac{4061}{216}\pi^2-\frac{761}{12}\zeta_{3}\right)
\nonumber &\\&
+\frac{1}{\e^2}\left(\frac{5340263}{5184}-\frac{483053}{5184}\pi^2-\frac{24763}{72}\zeta_{3}+\frac{16579}{8640}\pi^4\right)
\nonumber &\\&
+\frac{1}{\e}\Bigg(\frac{158279429}{31104}-\frac{14891159}{31104}\pi^2-\frac{252241}{144}\zeta_{3}+\frac{65657}{8640}\pi^4
\nonumber &\\&
\phantom{+\frac{1}{\e}\Bigg(}+\frac{23987}{144}\pi^2\zeta_{3}-\frac{51389}{60}\zeta_{5}\Bigg)
\nonumber &\\&
+\frac{4700501531}{186624}-\frac{452639681}{186624}\pi^2-\frac{7941865}{864}\zeta_{3}+\frac{906877}{23040}\pi^4
\nonumber &\\&
+\frac{175547}{216}\pi^2\zeta_{3}-\frac{1600871}{360}\zeta_{5}-\frac{719437}{1088640}\pi^6+\frac{19345}{12}\zeta_{3}^2 + \mathcal{O}(\e), &
\end{flalign}

\begin{flalign}
{\cal T}^{q\bar{q},(3)}_{q\bar{q}gg}\Big|_{N^{-2}} = &
+\frac{1}{\e^6}\left(-\frac{1}{2}\right)
+\frac{1}{\e^5}\left(-\frac{9}{4}\right)
+\frac{1}{\e^4}\left(-\frac{83}{8}+\frac{29}{24}\pi^2\right)
\nonumber &\\&
+\frac{1}{\e^3}\left(-\frac{1611}{32}+\frac{19}{4}\pi^2+\frac{137}{6}\zeta_{3}\right)
\nonumber &\\&
+\frac{1}{\e^2}\left(-\frac{977}{4}+\frac{2237}{96}\pi^2+\frac{177}{2}\zeta_{3}-\frac{169}{320}\pi^4\right)
\nonumber &\\&
+\frac{1}{\e}\left(-\frac{229351}{192}+\frac{133169}{1152}\pi^2+\frac{3517}{8}\zeta_{3}-\frac{4969}{2880}\pi^4-\frac{1463}{24}\pi^2\zeta_{3}+\frac{10469}{30}\zeta_{5}\right)
\nonumber &\\&
-\frac{753943}{128}+\frac{1325621}{2304}\pi^2+\frac{212593}{96}\zeta_{3}-\frac{104309}{11520}\pi^4
\nonumber &\\&
-\frac{851}{4}\pi^2\zeta_{3}+\frac{25639}{20}\zeta_{5}+\frac{51599}{311040}\pi^6-\frac{7667}{12}\zeta_{3}^2 + \mathcal{O}(\e), &
\end{flalign}

\begin{flalign}
{\cal T}^{q\bar{q},(3)}_{q\bar{q}gg}\Big|_{\NF N} = &
+\frac{1}{\e^5}\left(\frac{1}{2}\right)
+\frac{1}{\e^4}\left(\frac{65}{36}\right)
+\frac{1}{\e^3}\left(\frac{941}{108}-\frac{13}{18}\pi^2\right)
\nonumber &\\&
+\frac{1}{\e^2}\left(\frac{16667}{432}-\frac{589}{216}\pi^2-\frac{71}{6}\zeta_{3}\right)
\nonumber &\\&
+\frac{1}{\e}\left(\frac{1298723}{7776}-\frac{17005}{1296}\pi^2-\frac{1327}{27}\zeta_{3}+\frac{373}{2160}\pi^4\right)
\nonumber &\\&
+\frac{33149779}{46656}-\frac{56513}{972}\pi^2-\frac{76061}{324}\zeta_{3}+\frac{5207}{12960}\pi^4
\nonumber &\\&
+\frac{1891}{108}\pi^2\zeta_{3}-\frac{887}{10}\zeta_{5} + \mathcal{O}(\e), &
\end{flalign}

\begin{flalign}
{\cal T}^{q\bar{q},(3)}_{q\bar{q}gg}\Big|_{\NF N^{-1}} = &
+\frac{1}{\e^5}\left(-\frac{1}{3}\right)
+\frac{1}{\e^4}\left(-1\right)
+\frac{1}{\e^3}\left(-\frac{14}{3}+\frac{\pi^2}{2}\right)
\nonumber &\\&
+\frac{1}{\e^2}\left(-\frac{965}{48}+\frac{3}{2}\pi^2+\frac{80}{9}\zeta_{3}\right)
\nonumber &\\&
+\frac{1}{\e}\left(-\frac{8201}{96}+\frac{253}{36}\pi^2+\frac{80}{3}\zeta_{3}-\frac{17}{216}\pi^4\right)
\nonumber &\\&
-\frac{68803}{192}+\frac{8717}{288}\pi^2+\frac{2267}{18}\zeta_{3}-\frac{43}{180}\pi^4
\nonumber &\\&
-\frac{121}{9}\pi^2\zeta_{3}+\frac{424}{5}\zeta_{5} + \mathcal{O}(\e), &
\end{flalign}



\begin{flalign}
{\cal T}^{q\bar{q},(3)}_{q\bar{q}q\bar{q}}\Big|_{N} = &
+\frac{1}{\e^5}\left(\frac{13}{108}\right)
+\frac{1}{\e^4}\left(\frac{679}{648}\right)
+\frac{1}{\e^3}\left(\frac{11411}{1944}-\frac{425}{1296}\pi^2\right)
\nonumber &\\&
+\frac{1}{\e^2}\left(\frac{194443}{5832}-\frac{9217}{3888}\pi^2-\frac{265}{36}\zeta_{3}\right)
\nonumber &\\&
+\frac{1}{\e}\left(\frac{794069}{4374}-\frac{309253}{23328}\pi^2-\frac{593}{12}\zeta_{3}+\frac{3239}{51840}\pi^4\right)
\nonumber &\\&
+\frac{202587241}{209952}-\frac{10661983}{139968}\pi^2-\frac{18379}{72}\zeta_{3}+\frac{37319}{77760}\pi^4
\nonumber &\\&
+\frac{8261}{432}\pi^2\zeta_{3}-\frac{9121}{60}\zeta_{5} + \mathcal{O}(\e), &
\end{flalign}

\begin{flalign}
{\cal T}^{q\bar{q},(3)}_{q\bar{q}q\bar{q}}\Big|_{N^{0}} = &
+\frac{1}{\e^3}\left(-\frac{65}{48}+\frac{5}{24}\pi^2-\frac{5}{6}\zeta_{3}\right)
\nonumber &\\&
+\frac{1}{\e^2}\left(-\frac{2753}{144}+\frac{179}{144}\pi^2+\frac{187}{18}\zeta_{3}-\frac{101}{1080}\pi^4\right)
\nonumber &\\&
+\frac{1}{\e}\left(-\frac{135293}{864}+\frac{12803}{1728}\pi^2+\frac{3491}{54}\zeta_{3}+\frac{1469}{6480}\pi^4+\frac{169}{72}\pi^2\zeta_{3}-\frac{117}{2}\zeta_{5}\right)
\nonumber &\\&
-\frac{1353227}{1296}+\frac{609173}{10368}\pi^2+\frac{309323}{1296}\zeta_{3}+\frac{72677}{38880}\pi^4
\nonumber &\\&
-\frac{14069}{432}\pi^2\zeta_{3}+\frac{3187}{9}\zeta_{5}-\frac{8171}{90720}\pi^6+\frac{110}{3}\zeta_{3}^2 + \mathcal{O}(\e), &
\end{flalign}

\begin{flalign}
{\cal T}^{q\bar{q},(3)}_{q\bar{q}q\bar{q}}\Big|_{N^{-1}} = &
+\frac{1}{\e^5}\left(-\frac{11}{108}\right)
+\frac{1}{\e^4}\left(-\frac{425}{648}\right)
+\frac{1}{\e^3}\left(-\frac{7627}{1944}+\frac{133}{432}\pi^2\right)
\nonumber &\\&
+\frac{1}{\e^2}\left(-\frac{132731}{5832}+\frac{2213}{1296}\pi^2+\frac{889}{108}\zeta_{3}\right)
\nonumber &\\&
+\frac{1}{\e}\left(-\frac{4453349}{34992}+\frac{78287}{7776}\pi^2+\frac{12995}{324}\zeta_{3}-\frac{125}{10368}\pi^4\right)
\nonumber &\\&
-\frac{145744877}{209952}+\frac{2720603}{46656}\pi^2+\frac{425795}{1944}\zeta_{3}-\frac{22771}{77760}\pi^4
\nonumber &\\&
-\frac{10375}{432}\pi^2\zeta_{3}+\frac{34661}{180}\zeta_{5} + \mathcal{O}(\e), &
\end{flalign}

\begin{flalign}
{\cal T}^{q\bar{q},(3)}_{q\bar{q}q\bar{q}}\Big|_{N^{-2}} = &
+\frac{1}{\e^3}\left(\frac{65}{48}-\frac{5}{24}\pi^2+\frac{5}{6}\zeta_{3}\right)
\nonumber &\\&
+\frac{1}{\e^2}\left(\frac{71}{4}-\frac{47}{48}\pi^2-\frac{21}{2}\zeta_{3}+\frac{7}{90}\pi^4\right)
\nonumber &\\&
+\frac{1}{\e}\left(\frac{29285}{192}-\frac{4243}{576}\pi^2-\frac{289}{6}\zeta_{3}-\frac{17}{90}\pi^4-\frac{229}{72}\pi^2\zeta_{3}+\frac{277}{6}\zeta_{5}\right)
\nonumber &\\&
+\frac{35043}{32}-\frac{70655}{1152}\pi^2-\frac{11251}{48}\zeta_{3}-\frac{8231}{8640}\pi^4
\nonumber &\\&
+\frac{1535}{48}\pi^2\zeta_{3}-\frac{1235}{4}\zeta_{5}-\frac{11}{336}\pi^6-76 \zeta_{3}^2 + \mathcal{O}(\e), &
\end{flalign}

\begin{flalign}
{\cal T}^{q\bar{q},(3)}_{q\bar{q}q\bar{q}}\Big|_{\NF} = &
+\frac{1}{\e^4}\left(-\frac{7}{162}\right)
+\frac{1}{\e^3}\left(-\frac{79}{486}\right)
+\frac{1}{\e^2}\left(-\frac{53}{81}+\frac{\pi^2}{18}\right)
\nonumber &\\&
+\frac{1}{\e}\left(-\frac{16639}{8748}+\frac{73}{648}\pi^2+\frac{76}{81}\zeta_{3}\right)
\nonumber &\\&
-\frac{17879}{13122}+\frac{143}{3888}\pi^2+\frac{239}{486}\zeta_{3}+\frac{41}{38880}\pi^4 + \mathcal{O}(\e), &
\end{flalign}

\begin{flalign}
{\cal T}^{q\bar{q},(3)}_{q\bar{q}q\bar{q}}\Big|_{\NF N^{-1}} = &
+\frac{1}{\e^2}\left(\frac{13}{72}-\frac{\pi^2}{36}+\frac{\zeta_{3}}{9}\right)
+\frac{1}{\e}\left(\frac{19}{108}+\frac{11}{216}\pi^2-\frac{28}{27}\zeta_{3}+\frac{17}{3240}\pi^4\right)
\nonumber &\\&
-\frac{30701}{2592}+\frac{1001}{1296}\pi^2+\frac{833}{162}\zeta_{3}-\frac{1087}{19440}\pi^4
\nonumber &\\&
+\frac{8}{27}\pi^2\zeta_{3}-\frac{19}{9}\zeta_{5} + \mathcal{O}(\e), &
\end{flalign}



\begin{flalign}
{\cal T}^{q\bar{q},(3)}_{q\bar{q}q'\bar{q}'}\Big|_{(\NF-1) N} = &
+\frac{1}{\e^5}\left(\frac{13}{108}\right)
+\frac{1}{\e^4}\left(\frac{679}{648}\right)
+\frac{1}{\e^3}\left(\frac{11411}{1944}-\frac{425}{1296}\pi^2\right)
\nonumber &\\&
+\frac{1}{\e^2}\left(\frac{194443}{5832}-\frac{9217}{3888}\pi^2-\frac{265}{36}\zeta_{3}\right)
\nonumber &\\&
+\frac{1}{\e}\left(\frac{794069}{4374}-\frac{309253}{23328}\pi^2-\frac{593}{12}\zeta_{3}+\frac{3239}{51840}\pi^4\right)
\nonumber &\\&
+\frac{202587241}{209952}-\frac{10661983}{139968}\pi^2-\frac{18379}{72}\zeta_{3}+\frac{37319}{77760}\pi^4
\nonumber &\\&
+\frac{8261}{432}\pi^2\zeta_{3}-\frac{9121}{60}\zeta_{5} + \mathcal{O}(\e), &
\end{flalign}

\begin{flalign}
{\cal T}^{q\bar{q},(3)}_{q\bar{q}q'\bar{q}'}\Big|_{(\NF-1) N^{-1}} = &
+\frac{1}{\e^5}\left(-\frac{11}{108}\right)
+\frac{1}{\e^4}\left(-\frac{425}{648}\right)
+\frac{1}{\e^3}\left(-\frac{7627}{1944}+\frac{133}{432}\pi^2\right)
\nonumber &\\&
+\frac{1}{\e^2}\left(-\frac{132731}{5832}+\frac{2213}{1296}\pi^2+\frac{889}{108}\zeta_{3}\right)
\nonumber &\\&
+\frac{1}{\e}\left(-\frac{4453349}{34992}+\frac{78287}{7776}\pi^2+\frac{12995}{324}\zeta_{3}-\frac{125}{10368}\pi^4\right)
\nonumber &\\&
-\frac{145744877}{209952}+\frac{2720603}{46656}\pi^2+\frac{425795}{1944}\zeta_{3}-\frac{22771}{77760}\pi^4
\nonumber &\\&
-\frac{10375}{432}\pi^2\zeta_{3}+\frac{34661}{180}\zeta_{5} + \mathcal{O}(\e), &
\end{flalign}

\begin{flalign}
{\cal T}^{q\bar{q},(3)}_{q\bar{q}q'\bar{q}'}\Big|_{\NF (\NF-1)} = &
+\frac{1}{\e^4}\left(-\frac{7}{162}\right)
+\frac{1}{\e^3}\left(-\frac{79}{486}\right)
+\frac{1}{\e^2}\left(-\frac{53}{81}+\frac{\pi^2}{18}\right)
\nonumber &\\&
+\frac{1}{\e}\left(-\frac{16639}{8748}+\frac{73}{648}\pi^2+\frac{76}{81}\zeta_{3}\right)
\nonumber &\\&
-\frac{17879}{13122}+\frac{143}{3888}\pi^2+\frac{239}{486}\zeta_{3}+\frac{41}{38880}\pi^4 + \mathcal{O}(\e), &
\end{flalign}


\subsubsection{Five-particle final states}


\begin{flalign}
{\cal T}^{q\bar{q},(3)}_{q\bar{q}ggg}\Big|_{N^{2}} = &
+\frac{1}{\e^6}\left(\frac{1}{2}\right)
+\frac{1}{\e^5}\left(\frac{331}{108}\right)
+\frac{1}{\e^4}\left(\frac{12653}{648}-\frac{31}{24}\pi^2\right)
\nonumber &\\&
+\frac{1}{\e^3}\left(\frac{296155}{2592}-\frac{10745}{1296}\pi^2-\frac{439}{18}\zeta_{3}\right)
\nonumber &\\&
+\frac{1}{\e^2}\left(\frac{939217}{1458}-\frac{414311}{7776}\pi^2-\frac{6239}{36}\zeta_{3}+\frac{21853}{25920}\pi^4\right)
\nonumber &\\&
+\frac{1}{\e}\Bigg(\frac{990443209}{279936}-\frac{29177653}{93312}\pi^2-\frac{183905}{162}\zeta_{3}+\frac{75767}{17280}\pi^4
\nonumber &\\&
\phantom{+\frac{1}{\e}\Bigg(}+\frac{13993}{216}\pi^2\zeta_{3}-\frac{10946}{45}\zeta_{5}\Bigg)
\nonumber &\\&
+\frac{889159591}{46656}-\frac{246959419}{139968}\pi^2-\frac{52077425}{7776}\zeta_{3}+\frac{8315387}{311040}\pi^4
\nonumber &\\&
+\frac{67895}{144}\pi^2\zeta_{3}-\frac{103894}{45}\zeta_{5}-\frac{93257}{1306368}\pi^6+\frac{7861}{12}\zeta_{3}^2 + \mathcal{O}(\e), &
\end{flalign}

\begin{flalign}
{\cal T}^{q\bar{q},(3)}_{q\bar{q}ggg}\Big|_{N^{0}} = &
+\frac{1}{\e^6}\left(-\frac{7}{12}\right)
+\frac{1}{\e^5}\left(-\frac{37}{12}\right)
+\frac{1}{\e^4}\left(-\frac{167}{9}+\frac{25}{16}\pi^2\right)
\nonumber &\\&
+\frac{1}{\e^3}\left(-\frac{45169}{432}+\frac{76}{9}\pi^2+\frac{63}{2}\zeta_{3}\right)
\nonumber &\\&
+\frac{1}{\e^2}\left(-\frac{372211}{648}+\frac{44065}{864}\pi^2+\frac{12907}{72}\zeta_{3}-\frac{15811}{17280}\pi^4\right)
\nonumber &\\&
+\frac{1}{\e}\Bigg(-\frac{48151867}{15552}+\frac{186509}{648}\pi^2+\frac{472405}{432}\zeta_{3}-\frac{76187}{17280}\pi^4
\nonumber &\\&
\phantom{+\frac{1}{\e}\Bigg(}-\frac{1541}{18}\pi^2\zeta_{3}+\frac{5774}{15}\zeta_{5}\Bigg)
\nonumber &\\&
-\frac{95958199}{5832}+\frac{49202785}{31104}\pi^2+\frac{16086553}{2592}\zeta_{3}-\frac{668519}{25920}\pi^4
\nonumber &\\&
-\frac{142025}{288}\pi^2\zeta_{3}+\frac{218759}{90}\zeta_{5}+\frac{916243}{4354560}\pi^6-\frac{21473}{24}\zeta_{3}^2 + \mathcal{O}(\e), &
\end{flalign}

\begin{flalign}
{\cal T}^{q\bar{q},(3)}_{q\bar{q}ggg}\Big|_{N^{-2}} = &
+\frac{1}{\e^6}\left(\frac{1}{6}\right)
+\frac{1}{\e^5}\left(\frac{3}{4}\right)
+\frac{1}{\e^4}\left(\frac{35}{8}-\frac{11}{24}\pi^2\right)
+\frac{1}{\e^3}\left(\frac{771}{32}-\frac{33}{16}\pi^2-\frac{59}{6}\zeta_{3}\right)
\nonumber &\\&
+\frac{1}{\e^2}\left(\frac{1045}{8}-\frac{1159}{96}\pi^2-\frac{177}{4}\zeta_{3}+\frac{659}{2880}\pi^4\right)
\nonumber &\\&
+\frac{1}{\e}\left(\frac{267997}{384}-\frac{76609}{1152}\pi^2-\frac{1041}{4}\zeta_{3}+\frac{1981}{1920}\pi^4+\frac{1981}{72}\pi^2\zeta_{3}-\frac{1451}{10}\zeta_{5}\right)
\nonumber &\\&
+\frac{707891}{192}-\frac{103841}{288}\pi^2-\frac{137429}{96}\zeta_{3}+\frac{41219}{6912}\pi^4
\nonumber &\\&
+\frac{2939}{24}\pi^2\zeta_{3}-\frac{25533}{40}\zeta_{5}-\frac{197047}{2177280}\pi^6+\frac{1831}{6}\zeta_{3}^2 + \mathcal{O}(\e), &
\end{flalign}



\begin{flalign}
{\cal T}^{q\bar{q},(3)}_{q\bar{q}q\bar{q}g}\Big|_{N} = &
+\frac{1}{\e^5}\left(-\frac{7}{54}\right)
+\frac{1}{\e^4}\left(-\frac{101}{108}\right)
+\frac{1}{\e^3}\left(-\frac{12461}{1944}+\frac{247}{648}\pi^2\right)
\nonumber &\\&
+\frac{1}{\e^2}\left(-\frac{473479}{11664}+\frac{10579}{3888}\pi^2+\frac{493}{54}\zeta_{3}\right)
\nonumber &\\&
+\frac{1}{\e}\left(-\frac{5692453}{23328}+\frac{426841}{23328}\pi^2+\frac{21035}{324}\zeta_{3}-\frac{3613}{25920}\pi^4\right)
\nonumber &\\&
-\frac{592211495}{419904}+\frac{15998635}{139968}\pi^2+\frac{273361}{648}\zeta_{3}-\frac{148181}{155520}\pi^4
\nonumber &\\&
-\frac{205}{8}\pi^2\zeta_{3}+\frac{13757}{90}\zeta_{5} + \mathcal{O}(\e), &
\end{flalign}

\begin{flalign}
{\cal T}^{q\bar{q},(3)}_{q\bar{q}q\bar{q}g}\Big|_{N^{0}} = &
+\frac{1}{\e^3}\left(\frac{65}{48}-\frac{5}{24}\pi^2+\frac{5}{6}\zeta_{3}\right)
\nonumber &\\&
+\frac{1}{\e^2}\left(\frac{145}{8}-\frac{157}{144}\pi^2-11 \zeta_{3}+\frac{101}{1080}\pi^4\right)
\nonumber &\\&
+\frac{1}{\e}\left(\frac{30593}{192}-\frac{1613}{192}\pi^2-\frac{451}{8}\zeta_{3}-\frac{19}{80}\pi^4-\frac{19}{8}\pi^2\zeta_{3}+\frac{176}{3}\zeta_{5}\right)
\nonumber &\\&
+\frac{55511}{48}-\frac{39751}{576}\pi^2-295 \zeta_{3}-\frac{19237}{17280}\pi^4
\nonumber &\\&
+\frac{1627}{48}\pi^2\zeta_{3}-\frac{1437}{4}\zeta_{5}+\frac{167}{1701}\pi^6-\frac{511}{12}\zeta_{3}^2 + \mathcal{O}(\e), &
\end{flalign}

\begin{flalign}
{\cal T}^{q\bar{q},(3)}_{q\bar{q}q\bar{q}g}\Big|_{N^{-1}} = &
+\frac{1}{\e^5}\left(\frac{11}{108}\right)
+\frac{1}{\e^4}\left(\frac{425}{648}\right)
+\frac{1}{\e^3}\left(\frac{16955}{3888}-\frac{47}{144}\pi^2\right)
\nonumber &\\&
+\frac{1}{\e^2}\left(\frac{629825}{23328}-\frac{1685}{864}\pi^2-\frac{979}{108}\zeta_{3}\right)
\nonumber &\\&
+\frac{1}{\e}\left(\frac{22310435}{139968}-\frac{65447}{5184}\pi^2-\frac{31345}{648}\zeta_{3}+\frac{1777}{51840}\pi^4\right)
\nonumber &\\&
+\frac{762800561}{839808}-\frac{2391413}{31104}\pi^2-\frac{1163287}{3888}\zeta_{3}+\frac{35567}{62208}\pi^4
\nonumber &\\&
+\frac{1261}{48}\pi^2\zeta_{3}-\frac{34651}{180}\zeta_{5} + \mathcal{O}(\e), &
\end{flalign}

\begin{flalign}
{\cal T}^{q\bar{q},(3)}_{q\bar{q}q\bar{q}g}\Big|_{N^{-2}} = &
+\frac{1}{\e^3}\left(-\frac{65}{48}+\frac{5}{24}\pi^2-\frac{5}{6}\zeta_{3}\right)
\nonumber &\\&
+\frac{1}{\e^2}\left(-\frac{71}{4}+\frac{47}{48}\pi^2+\frac{21}{2}\zeta_{3}-\frac{7}{90}\pi^4\right)
\nonumber &\\&
+\frac{1}{\e}\left(-\frac{29933}{192}+\frac{4651}{576}\pi^2+\frac{142}{3}\zeta_{3}+\frac{17}{144}\pi^4+\frac{27}{8}\pi^2\zeta_{3}-44 \zeta_{5}\right)
\nonumber &\\&
-\frac{109309}{96}+\frac{39031}{576}\pi^2+\frac{4449}{16}\zeta_{3}+\frac{4751}{17280}\pi^4
\nonumber &\\&
-\frac{1559}{48}\pi^2\zeta_{3}+\frac{1025}{4}\zeta_{5}+\frac{22229}{272160}\pi^6+\frac{345}{4}\zeta_{3}^2 + \mathcal{O}(\e), &
\end{flalign}



\begin{flalign}
{\cal T}^{q\bar{q},(3)}_{q\bar{q}q'\bar{q}'g}\Big|_{(\NF-1) N} = &
+\frac{1}{\e^5}\left(-\frac{7}{54}\right)
+\frac{1}{\e^4}\left(-\frac{101}{108}\right)
+\frac{1}{\e^3}\left(-\frac{12461}{1944}+\frac{247}{648}\pi^2\right)
\nonumber &\\&
+\frac{1}{\e^2}\left(-\frac{473479}{11664}+\frac{10579}{3888}\pi^2+\frac{493}{54}\zeta_{3}\right)
\nonumber &\\&
+\frac{1}{\e}\left(-\frac{5692453}{23328}+\frac{426841}{23328}\pi^2+\frac{21035}{324}\zeta_{3}-\frac{3613}{25920}\pi^4\right)
\nonumber &\\&
-\frac{592211495}{419904}+\frac{15998635}{139968}\pi^2+\frac{273361}{648}\zeta_{3}-\frac{148181}{155520}\pi^4
\nonumber &\\&
-\frac{205}{8}\pi^2\zeta_{3}+\frac{13757}{90}\zeta_{5} + \mathcal{O}(\e), &
\end{flalign}

\begin{flalign}\label{eq:Hbbend}
{\cal T}^{q\bar{q},(3)}_{q\bar{q}q'\bar{q}'g}\Big|_{(\NF-1) N^{-1}} = &
+\frac{1}{\e^5}\left(\frac{11}{108}\right)
+\frac{1}{\e^4}\left(\frac{425}{648}\right)
+\frac{1}{\e^3}\left(\frac{16955}{3888}-\frac{47}{144}\pi^2\right)
\nonumber &\\&
+\frac{1}{\e^2}\left(\frac{629825}{23328}-\frac{1685}{864}\pi^2-\frac{979}{108}\zeta_{3}\right)
\nonumber &\\&
+\frac{1}{\e}\left(\frac{22310435}{139968}-\frac{65447}{5184}\pi^2-\frac{31345}{648}\zeta_{3}+\frac{1777}{51840}\pi^4\right)
\nonumber &\\&
+\frac{762800561}{839808}-\frac{2391413}{31104}\pi^2-\frac{1163287}{3888}\zeta_{3}+\frac{35567}{62208}\pi^4
\nonumber &\\&
+\frac{1261}{48}\pi^2\zeta_{3}-\frac{34651}{180}\zeta_{5} + \mathcal{O}(\e), &
\end{flalign}


\subsection{Higgs to gluons}

\subsubsection{Two-particle final states}


\begin{flalign}\label{eq:Hggstart}
{\cal T}^{gg,(3,\left[2\times 1\right])}_{gg}\Big|_{N^{3}} = &
+\frac{1}{\e^6}\left(-1\right)
+\frac{1}{\e^5}\left(-\frac{33}{4}\right)
+\frac{1}{\e^4}\left(-\frac{133}{8}+\frac{2}{3}\pi^2\right)
\nonumber &\\&
+\frac{1}{\e^3}\left(-\frac{1189}{216}+\frac{583}{144}\pi^2+\frac{13}{2}\zeta_{3}\right)
\nonumber &\\&
+\frac{1}{\e^2}\left(-\frac{1445}{54}+\frac{599}{108}\pi^2+\frac{352}{9}\zeta_{3}-\frac{59}{1440}\pi^4\right)
\nonumber &\\&
+\frac{1}{\e}\left(-\frac{14669}{81}-\frac{14795}{1296}\pi^2+\frac{3385}{54}\zeta_{3}-\frac{11429}{17280}\pi^4-\frac{37}{8}\pi^2\zeta_{3}-\frac{9}{10}\zeta_{5}\right)
\nonumber &\\&
-\frac{592597}{972}+\frac{15629}{972}\pi^2+\frac{63739}{324}\zeta_{3}+\frac{50081}{17280}\pi^4-\frac{2585}{144}\pi^2\zeta_{3}+\frac{913}{15}\zeta_{5}
\nonumber &\\&
-\frac{6647}{60480}\pi^6-\frac{125}{2}\zeta_{3}^2 + \mathcal{O}(\e), &
\end{flalign}

\begin{flalign}
{\cal T}^{gg,(3,\left[2\times 1\right])}_{gg}\Big|_{\NF N^{2}} = &
+\frac{1}{\e^5}\left(\frac{3}{2}\right)
+\frac{1}{\e^4}\left(\frac{58}{9}\right)
+\frac{1}{\e^3}\left(\frac{449}{108}-\frac{53}{72}\pi^2\right)
\nonumber &\\&
+\frac{1}{\e^2}\left(\frac{1001}{162}-\frac{529}{216}\pi^2-\frac{46}{9}\zeta_{3}\right)
\nonumber &\\&
+\frac{1}{\e}\left(\frac{70685}{972}+\frac{4811}{1296}\pi^2-\frac{479}{54}\zeta_{3}+\frac{151}{960}\pi^4\right)
\nonumber &\\&
+\frac{820825}{2916}-\frac{65887}{7776}\pi^2-\frac{1295}{324}\zeta_{3}-\frac{1813}{2880}\pi^4+\frac{37}{24}\pi^2\zeta_{3}-\frac{46}{15}\zeta_{5}
\nonumber &\\&
+ \mathcal{O}(\e), &
\end{flalign}

\begin{flalign}
{\cal T}^{gg,(3,\left[2\times 1\right])}_{gg}\Big|_{\NF} = &
+\frac{1}{\e^3}\left(\frac{1}{4}\right)
+\frac{1}{\e^2}\left(-\frac{7}{3}+2 \zeta_{3}\right)
\nonumber &\\&
+\frac{1}{\e}\left(-\frac{691}{36}-\frac{7}{144}\pi^2+\frac{34}{3}\zeta_{3}+\frac{\pi^4}{27}\right)
\nonumber &\\&
-\frac{17393}{216}+\frac{2771}{864}\pi^2+\frac{1609}{36}\zeta_{3}+\frac{17}{81}\pi^4-\frac{31}{18}\pi^2\zeta_{3}+8 \zeta_{5} + \mathcal{O}(\e), &
\end{flalign}

\begin{flalign}
{\cal T}^{gg,(3,\left[2\times 1\right])}_{gg}\Big|_{\NF^2 N} = &
+\frac{1}{\e^4}\left(-\frac{11}{18}\right)
+\frac{1}{\e^3}\left(-\frac{61}{54}\right)
+\frac{1}{\e^2}\left(\frac{155}{324}+\frac{\pi^2}{4}\right)
\nonumber &\\&
+\frac{1}{\e}\left(-\frac{6337}{1944}-\frac{25}{108}\pi^2+\frac{23}{27}\zeta_{3}\right)
\nonumber &\\&
-\frac{170581}{11664}+\frac{305}{648}\pi^2-\frac{95}{81}\zeta_{3}+\frac{77}{1620}\pi^4 + \mathcal{O}(\e), &
\end{flalign}

\begin{flalign}
{\cal T}^{gg,(3,\left[2\times 1\right])}_{gg}\Big|_{\NF^2 N^{-1}} = &
+\frac{1}{\e^2}\left(-\frac{1}{12}\right)
+\frac{1}{\e}\left(\frac{67}{72}-\frac{2}{3}\zeta_{3}\right)
\nonumber &\\&
+\frac{2027}{432}-\frac{43}{216}\pi^2-\frac{23}{9}\zeta_{3}-\frac{\pi^4}{81} + \mathcal{O}(\e), &
\end{flalign}

\begin{flalign}
{\cal T}^{gg,(3,\left[2\times 1\right])}_{gg}\Big|_{\NF^3 N^{0}} = &
+\frac{1}{\e^3}\left(\frac{2}{27}\right) &
\end{flalign}



\begin{flalign}
{\cal T}^{gg,(3,\left[3\times 0\right])}_{gg}\Big|_{N^{3}} = &
+\frac{1}{\e^6}\left(-\frac{1}{3}\right)
+\frac{1}{\e^5}\left(-\frac{55}{12}\right)
+\frac{1}{\e^4}\left(-\frac{9079}{648}+\frac{3}{2}\pi^2\right)
\nonumber &\\&
+\frac{1}{\e^3}\left(\frac{5453}{1944}+\frac{8987}{1296}\pi^2+\frac{11}{6}\zeta_{3}\right)
\nonumber &\\&
+\frac{1}{\e^2}\left(-\frac{4277}{972}-\frac{2785}{486}\pi^2+\frac{1133}{54}\zeta_{3}-\frac{14333}{12960}\pi^4\right)
\nonumber &\\&
+\frac{1}{\e}\left(-\frac{326926}{2187}-\frac{126337}{11664}\pi^2+\frac{907}{18}\zeta_{3}+\frac{28721}{51840}\pi^4
-\frac{1867}{216}\pi^2\zeta_{3}-\frac{439}{30}\zeta_{5}\right)
\nonumber &\\&
-\frac{23496187}{104976}+\frac{3476911}{34992}\pi^2+\frac{24893}{972}\zeta_{3}+\frac{1669751}{155520}\pi^4
\nonumber &\\&
-\frac{17479}{432}\pi^2\zeta_{3}+\frac{6941}{90}\zeta_{5}+\frac{18101}{116640}\pi^6-\frac{883}{18}\zeta_{3}^2 + \mathcal{O}(\e), &
\end{flalign}

\begin{flalign}
{\cal T}^{gg,(3,\left[3\times 0\right])}_{gg}\Big|_{\NF N^{2}} = &
+\frac{1}{\e^5}\left(\frac{5}{6}\right)
+\frac{1}{\e^4}\left(\frac{445}{81}\right)
+\frac{1}{\e^3}\left(\frac{1885}{972}-\frac{817}{648}\pi^2\right)
\nonumber &\\&
+\frac{1}{\e^2}\left(\frac{775}{1944}+\frac{223}{1944}\pi^2-\frac{43}{27}\zeta_{3}\right)
\nonumber &\\&
+\frac{1}{\e}\left(\frac{2445575}{34992}+\frac{6613}{11664}\pi^2-\frac{449}{162}\zeta_{3}-\frac{311}{5184}\pi^4\right)
\nonumber &\\&
+\frac{16710463}{209952}-\frac{372013}{8748}\pi^2+\frac{7525}{324}\zeta_{3}-\frac{2615}{1728}\pi^4
\nonumber &\\&
-\frac{787}{216}\pi^2\zeta_{3}+\frac{233}{15}\zeta_{5} + \mathcal{O}(\e), &
\end{flalign}

\begin{flalign}
{\cal T}^{gg,(3,\left[3\times 0\right])}_{gg}\Big|_{\NF} = &
+\frac{1}{\e^3}\left(\frac{17}{36}\right)
+\frac{1}{\e^2}\left(-\frac{427}{216}+\frac{20}{9}\zeta_{3}\right)
\nonumber &\\&
+\frac{1}{\e}\left(-\frac{3407}{162}+\frac{139}{144}\pi^2+\frac{325}{27}\zeta_{3}+\frac{11}{270}\pi^4\right)
\nonumber &\\&
-\frac{317761}{7776}+\frac{523}{36}\pi^2+\frac{3203}{324}\zeta_{3}+\frac{29}{135}\pi^4
\nonumber &\\&
-11 \pi^2\zeta_{3}+\frac{176}{9}\zeta_{5} + \mathcal{O}(\e), &
\end{flalign}

\begin{flalign}
{\cal T}^{gg,(3,\left[3\times 0\right])}_{gg}\Big|_{\NF N^{-2}} = &
+\frac{1}{\e}\left(-\frac{1}{48}\right)
+\frac{19}{9}+\frac{37}{6}\zeta_{3}-10 \zeta_{5} + \mathcal{O}(\e), &
\end{flalign}

\begin{flalign}
{\cal T}^{gg,(3,\left[3\times 0\right])}_{gg}\Big|_{\NF^2 N} = &
+\frac{1}{\e^4}\left(-\frac{85}{162}\right)
+\frac{1}{\e^3}\left(-\frac{499}{486}\right)
+\frac{1}{\e^2}\left(\frac{113}{108}+\frac{37}{324}\pi^2\right)
\nonumber &\\&
+\frac{1}{\e}\left(-\frac{54305}{17496}-\frac{215}{972}\pi^2+\frac{13}{81}\zeta_{3}\right)
\nonumber &\\&
+\frac{483479}{104976}-\frac{53}{648}\pi^2-\frac{275}{243}\zeta_{3}+\frac{59}{19440}\pi^4 + \mathcal{O}(\e), &
\end{flalign}

\begin{flalign}
{\cal T}^{gg,(3,\left[3\times 0\right])}_{gg}\Big|_{\NF^2 N^{-1}} = &
+\frac{1}{\e^2}\left(-\frac{7}{36}\right)
+\frac{1}{\e}\left(\frac{53}{54}-\frac{2}{3}\zeta_{3}\right)
\nonumber &\\&
-\frac{2881}{1296}+\frac{\pi^2}{72}+\frac{19}{9}\zeta_{3}-\frac{\pi^4}{90} + \mathcal{O}(\e), &
\end{flalign}

\begin{flalign}
{\cal T}^{gg,(3,\left[3\times 0\right])}_{gg}\Big|_{\NF^3} = &
+\frac{1}{\e^3}\left(\frac{2}{27}\right) &
\end{flalign}


\subsubsection{Three-particle final states}


\begin{flalign}
{\cal T}^{gg,(3,\left[1\times 1\right])}_{ggg}\Big|_{N^{3}} = &
+\frac{1}{\e^6}\left(\frac{23}{9}\right)
+\frac{1}{\e^5}\left(\frac{1309}{72}\right)
+\frac{1}{\e^4}\left(\frac{1625}{24}-\frac{301}{108}\pi^2\right)
\nonumber &\\&
+\frac{1}{\e^3}\left(\frac{171727}{648}-\frac{187}{8}\pi^2-\frac{607}{9}\zeta_{3}\right)
\nonumber &\\&
+\frac{1}{\e^2}\left(\frac{4600213}{3888}-\frac{52463}{648}\pi^2-\frac{4543}{12}\zeta_{3}-\frac{3311}{4320}\pi^4\right)
\nonumber &\\&
+\frac{1}{\e}\bigg(\frac{10480321}{1944}-\frac{934837}{2592}\pi^2-\frac{85027}{54}\zeta_{3}
\nonumber &\\& \phantom{+\frac{1}{\e}\bigg(}
+\frac{79387}{51840}\pi^4+\frac{10253}{108}\pi^2\zeta_{3}-\frac{13393}{15}\zeta_{5}\bigg)
\nonumber &\\&
+\frac{3526701253}{139968}-\frac{13179161}{7776}\pi^2-\frac{1661833}{216}\zeta_{3}-\frac{275233}{38880}\pi^4
\nonumber &\\&
+\frac{242561}{432}\pi^2\zeta_{3}-\frac{88407}{20}\zeta_{5}-\frac{101317}{120960}\pi^6+\frac{23447}{18}\zeta_{3}^2 + \mathcal{O}(\e), &
\end{flalign}

\begin{flalign}
{\cal T}^{gg,(3,\left[1\times 1\right])}_{ggg}\Big|_{\NF N^{2}} = &
+\frac{1}{\e^5}\left(-\frac{9}{4}\right)
+\frac{1}{\e^4}\left(-\frac{121}{12}\right)
+\frac{1}{\e^3}\left(-\frac{257}{9}+\frac{71}{24}\pi^2\right)
\nonumber &\\&
+\frac{1}{\e^2}\left(-\frac{988}{9}+\frac{341}{36}\pi^2+36 \zeta_{3}\right)
\nonumber &\\&
+\frac{1}{\e}\left(-\frac{179999}{432}+\frac{13967}{432}\pi^2+\frac{1133}{9}\zeta_{3}-\frac{199}{288}\pi^4\right)
\nonumber &\\&
-\frac{12462217}{7776}+\frac{113855}{864}\pi^2+\frac{5365}{12}\zeta_{3}-\frac{1991}{1728}\pi^4
\nonumber &\\&
-\frac{470}{9}\pi^2\zeta_{3}+\frac{1612}{5}\zeta_{5} + \mathcal{O}(\e), &
\end{flalign}

\begin{flalign}
{\cal T}^{gg,(3,\left[1\times 1\right])}_{ggg}\Big|_{\NF^2 N} = &
+\frac{1}{\e^4}\left(\frac{1}{2}\right)
+\frac{1}{\e^3}\left(\frac{11}{12}\right)
+\frac{1}{\e^2}\left(\frac{25}{8}-\frac{7}{24}\pi^2\right)
\nonumber &\\&
+\frac{1}{\e}\left(\frac{13037}{1296}-\frac{77}{144}\pi^2-\frac{25}{6}\zeta_{3}\right)
\nonumber &\\&
+\frac{83119}{2592}-\frac{547}{288}\pi^2-\frac{275}{36}\zeta_{3}-\frac{71}{2880}\pi^4 + \mathcal{O}(\e), &
\end{flalign}



\begin{flalign}
{\cal T}^{gg,(3,\left[2\times 0\right])}_{ggg}\Big|_{N^{3}} = &
+\frac{1}{\e^6}\left(\frac{23}{9}\right)
+\frac{1}{\e^5}\left(\frac{5291}{216}\right)
+\frac{1}{\e^4}\left(\frac{14053}{162}-\frac{139}{18}\pi^2\right)
\nonumber &\\&
+\frac{1}{\e^3}\left(\frac{1204027}{3888}-\frac{50765}{1296}\pi^2-\frac{1195}{18}\zeta_{3}\right)
\nonumber &\\&
+\frac{1}{\e^2}\left(\frac{32569141}{23328}-\frac{1065509}{7776}\pi^2-\frac{17017}{36}\zeta_{3}+\frac{15613}{2592}\pi^4\right)
\nonumber &\\&
+\frac{1}{\e}\bigg(\frac{864831877}{139968}-\frac{29419889}{46656}\pi^2-\frac{197453}{108}\zeta_{3}
\nonumber &\\&\phantom{+\frac{1}{\e}\bigg(}
+\frac{97097}{6480}\pi^4+\frac{48313}{216}\pi^2\zeta_{3}-\frac{77033}{90}\zeta_{5}\bigg)
\nonumber &\\&
+\frac{23345187541}{839808}-\frac{859773641}{279936}\pi^2-\frac{695786}{81}\zeta_{3}+\frac{18651233}{311040}\pi^4
\nonumber &\\&
+\frac{416647}{432}\pi^2\zeta_{3}-\frac{893563}{180}\zeta_{5}-\frac{716087}{816480}\pi^6+\frac{24409}{18}\zeta_{3}^2 + \mathcal{O}(\e), &
\end{flalign}

\begin{flalign}
{\cal T}^{gg,(3,\left[2\times 0\right])}_{ggg}\Big|_{\NF N^{2}} = &
+\frac{1}{\e^5}\left(-\frac{367}{108}\right)
+\frac{1}{\e^4}\left(-\frac{1256}{81}\right)
+\frac{1}{\e^3}\left(-\frac{76247}{1944}+\frac{613}{162}\pi^2\right)
\nonumber &\\&
+\frac{1}{\e^2}\left(-\frac{1986005}{11664}+\frac{44611}{3888}\pi^2+\frac{99}{2}\zeta_{3}\right)
\nonumber &\\&
+\frac{1}{\e}\left(-\frac{46628939}{69984}+\frac{922633}{23328}\pi^2+\frac{6091}{36}\zeta_{3}-\frac{971}{8640}\pi^4\right)
\nonumber &\\&
-\frac{1069628111}{419904}+\frac{26618059}{139968}\pi^2+\frac{40385}{72}\zeta_{3}+\frac{324821}{155520}\pi^4
\nonumber &\\&
-\frac{10991}{216}\pi^2\zeta_{3}+\frac{37241}{90}\zeta_{5} + \mathcal{O}(\e), &
\end{flalign}

\begin{flalign}
{\cal T}^{gg,(3,\left[2\times 0\right])}_{ggg}\Big|_{\NF} = &
+\frac{1}{\e^3}\left(-\frac{3}{4}\right)
+\frac{1}{\e^2}\left(\frac{101}{24}-4 \zeta_{3}\right)
\nonumber &\\&
+\frac{1}{\e}\left(\frac{4883}{144}-\frac{109}{144}\pi^2-\frac{68}{3}\zeta_{3}-\frac{2}{27}\pi^4\right)
\nonumber &\\&
+\frac{52751}{288}-\frac{14305}{864}\pi^2-\frac{4019}{36}\zeta_{3}-\frac{34}{81}\pi^4
\nonumber &\\&
+\frac{103}{9}\pi^2\zeta_{3}-16 \zeta_{5} + \mathcal{O}(\e), &
\end{flalign}

\begin{flalign}
{\cal T}^{gg,(3,\left[2\times 0\right])}_{ggg}\Big|_{\NF^2 N} = &
+\frac{1}{\e^4}\left(\frac{5}{6}\right)
+\frac{1}{\e^3}\left(\frac{55}{36}\right)
+\frac{1}{\e^2}\left(\frac{1117}{216}-\frac{35}{72}\pi^2\right)
\nonumber &\\&
+\frac{1}{\e}\left(\frac{2359}{144}-\frac{385}{432}\pi^2-\frac{125}{18}\zeta_{3}\right)
\nonumber &\\&
+\frac{14621}{288}-\frac{7877}{2592}\pi^2-\frac{1375}{108}\zeta_{3}-\frac{71}{1728}\pi^4 + \mathcal{O}(\e), &
\end{flalign}



\begin{flalign}
{\cal T}^{gg,(3,\left[1\times 1\right])}_{q\bar{q}g}\Big|_{\NF N^{2}} = &
+\frac{1}{\e^5}\left(-\frac{2}{3}\right)
+\frac{1}{\e^4}\left(-6\right)
+\frac{1}{\e^3}\left(-\frac{8939}{324}+\frac{20}{27}\pi^2\right)
\nonumber &\\&
+\frac{1}{\e^2}\left(-\frac{24287}{216}+\frac{1289}{162}\pi^2+\frac{154}{9}\zeta_{3}\right)
\nonumber &\\&
+\frac{1}{\e}\left(-\frac{960457}{1944}+\frac{23275}{648}\pi^2+\frac{3674}{27}\zeta_{3}+\frac{1087}{6480}\pi^4\right)
\nonumber &\\&
-\frac{80043379}{34992}+\frac{3689513}{23328}\pi^2+\frac{18431}{27}\zeta_{3}-\frac{8843}{38880}\pi^4
\nonumber &\\&
-\frac{1193}{54}\pi^2\zeta_{3}+\frac{6127}{30}\zeta_{5} + \mathcal{O}(\e), &
\end{flalign}

\begin{flalign}
{\cal T}^{gg,(3,\left[1\times 1\right])}_{q\bar{q}g}\Big|_{\NF} = &
+\frac{1}{\e^5}\left(\frac{1}{3}\right)
+\frac{1}{\e^4}\left(\frac{115}{36}\right)
+\frac{1}{\e^3}\left(\frac{1387}{72}-\frac{17}{36}\pi^2\right)
\nonumber &\\&
+\frac{1}{\e^2}\left(\frac{67231}{648}-\frac{173}{36}\pi^2-13 \zeta_{3}\right)
\nonumber &\\&
+\frac{1}{\e}\left(\frac{130232}{243}-\frac{39373}{1296}\pi^2-\frac{1033}{9}\zeta_{3}-\frac{799}{4320}\pi^4\right)
\nonumber &\\&
+\frac{5296811}{1944}-\frac{1315985}{7776}\pi^2-\frac{77185}{108}\zeta_{3}-\frac{27481}{25920}\pi^4
\nonumber &\\&
+\frac{265}{12}\pi^2\zeta_{3}-\frac{3238}{15}\zeta_{5} + \mathcal{O}(\e), &
\end{flalign}

\begin{flalign}\label{eq:HggX32_1}
{\cal T}^{gg,(3,\left[1\times 1\right])}_{q\bar{q}g}\Big|_{\NF N^{-2}} = &
+\frac{1}{\e^5}\left(-\frac{1}{18}\right)
+\frac{1}{\e^4}\left(-\frac{61}{108}\right)
+\frac{1}{\e^3}\left(-\frac{661}{162}+\frac{17}{216}\pi^2\right)
\nonumber &\\&
+\frac{1}{\e^2}\left(-\frac{49697}{1944}+\frac{1037}{1296}\pi^2+\frac{49}{18}\zeta_{3}\right)
\nonumber &\\&
+\frac{1}{\e}\left(-\frac{1736189}{11664}+\frac{5821}{972}\pi^2+\frac{2773}{108}\zeta_{3}+\frac{517}{8640}\pi^4\right)
\nonumber &\\&
-\frac{14505527}{17496}+\frac{898633}{23328}\pi^2+\frac{15074}{81}\zeta_{3}+\frac{26713}{51840}\pi^4
\nonumber &\\&
-\frac{1229}{216}\pi^2\zeta_{3}+\frac{306}{5}\zeta_{5} + \mathcal{O}(\e), &
\end{flalign}

\begin{flalign}
{\cal T}^{gg,(3,\left[1\times 1\right])}_{q\bar{q}g}\Big|_{\NF^2 N} = &
+\frac{1}{\e^4}\left(\frac{4}{9}\right)
+\frac{1}{\e^3}\left(2\right)
+\frac{1}{\e^2}\left(\frac{53}{27}-\frac{31}{54}\pi^2\right)
\nonumber &\\&
+\frac{1}{\e}\left(-\frac{13393}{972}-\frac{731}{648}\pi^2-\frac{8}{3}\zeta_{3}\right)
\nonumber &\\&
-\frac{135533}{972}+\frac{4727}{1296}\pi^2+\frac{757}{54}\zeta_{3}+\frac{133}{432}\pi^4 + \mathcal{O}(\e), &
\end{flalign}

\begin{flalign}
{\cal T}^{gg,(3,\left[1\times 1\right])}_{q\bar{q}g}\Big|_{\NF^2 N^{-1}} = &
+\frac{1}{\e^4}\left(-\frac{5}{54}\right)
+\frac{1}{\e^3}\left(-\frac{121}{324}\right)
+\frac{1}{\e^2}\left(-\frac{29}{54}+\frac{47}{324}\pi^2\right)
\nonumber &\\&
+\frac{1}{\e}\left(\frac{30473}{5832}+\frac{1213}{1944}\pi^2+\frac{25}{27}\zeta_{3}\right)
\nonumber &\\&
+\frac{2264881}{34992}+\frac{409}{324}\pi^2-\frac{889}{162}\zeta_{3}-\frac{299}{3240}\pi^4 + \mathcal{O}(\e), &
\end{flalign}

\begin{flalign}
{\cal T}^{gg,(3,\left[1\times 1\right])}_{q\bar{q}g}\Big|_{\NF^3} = &
+\frac{1}{\e^3}\left(-\frac{13}{162}\right)
+\frac{1}{\e^2}\left(-\frac{7}{108}\right)
+\frac{1}{\e}\left(\frac{575}{1944}-\frac{67}{1944}\pi^2\right)
\nonumber &\\&
+\frac{49445}{34992}-\frac{595}{1296}\pi^2+\frac{\zeta_{3}}{162} + \mathcal{O}(\e), &
\end{flalign}



\begin{flalign}
{\cal T}^{gg,(3,\left[2\times 0\right])}_{q\bar{q}g}\Big|_{\NF N^{2}} = &
+\frac{1}{\e^5}\left(-\frac{2}{3}\right)
+\frac{1}{\e^4}\left(-\frac{47}{6}\right)
+\frac{1}{\e^3}\left(-\frac{11705}{324}+\frac{109}{54}\pi^2\right)
\nonumber &\\&
+\frac{1}{\e^2}\left(-\frac{178273}{1296}+\frac{8857}{648}\pi^2+\frac{148}{9}\zeta_{3}\right)
\nonumber &\\&
+\frac{1}{\e}\left(-\frac{766405}{1296}+\frac{13855}{243}\pi^2+\frac{2917}{18}\zeta_{3}-\frac{3509}{2160}\pi^4\right)
\nonumber &\\&
-\frac{373115107}{139968}+\frac{11784379}{46656}\pi^2+\frac{65441}{81}\zeta_{3}-\frac{479149}{77760}\pi^4
\nonumber &\\&
-\frac{1424}{27}\pi^2\zeta_{3}+\frac{842}{5}\zeta_{5} + \mathcal{O}(\e), &
\end{flalign}

\begin{flalign}
{\cal T}^{gg,(3,\left[2\times 0\right])}_{q\bar{q}g}\Big|_{\NF} = &
+\frac{1}{\e^5}\left(\frac{1}{3}\right)
+\frac{1}{\e^4}\left(\frac{100}{27}\right)
+\frac{1}{\e^3}\left(\frac{1798}{81}-\frac{61}{54}\pi^2\right)
\nonumber &\\&
+\frac{1}{\e^2}\left(\frac{149993}{1296}-\frac{12755}{1296}\pi^2-\frac{233}{18}\zeta_{3}\right)
\nonumber &\\&
+\frac{1}{\e}\left(\frac{13320487}{23328}-\frac{149693}{2592}\pi^2-\frac{243}{2}\zeta_{3}+\frac{12101}{12960}\pi^4\right)
\nonumber &\\&
+\frac{385924205}{139968}-\frac{14238229}{46656}\pi^2-\frac{116785}{162}\zeta_{3}+\frac{1062851}{155520}\pi^4
\nonumber &\\&
+\frac{10391}{216}\pi^2\zeta_{3}-\frac{6941}{30}\zeta_{5} + \mathcal{O}(\e), &
\end{flalign}

\begin{flalign}\label{eq:HggX32_2}
{\cal T}^{gg,(3,\left[2\times 0\right])}_{q\bar{q}g}\Big|_{\NF N^{-2}} = &
+\frac{1}{\e^5}\left(-\frac{1}{18}\right)
+\frac{1}{\e^4}\left(-\frac{61}{108}\right)
+\frac{1}{\e^3}\left(-\frac{661}{162}+\frac{41}{216}\pi^2\right)
\nonumber &\\&
+\frac{1}{\e^2}\left(-\frac{99313}{3888}+\frac{2465}{1296}\pi^2+\frac{55}{18}\zeta_{3}\right)
\nonumber &\\&
+\frac{1}{\e}\left(-\frac{1734407}{11664}+\frac{13429}{972}\pi^2+\frac{3121}{108}\zeta_{3}-\frac{3313}{25920}\pi^4\right)
\nonumber &\\&
-\frac{116159965}{139968}+\frac{4060601}{46656}\pi^2+\frac{65825}{324}\zeta_{3}-\frac{207853}{155520}\pi^4
\nonumber &\\&
-\frac{877}{72}\pi^2\zeta_{3}+\frac{2191}{30}\zeta_{5} + \mathcal{O}(\e), &
\end{flalign}

\begin{flalign}
{\cal T}^{gg,(3,\left[2\times 0\right])}_{q\bar{q}g}\Big|_{\NF^2 N} = &
+\frac{1}{\e^4}\left(\frac{7}{9}\right)
+\frac{1}{\e^3}\left(\frac{637}{162}\right)
+\frac{1}{\e^2}\left(\frac{2321}{324}-\frac{119}{324}\pi^2\right)
\nonumber &\\&
+\frac{1}{\e}\left(\frac{3577}{324}+\frac{173}{162}\pi^2-\frac{19}{3}\zeta_{3}\right)
\nonumber &\\&
-\frac{400247}{34992}+\frac{1715}{108}\pi^2-\frac{260}{81}\zeta_{3}-\frac{24449}{38880}\pi^4 + \mathcal{O}(\e), &
\end{flalign}

\begin{flalign}
{\cal T}^{gg,(3,\left[2\times 0\right])}_{q\bar{q}g}\Big|_{\NF^2 N^{-1}} = &
+\frac{1}{\e^4}\left(-\frac{5}{27}\right)
+\frac{1}{\e^3}\left(-\frac{74}{81}\right)
+\frac{1}{\e^2}\left(-\frac{1685}{648}+\frac{23}{216}\pi^2\right)
\nonumber &\\&
+\frac{1}{\e}\left(-\frac{42589}{11664}-\frac{575}{1296}\pi^2+\frac{229}{54}\zeta_{3}\right)
\nonumber &\\&
+\frac{2252299}{69984}-\frac{22421}{2592}\pi^2+\frac{4547}{324}\zeta_{3}+\frac{15673}{77760}\pi^4 + \mathcal{O}(\e), &
\end{flalign}

\begin{flalign}
{\cal T}^{gg,(3,\left[2\times 0\right])}_{q\bar{q}g}\Big|_{\NF^3} = &
+\frac{1}{\e^3}\left(-\frac{23}{162}\right)
+\frac{1}{\e^2}\left(-\frac{7}{36}\right)
+\frac{1}{\e}\left(-\frac{235}{1944}+\frac{103}{1944}\pi^2\right)
\nonumber &\\&
-\frac{47153}{34992}+\frac{511}{1296}\pi^2+\frac{17}{54}\zeta_{3} + \mathcal{O}(\e), &
\end{flalign}


\subsubsection{Four-particle final states}


\begin{flalign}
{\cal T}^{gg,(3)}_{gggg}\Big|_{N^{3}} = &
+\frac{1}{\e^6}\left(-\frac{113}{18}\right)
+\frac{1}{\e^5}\left(-\frac{1661}{36}\right)
+\frac{1}{\e^4}\left(-\frac{147871}{648}+\frac{3233}{216}\pi^2\right)
\nonumber &\\&
+\frac{1}{\e^3}\left(-\frac{4615997}{3888}+\frac{122815}{1296}\pi^2+\frac{4685}{18}\zeta_{3}\right)
\nonumber &\\&
+\frac{1}{\e^2}\left(-\frac{15593323}{2592}+\frac{3877639}{7776}\pi^2+\frac{179729}{108}\zeta_{3}-\frac{200657}{25920}\pi^4\right)
\nonumber &\\&
+\frac{1}{\e}\bigg(-\frac{4232245261}{139968}+\frac{125073985}{46656}\pi^2+\frac{5912125}{648}\zeta_{3}
\nonumber &\\& \phantom{+\frac{1}{\e}\bigg(}
-\frac{2118721}{51840}\pi^4-\frac{48431}{72}\pi^2\zeta_{3}+\frac{316459}{90}\zeta_{5}\bigg)
\nonumber &\\&
-\frac{127277031491}{839808}+\frac{3914403535}{279936}\pi^2+\frac{65526025}{1296}\zeta_{3}-\frac{66797881}{311040}\pi^4
\nonumber &\\&
-\frac{1657051}{432}\pi^2\zeta_{3}+\frac{3630913}{180}\zeta_{5}+\frac{15983183}{6531840}\pi^6-\frac{234529}{36}\zeta_{3}^2 + \mathcal{O}(\e), &
\end{flalign}

\begin{flalign}
{\cal T}^{gg,(3)}_{gggg}\Big|_{\NF N^{2}} = &
+\frac{1}{\e^5}\left(\frac{10}{3}\right)
+\frac{1}{\e^4}\left(\frac{121}{9}\right)
+\frac{1}{\e^3}\left(\frac{1171}{18}-\frac{44}{9}\pi^2\right)
\nonumber &\\&
+\frac{1}{\e^2}\left(\frac{31807}{108}-\frac{1067}{54}\pi^2-\frac{758}{9}\zeta_{3}\right)
\nonumber &\\&
+\frac{1}{\e}\left(\frac{5025205}{3888}-\frac{94729}{972}\pi^2-\frac{9196}{27}\zeta_{3}+\frac{41}{45}\pi^4\right)
\nonumber &\\&
+\frac{130080583}{23328}-\frac{869467}{1944}\pi^2-\frac{281887}{162}\zeta_{3}+\frac{36811}{9720}\pi^4
\nonumber &\\&
+\frac{3415}{27}\pi^2\zeta_{3}-\frac{6596}{9}\zeta_{5} + \mathcal{O}(\e), &
\end{flalign}



\begin{flalign}
{\cal T}^{gg,(3)}_{q\bar{q}gg}\Big|_{\NF N^{2}} = &
+\frac{1}{\e^5}\left(\frac{92}{27}\right)
+\frac{1}{\e^4}\left(\frac{2539}{81}\right)
+\frac{1}{\e^3}\left(\frac{173873}{972}-\frac{658}{81}\pi^2\right)
\nonumber &\\&
+\frac{1}{\e^2}\left(\frac{1406861}{1458}-\frac{32485}{486}\pi^2-\frac{410}{3}\zeta_{3}\right)
\nonumber &\\&
+\frac{1}{\e}\left(\frac{176604815}{34992}-\frac{4580383}{11664}\pi^2-\frac{10718}{9}\zeta_{3}+\frac{2981}{648}\pi^4\right)
\nonumber &\\&
+\frac{5457831455}{209952}-\frac{76313957}{34992}\pi^2-\frac{2365255}{324}\zeta_{3}+\frac{144191}{4860}\pi^4
\nonumber &\\&
+\frac{18689}{54}\pi^2\zeta_{3}-\frac{72307}{45}\zeta_{5} + \mathcal{O}(\e), &
\end{flalign}

\begin{flalign}
{\cal T}^{gg,(3)}_{q\bar{q}gg}\Big|_{\NF} = &
+\frac{1}{\e^5}\left(-\frac{77}{54}\right)
+\frac{1}{\e^4}\left(-\frac{4559}{324}\right)
+\frac{1}{\e^3}\left(-\frac{181577}{1944}+\frac{787}{216}\pi^2\right)
\nonumber &\\&
+\frac{1}{\e^2}\left(-\frac{6463727}{11664}+\frac{44281}{1296}\pi^2+\frac{3775}{54}\zeta_{3}\right)
\nonumber &\\&
+\frac{1}{\e}\left(-\frac{218284493}{69984}+\frac{1789471}{7776}\pi^2+\frac{214975}{324}\zeta_{3}-\frac{42967}{25920}\pi^4\right)
\nonumber &\\&
-\frac{7149077831}{419904}+\frac{64782577}{46656}\pi^2+\frac{8782471}{1944}\zeta_{3}-\frac{2218093}{155520}\pi^4
\nonumber &\\&
-\frac{40945}{216}\pi^2\zeta_{3}+\frac{33079}{30}\zeta_{5} + \mathcal{O}(\e), &
\end{flalign}

\begin{flalign}\label{eq:HggX41}
{\cal T}^{gg,(3)}_{q\bar{q}gg}\Big|_{\NF N^{-2}} = &
+\frac{1}{\e^5}\left(\frac{2}{9}\right)
+\frac{1}{\e^4}\left(\frac{61}{27}\right)
+\frac{1}{\e^3}\left(\frac{5387}{324}-\frac{31}{54}\pi^2\right)
\nonumber &\\&
+\frac{1}{\e^2}\left(\frac{103885}{972}-\frac{1873}{324}\pi^2-\frac{118}{9}\zeta_{3}\right)
\nonumber &\\&
+\frac{1}{\e}\left(\frac{3730075}{5832}-\frac{166547}{3888}\pi^2-\frac{6757}{54}\zeta_{3}+\frac{833}{6480}\pi^4\right)
\nonumber &\\&
+\frac{7996690}{2187}-\frac{3227527}{11664}\pi^2-\frac{293857}{324}\zeta_{3}+\frac{69983}{38880}\pi^4
\nonumber &\\&
+\frac{701}{18}\pi^2\zeta_{3}-\frac{13436}{45}\zeta_{5} + \mathcal{O}(\e), &
\end{flalign}

\begin{flalign}
{\cal T}^{gg,(3)}_{q\bar{q}gg}\Big|_{\NF^2 N} = &
+\frac{1}{\e^4}\left(-\frac{124}{81}\right)
+\frac{1}{\e^3}\left(-\frac{1717}{243}\right)
+\frac{1}{\e^2}\left(-\frac{745}{27}+\frac{277}{162}\pi^2\right)
\nonumber &\\&
+\frac{1}{\e}\left(-\frac{763547}{8748}+\frac{605}{108}\pi^2+\frac{2246}{81}\zeta_{3}\right)
\nonumber &\\&
-\frac{8813443}{52488}+\frac{2413}{324}\pi^2+\frac{21857}{243}\zeta_{3}+\frac{559}{6480}\pi^4 + \mathcal{O}(\e), &
\end{flalign}

\begin{flalign}
{\cal T}^{gg,(3)}_{q\bar{q}gg}\Big|_{\NF^2 N^{-1}} = &
+\frac{1}{\e^4}\left(\frac{8}{27}\right)
+\frac{1}{\e^3}\left(\frac{122}{81}\right)
+\frac{1}{\e^2}\left(\frac{430}{81}-\frac{25}{81}\pi^2\right)
\nonumber &\\&
+\frac{1}{\e}\left(\frac{31837}{2916}-\frac{208}{243}\pi^2-\frac{160}{27}\zeta_{3}\right)
\nonumber &\\&
-\frac{128623}{4374}+\frac{178}{81}\pi^2-\frac{1240}{81}\zeta_{3}-\frac{97}{1080}\pi^4 + \mathcal{O}(\e), &
\end{flalign}



\begin{flalign}
{\cal T}^{gg,(3)}_{q\bar{q}q\bar{q}}\Big|_{\NF N} = &
+\frac{1}{\e^4}\left(-\frac{2}{9}\right)
+\frac{1}{\e^3}\left(-\frac{64}{27}\right)
+\frac{1}{\e^2}\left(-\frac{265}{18}+\frac{29}{54}\pi^2\right)
\nonumber &\\&
+\frac{1}{\e}\left(-\frac{39211}{486}+\frac{829}{162}\pi^2+\frac{86}{9}\zeta_{3}\right)
\nonumber &\\&
-\frac{819503}{1944}+\frac{20695}{648}\pi^2+\frac{7640}{81}\zeta_{3}-\frac{325}{1296}\pi^4 + \mathcal{O}(\e), &
\end{flalign}

\begin{flalign}
{\cal T}^{gg,(3)}_{q\bar{q}q\bar{q}}\Big|_{\NF} = &
+\frac{1}{\e^2}\left(\frac{5}{6}-\frac{2}{3}\zeta_{3}\right)
+\frac{1}{\e}\left(\frac{461}{36}-\frac{77}{18}\zeta_{3}-\frac{7}{90}\pi^4\right)
\nonumber &\\&
+\frac{48805}{432}-\frac{55}{24}\pi^2-\frac{1271}{108}\zeta_{3}-\frac{19}{40}\pi^4+\frac{11}{6}\pi^2\zeta_{3}-\frac{161}{3}\zeta_{5} + \mathcal{O}(\e), &
\end{flalign}

\begin{flalign}
{\cal T}^{gg,(3)}_{q\bar{q}q\bar{q}}\Big|_{\NF N^{-1}} = &
+\frac{1}{\e^4}\left(\frac{1}{9}\right)
+\frac{1}{\e^3}\left(\frac{31}{27}\right)
+\frac{1}{\e^2}\left(\frac{77}{9}-\frac{11}{36}\pi^2\right)
\nonumber &\\&
+\frac{1}{\e}\left(\frac{54263}{972}-\frac{341}{108}\pi^2-\frac{61}{9}\zeta_{3}\right)
\nonumber &\\&
+\frac{989915}{2916}-\frac{2551}{108}\pi^2-\frac{1819}{27}\zeta_{3}+\frac{1153}{12960}\pi^4 + \mathcal{O}(\e), &
\end{flalign}

\begin{flalign}
{\cal T}^{gg,(3)}_{q\bar{q}q\bar{q}}\Big|_{\NF N^{-2}} = &
+\frac{1}{\e^2}\left(-\frac{5}{6}+\frac{2}{3}\zeta_{3}\right)
+\frac{1}{\e}\left(-\frac{223}{18}+\frac{95}{18}\zeta_{3}+\frac{\pi^4}{18}\right)
\nonumber &\\&
-\frac{12643}{108}+\frac{55}{24}\pi^2+\frac{3785}{108}\zeta_{3}+\frac{511}{1080}\pi^4-\frac{11}{6}\pi^2\zeta_{3}+\frac{59}{3}\zeta_{5} + \mathcal{O}(\e), &
\end{flalign}

\begin{flalign}
{\cal T}^{gg,(3)}_{q\bar{q}q\bar{q}}\Big|_{\NF^2} = &
+\frac{1}{\e^3}\left(\frac{2}{27}\right)
+\frac{1}{\e^2}\left(\frac{7}{27}\right)
+\frac{1}{\e}\left(-\frac{85}{486}-\frac{\pi^2}{54}\right)
\nonumber &\\&
-\frac{28699}{2916}+\frac{7}{12}\pi^2-\frac{2}{81}\zeta_{3} + \mathcal{O}(\e), &
\end{flalign}

\begin{flalign}
{\cal T}^{gg,(3)}_{q\bar{q}q\bar{q}}\Big|_{\NF^2 N^{-1}} = &
+\frac{1}{\e}\left(-\frac{5}{9}+\frac{4}{9}\zeta_{3}\right)
-\frac{313}{54}+\frac{46}{27}\zeta_{3}+\frac{\pi^4}{27} + \mathcal{O}(\e), &
\end{flalign}



\begin{flalign}
{\cal T}^{gg,(3)}_{q\bar{q}q'\bar{q}'}\Big|_{\NF (\NF-1) N} = &
+\frac{1}{\e^4}\left(-\frac{2}{9}\right)
+\frac{1}{\e^3}\left(-\frac{64}{27}\right)
+\frac{1}{\e^2}\left(-\frac{265}{18}+\frac{29}{54}\pi^2\right)
\nonumber &\\&
+\frac{1}{\e}\left(-\frac{39211}{486}+\frac{829}{162}\pi^2+\frac{86}{9}\zeta_{3}\right)
\nonumber &\\&
-\frac{819503}{1944}+\frac{20695}{648}\pi^2+\frac{7640}{81}\zeta_{3}-\frac{325}{1296}\pi^4 + \mathcal{O}(\e), &
\end{flalign}

\begin{flalign}
{\cal T}^{gg,(3)}_{q\bar{q}q'\bar{q}'}\Big|_{\NF (\NF-1) N^{-1}} = &
+\frac{1}{\e^4}\left(\frac{1}{9}\right)
+\frac{1}{\e^3}\left(\frac{31}{27}\right)
+\frac{1}{\e^2}\left(\frac{77}{9}-\frac{11}{36}\pi^2\right)
\nonumber &\\&
+\frac{1}{\e}\left(\frac{54263}{972}-\frac{341}{108}\pi^2-\frac{61}{9}\zeta_{3}\right)
\nonumber &\\&
+\frac{989915}{2916}-\frac{2551}{108}\pi^2-\frac{1819}{27}\zeta_{3}+\frac{1153}{12960}\pi^4 + \mathcal{O}(\e), &
\end{flalign}


\subsubsection{Five-particle final states}


\begin{flalign}
{\cal T}^{gg,(3)}_{ggggg}\Big|_{N^{3}} = &
+\frac{1}{\e^6}\left(\frac{5}{2}\right)
+\frac{1}{\e^5}\left(\frac{440}{27}\right)
+\frac{1}{\e^4}\left(\frac{16909}{162}-\frac{53}{8}\pi^2\right)
\nonumber &\\&
+\frac{1}{\e^3}\left(\frac{99671}{162}-\frac{27995}{648}\pi^2-\frac{1213}{9}\zeta_{3}\right)
\nonumber &\\&
+\frac{1}{\e^2}\left(\frac{10112047}{2916}-\frac{1090571}{3888}\pi^2-\frac{31427}{36}\zeta_{3}+\frac{94121}{25920}\pi^4\right)
\nonumber &\\&
+\frac{1}{\e}\bigg(\frac{41548462}{2187}-\frac{38894531}{23328}\pi^2-\frac{3780355}{648}\zeta_{3}
\nonumber &\\&\phantom{+\frac{1}{\e}\bigg(}
+\frac{105677}{4320}\pi^4+\frac{19835}{54}\pi^2\zeta_{3}-\frac{15767}{9}\zeta_{5}\bigg)
\nonumber &\\&
+\frac{1190169835}{11664}-\frac{1320846413}{139968}\pi^2-\frac{137338601}{3888}\zeta_{3}+\frac{480287}{3240}\pi^4
\nonumber &\\&
+\frac{1023077}{432}\pi^2\zeta_{3}-\frac{388685}{36}\zeta_{5}-\frac{5079149}{6531840}\pi^6+\frac{47611}{12}\zeta_{3}^2 + \mathcal{O}(\e), &
\end{flalign}



\begin{flalign}
{\cal T}^{gg,(3)}_{q\bar{q}ggg}\Big|_{\NF N^{2}} = &
+\frac{1}{\e^5}\left(-\frac{113}{54}\right)
+\frac{1}{\e^4}\left(-\frac{623}{36}\right)
+\frac{1}{\e^3}\left(-\frac{230443}{1944}+\frac{3569}{648}\pi^2\right)
\nonumber &\\&
+\frac{1}{\e^2}\left(-\frac{4292179}{5832}+\frac{180263}{3888}\pi^2+\frac{5861}{54}\zeta_{3}\right)
\nonumber &\\&
+\frac{1}{\e}\left(-\frac{100630897}{23328}+\frac{7489607}{23328}\pi^2+\frac{307771}{324}\zeta_{3}-\frac{17359}{5184}\pi^4\right)
\nonumber &\\&
-\frac{5114927471}{209952}+\frac{140289241}{69984}\pi^2+\frac{4373629}{648}\zeta_{3}-\frac{1340599}{51840}\pi^4
\nonumber &\\&
-\frac{63187}{216}\pi^2\zeta_{3}+\frac{54679}{45}\zeta_{5} + \mathcal{O}(\e), &
\end{flalign}

\begin{flalign}
{\cal T}^{gg,(3)}_{q\bar{q}ggg}\Big|_{\NF} = &
+\frac{1}{\e^5}\left(\frac{41}{54}\right)
+\frac{1}{\e^4}\left(\frac{581}{81}\right)
+\frac{1}{\e^3}\left(\frac{50515}{972}-\frac{49}{24}\pi^2\right)
\nonumber &\\&
+\frac{1}{\e^2}\left(\frac{1953463}{5832}-\frac{12667}{648}\pi^2-\frac{1190}{27}\zeta_{3}\right)
\nonumber &\\&
+\frac{1}{\e}\left(\frac{70758217}{34992}-\frac{34637}{243}\pi^2-\frac{138877}{324}\zeta_{3}+\frac{23831}{25920}\pi^4\right)
\nonumber &\\&
+\frac{2455817875}{209952}-\frac{3581371}{3888}\pi^2-\frac{6116899}{1944}\zeta_{3}+\frac{16571}{1944}\pi^4
\nonumber &\\&
+\frac{13019}{108}\pi^2\zeta_{3}-\frac{29248}{45}\zeta_{5} + \mathcal{O}(\e), &
\end{flalign}

\begin{flalign}\label{eq:HggX50}
{\cal T}^{gg,(3)}_{q\bar{q}ggg}\Big|_{\NF N^{-2}} = &
+\frac{1}{\e^5}\left(-\frac{1}{9}\right)
+\frac{1}{\e^4}\left(-\frac{61}{54}\right)
+\frac{1}{\e^3}\left(-\frac{2743}{324}+\frac{11}{36}\pi^2\right)
\nonumber &\\&
+\frac{1}{\e^2}\left(-\frac{217535}{3888}+\frac{671}{216}\pi^2+\frac{65}{9}\zeta_{3}\right)
\nonumber &\\&
+\frac{1}{\e}\left(-\frac{4031863}{11664}+\frac{30311}{1296}\pi^2+\frac{1915}{27}\zeta_{3}-\frac{913}{12960}\pi^4\right)
\nonumber &\\&
-\frac{285618695}{139968}+\frac{2404825}{15552}\pi^2+\frac{85813}{162}\zeta_{3}-\frac{81829}{77760}\pi^4
\nonumber &\\&
-\frac{2221}{108}\pi^2\zeta_{3}+\frac{13901}{90}\zeta_{5} + \mathcal{O}(\e), &
\end{flalign}



\begin{flalign}
{\cal T}^{gg,(3)}_{q\bar{q}q\bar{q}g}\Big|_{\NF N} = &
+\frac{1}{\e^4}\left(\frac{1}{3}\right)
+\frac{1}{\e^3}\left(\frac{521}{162}\right)
+\frac{1}{\e^2}\left(\frac{7571}{324}-\frac{289}{324}\pi^2\right)
\nonumber &\\&
+\frac{1}{\e}\left(\frac{146407}{972}-\frac{17069}{1944}\pi^2-\frac{491}{27}\zeta_{3}\right)
\nonumber &\\&
+\frac{31759967}{34992}-\frac{83377}{1296}\pi^2-\frac{30233}{162}\zeta_{3}+\frac{3893}{7776}\pi^4 + \mathcal{O}(\e), &
\end{flalign}

\begin{flalign}
{\cal T}^{gg,(3)}_{q\bar{q}q\bar{q}g}\Big|_{\NF} = &
+\frac{1}{\e^2}\left(-\frac{73}{72}+\frac{\pi^2}{36}+\frac{5}{9}\zeta_{3}\right)
\nonumber &\\&
+\frac{1}{\e}\left(-\frac{3563}{216}+\frac{83}{216}\pi^2+\frac{269}{54}\zeta_{3}+\frac{103}{1620}\pi^4\right)
\nonumber &\\&
-\frac{444697}{2592}+\frac{15961}{2592}\pi^2+\frac{5827}{162}\zeta_{3}+\frac{2111}{4860}\pi^4
\nonumber &\\&
-\frac{187}{108}\pi^2\zeta_{3}+\frac{140}{3}\zeta_{5} + \mathcal{O}(\e), &
\end{flalign}

\begin{flalign}
{\cal T}^{gg,(3)}_{q\bar{q}q\bar{q}g}\Big|_{\NF N^{-1}} = &
+\frac{1}{\e^4}\left(-\frac{7}{54}\right)
+\frac{1}{\e^3}\left(-\frac{443}{324}\right)
\nonumber &\\&
+\frac{1}{\e^2}\left(-\frac{2257}{216}+\frac{235}{648}\pi^2\right)
\nonumber &\\&
+\frac{1}{\e}\left(-\frac{812683}{11664}+\frac{14903}{3888}\pi^2+\frac{455}{54}\zeta_{3}\right)
\nonumber &\\&
-\frac{30173765}{69984}+\frac{76093}{2592}\pi^2+\frac{28471}{324}\zeta_{3}-\frac{9487}{77760}\pi^4 + \mathcal{O}(\e), &
\end{flalign}

\begin{flalign}
{\cal T}^{gg,(3)}_{q\bar{q}q\bar{q}g}\Big|_{\NF N^{-2}} = &
+\frac{1}{\e^2}\left(\frac{73}{72}-\frac{\pi^2}{36}-\frac{5}{9}\zeta_{3}\right)
+\frac{1}{\e}\left(\frac{433}{27}-\frac{77}{216}\pi^2-\frac{305}{54}\zeta_{3}-\frac{37}{810}\pi^4\right)
\nonumber &\\&
+\frac{415681}{2592}-\frac{14953}{2592}\pi^2-\frac{15299}{324}\zeta_{3}-\frac{971}{2430}\pi^4
\nonumber &\\&
+\frac{143}{108}\pi^2\zeta_{3}-\frac{88}{9}\zeta_{5} + \mathcal{O}(\e), &
\end{flalign}



\begin{flalign}
{\cal T}^{gg,(3)}_{q\bar{q}q'\bar{q}'g}\Big|_{\NF (\NF-1) N} = &
+\frac{1}{\e^4}\left(\frac{1}{3}\right)
+\frac{1}{\e^3}\left(\frac{521}{162}\right)
+\frac{1}{\e^2}\left(\frac{7571}{324}-\frac{289}{324}\pi^2\right)
\nonumber &\\&
+\frac{1}{\e}\left(\frac{146407}{972}-\frac{17069}{1944}\pi^2-\frac{491}{27}\zeta_{3}\right)
\nonumber &\\&
+\frac{31759967}{34992}-\frac{83377}{1296}\pi^2-\frac{30233}{162}\zeta_{3}+\frac{3893}{7776}\pi^4 + \mathcal{O}(\e), &
\end{flalign}

\begin{flalign}\label{eq:Hggend}
{\cal T}^{gg,(3)}_{q\bar{q}q'\bar{q}'g}\Big|_{\NF (\NF-1) N^{-1}} = &
+\frac{1}{\e^4}\left(-\frac{7}{54}\right)
+\frac{1}{\e^3}\left(-\frac{443}{324}\right)
+\frac{1}{\e^2}\left(-\frac{2257}{216}+\frac{235}{648}\pi^2\right)
\nonumber &\\&
+\frac{1}{\e}\left(-\frac{812683}{11664}+\frac{14903}{3888}\pi^2+\frac{455}{54}\zeta_{3}\right)
\nonumber &\\&
-\frac{30173765}{69984}+\frac{76093}{2592}\pi^2+\frac{28471}{324}\zeta_{3}-\frac{9487}{77760}\pi^4 + \mathcal{O}(\e), &
\end{flalign}


\section{Lower order results}\label{app:exprlower}

\subsection{Higgs to bottom quarks}

\subsubsection{NLO}

\begin{flalign}
{\cal T}^{q\bar{q},(1)}_{q\bar{q}}\Big|_{N^0} = 
&- \frac{1}{\epsilon^2}
+ \frac{1}{\epsilon}\left(-\frac{3}{2}\right)
+ \left(-1+\frac{7}{12}\pi^2\right)
+ \epsilon\left(-2+\frac{7}{3}\zeta_3\right)\nonumber &\\
&+ \epsilon^2\left(-4+\frac{7}{12}\pi^2-\frac{73}{1440}\pi^4\right)+ \epsilon^3\left(-8+\frac{7}{6}\pi^2+\frac{7}{3}\zeta_3-\frac{49}{36}\pi^2\zeta_3+\frac{31}{5}\zeta_5\right)\nonumber &\\
&+ \epsilon^4\left(-16+\frac{7}{3}\pi^2+\frac{14}{3}\zeta_3-\frac{73}{1440}\pi^4-\frac{437}{120960}\pi^6-\frac{49}{18}\zeta_3^{2}\right)+ \mathcal{O}(\epsilon^5), &
\end{flalign}
\begin{flalign}
{\cal T}^{q\bar{q},(1)}_{q\bar{q}g}\Big|_{N^0} = 
&+ \frac{1}{\epsilon^2} 
+ \frac{1}{\epsilon}\left(\frac{3}{2}\right)
+ \left(\frac{21}{4}-\frac{7}{12}\pi^2\right)
+ \epsilon\left(\frac{127}{8}-\frac{7}{8}\pi^2-\frac{25}{3}\zeta_3\right)\nonumber &\\
&+ \epsilon^2\left(\frac{765}{16}-\frac{49}{16}\pi^2-\frac{25}{2}\zeta_3-\frac{71}{1440}\pi^4\right)\nonumber &\\
&+ \epsilon^3\left(\frac{4599}{32}-\frac{889}{96}\pi^2-\frac{175}{4}\zeta_3-\frac{71}{960}\pi^4+\frac{175}{36}\pi^2\zeta_3-\frac{241}{5}\zeta_5\right)\nonumber &\\
&+ \epsilon^4\Bigg(+\frac{27621}{64}-\frac{1785}{64}\pi^2-\frac{3175}{24}\zeta_3-\frac{497}{1920}\pi^4
\nonumber &\\
&\phantom{+ \epsilon^2 \quad\,,}+\frac{175}{24}\pi^2\zeta_3-\frac{723}{10}\zeta_5-\frac{4027}{120960}\pi^6+\frac{625}{18}\zeta_3^{2}\Bigg)+ \mathcal{O}(\epsilon^5), &
\end{flalign}

\subsubsection{NNLO}

\begin{flalign}\label{eq:HbbX22_1}
{\cal T}^{q\bar{q},(2, \left[1\times 1\right])}_{q\bar{q}}\Big|_{N^{-1}} = 
&+ \frac{1}{\epsilon^4}\left(-\frac{1}{4}\right)
+ \frac{1}{\epsilon^3}\left(-\frac{3}{4}\right)
+ \frac{1}{\epsilon^2}\left(-\frac{17}{16}+\frac{1}{24}\pi^2\right)\nonumber  &\\
&+ \frac{1}{\epsilon}\left(-\frac{7}{4}+\frac{7}{16}\pi^2+\frac{7}{6}\zeta_3\right)
+ \left(-\frac{15}{4}+\frac{1}{12}\pi^2+\frac{7}{4}\zeta_3+\frac{7}{480}\pi^4\right)\nonumber &\\
&
+ \epsilon\left(-8+\frac{29}{48}\pi^2+\frac{7}{3}\zeta_3-\frac{73}{1920}\pi^4-\frac{7}{36}\pi^2\zeta_3+\frac{31}{10}\zeta_5\right)\nonumber &\\
&+ \epsilon^2\Bigg(-17+\frac{5}{4}\pi^2+\frac{77}{12}\zeta_3+\frac{7}{240}\pi^4\nonumber &\\
&\phantom{+ \epsilon^2 \quad\,,}-\frac{49}{48}\pi^2\zeta_3+\frac{93}{20}\zeta_5+\frac{31}{12096}\pi^6-\frac{49}{18}\zeta_3^{2}\Bigg)+ \mathcal{O}(\epsilon^3), &
\end{flalign}
\begin{flalign}\label{eq:HbbX22_2}
{\cal T}^{q\bar{q},(2,\left[1\times 1\right])}_{q\bar{q}}\Big|_{N} = 
&+ \frac{1}{\epsilon^4}\left(\frac{1}{4}\right)
+ \frac{1}{\epsilon^3}\left(\frac{3}{4}\right)
+ \frac{1}{\epsilon^2}\left(\frac{17}{16}-\frac{1}{24}\pi^2\right)
+ \frac{1}{\epsilon}\left(\frac{7}{4}-\frac{7}{16}\pi^2-\frac{7}{6}\zeta_3\right)\nonumber &\\
&+ \left(\frac{15}{4}-\frac{1}{12}\pi^2-\frac{7}{4}\zeta_3-\frac{7}{480}\pi^4\right)\nonumber &\\
&
+ \epsilon\left(8-\frac{29}{48}\pi^2-\frac{7}{3}\zeta_3+\frac{73}{1920}\pi^4+\frac{7}{36}\pi^2\zeta_3-\frac{31}{10}\zeta_5\right)\nonumber &\\
&+ \epsilon^2\Bigg(+17-\frac{5}{4}\pi^2-\frac{77}{12}\zeta_3-\frac{7}{240}\pi^4\nonumber &\\
&\phantom{+ \epsilon^2 \quad\,,}+\frac{49}{48}\pi^2\zeta_3-\frac{93}{20}\zeta_5-\frac{31}{12096}\pi^6+\frac{49}{18}\zeta_3^{2}\Bigg)+ \mathcal{O}(\epsilon^3), &
\end{flalign}
\begin{flalign}\label{eq:HbbX22_3}
{\cal T}^{q\bar{q},(2,\left[2\times 0\right])}_{q\bar{q}}\Big|_{N^{-1}} = 
&+ \frac{1}{\epsilon^4}\left(-\frac{1}{4}\right)
+ \frac{1}{\epsilon^3}\left(-\frac{3}{4}\right)
+ \frac{1}{\epsilon^2}\left(-\frac{17}{16}+\frac{13}{24}\pi^2\right)\nonumber &\\
&+ \frac{1}{\epsilon}\left(-\frac{53}{32}+\frac{5}{16}\pi^2+\frac{8}{3}\zeta_3\right)
+ \left(-\frac{17}{4}+\frac{3}{4}\pi^2+\frac{11}{2}\zeta_3-\frac{59}{288}\pi^4\right)\nonumber &\\
&
+ \epsilon\left(-11+\frac{27}{16}\pi^2+\frac{101}{6}\zeta_3+\frac{107}{384}\pi^4-\frac{55}{9}\pi^2\zeta_3+\frac{23}{5}\zeta_5\right)\nonumber &\\
&+ \epsilon^2\Bigg(-30+\frac{25}{6}\pi^2+\frac{893}{12}\zeta_3+\frac{37}{80}\pi^4\nonumber &\\
&\phantom{+ \epsilon^2 \quad\,,}-\frac{137}{16}\pi^2\zeta_3+\frac{102}{5}\zeta_5-\frac{571}{8640}\pi^6-\frac{326}{9}\zeta_3^{2}\Bigg)+ \mathcal{O}(\epsilon^3), &
\end{flalign}
\begin{flalign}
{\cal T}^{q\bar{q},(2,\left[2\times 0\right])}_{q\bar{q}}\Big|_{N_{F}} = 
&+ \frac{1}{\epsilon^3}\left(-\frac{1}{4}\right)
+ \frac{1}{\epsilon^2}\left(-\frac{1}{9}\right)
+ \frac{1}{\epsilon}\left(\frac{65}{216}+\frac{1}{24}\pi^2\right)
+ \left(\frac{50}{81}-\frac{55}{216}\pi^2+\frac{1}{18}\zeta_3\right)\nonumber &\\
&+ \epsilon\left(\frac{613}{243}-\frac{95}{162}\pi^2-\frac{65}{54}\zeta_3+\frac{1}{720}\pi^4\right)\nonumber &\\
&
+ \epsilon^2\left(\frac{6491}{729}-\frac{955}{486}\pi^2-\frac{236}{81}\zeta_3+\frac{79}{2592}\pi^4+\frac{47}{54}\pi^2\zeta_3-\frac{59}{30}\zeta_5\right)
\nonumber &\\&+ \mathcal{O}(\epsilon^3), &
\end{flalign}
\begin{flalign}
{\cal T}^{q\bar{q},(2,\left[2\times 0\right])}_{q\bar{q}}\Big|_{N} = 
&+ \frac{1}{\epsilon^4}\left(\frac{1}{4}\right)
+ \frac{1}{\epsilon^3}\left(\frac{17}{8}\right)
+ \frac{1}{\epsilon^2}\left(\frac{217}{144}-\frac{1}{2}\pi^2\right)\nonumber &\\
&+ \frac{1}{\epsilon}\left(-\frac{491}{864}-\frac{13}{24}\pi^2+\frac{7}{12}\zeta_3\right) 
+ \left(\frac{455}{162}+\frac{377}{432}\pi^2-\frac{47}{36}\zeta_3+\frac{263}{1440}\pi^4\right)\nonumber &\\
&
+ \epsilon\left(\frac{2557}{486}+\frac{727}{1296}\pi^2+\frac{1105}{108}\zeta_3-\frac{1169}{5760}\pi^4-\frac{13}{8}\pi^2\zeta_3+\frac{163}{20}\zeta_5\right)\nonumber &\\
&+ \epsilon^2\Bigg(+\frac{15239}{1458}+\frac{2869}{1944}\pi^2+\frac{5171}{324}\zeta_3-\frac{3889}{25920}\pi^4\nonumber &\\
&\phantom{+ \epsilon^2 \quad\,,}-\frac{2797}{432}\pi^2\zeta_3+\frac{101}{12}\zeta_5-\frac{631}{15120}\pi^6-\frac{403}{36}\zeta_3^{2}\Bigg)+ \mathcal{O}(\epsilon^3), &
\end{flalign}
\begin{flalign}\label{eq:HbbX31}
{\cal T}^{q\bar{q},(2)}_{q\bar{q}g}\Big|_{N^{-1}} = 
&+ \frac{1}{\epsilon^4}
+ \frac{3}{\epsilon^3}
+ \frac{1}{\epsilon^2}\left(\frac{73}{8}-\frac{4}{3}\pi^2\right)
+ \frac{1}{\epsilon}\left(\frac{131}{4}-\frac{23}{8}\pi^2-\frac{53}{3}\zeta_3\right)\nonumber &\\
&+ \left(\frac{1879}{16}-\frac{85}{8}\pi^2-39\zeta_3+\frac{19}{72}\pi^4\right)\nonumber &\\
&
+ \epsilon\left(\frac{13763}{32}-\frac{1261}{32}\pi^2-\frac{863}{6}\zeta_3+\frac{103}{320}\pi^4+\frac{493}{18}\pi^2\zeta_3-\frac{897}{5}\zeta_5\right)\nonumber &\\
&+ \epsilon^2\Bigg(+\frac{102725}{64}-\frac{9461}{64}\pi^2-\frac{13385}{24}\zeta_3+\frac{9979}{5760}\pi^4\nonumber &\\
&\phantom{+ \epsilon^2 \quad\,,}+\frac{387}{8}\pi^2\zeta_3-\frac{1707}{5}\zeta_5-\frac{827}{11340}\pi^6+\frac{1931}{9}\zeta_3^{2}\Bigg)+ \mathcal{O}(\epsilon^3), &
\end{flalign}
\begin{flalign}
{\cal T}^{q\bar{q},(2)}_{q\bar{q}g}\Big|_{N_{F}} = 
&+ \frac{1}{\epsilon^3}\left(\frac{1}{3}\right)
+ \frac{1}{\epsilon^2}\left(\frac{1}{2}\right)
+ \frac{1}{\epsilon}\left(\frac{7}{4}-\frac{7}{36}\pi^2\right)
+ \left(\frac{127}{24}-\frac{7}{24}\pi^2-\frac{25}{9}\zeta_3\right)\nonumber &\\
&+ \epsilon\left(\frac{255}{16}-\frac{49}{48}\pi^2-\frac{25}{6}\zeta_3-\frac{71}{4320}\pi^4\right)\nonumber &\\
&+ \epsilon^2\left(\frac{1533}{32}-\frac{889}{288}\pi^2-\frac{175}{12}\zeta_3-\frac{71}{2880}\pi^4+\frac{175}{108}\pi^2\zeta_3-\frac{241}{15}\zeta_5\right)\nonumber &\\&+ \mathcal{O}(\epsilon^3), &
\end{flalign}
\begin{flalign}
{\cal T}^{q\bar{q},(2)}_{q\bar{q}g}\Big|_{N} = 
&+ \frac{1}{\epsilon^4}\left(-\frac{5}{4}\right)
+ \frac{1}{\epsilon^3}\left(-\frac{67}{12}\right)
+ \frac{1}{\epsilon^2}\left(-\frac{125}{8}+\frac{13}{8}\pi^2\right)\nonumber &\\
&+ \frac{1}{\epsilon}\left(-\frac{471}{8}+\frac{365}{72}\pi^2+\frac{55}{3}\zeta_3\right) 
+ \left(-\frac{5203}{24}+\frac{865}{48}\pi^2+\frac{601}{9}\zeta_3-\frac{41}{96}\pi^4\right)\nonumber &\\
&
+ \epsilon\left(-812+\frac{3389}{48}\pi^2+\frac{945}{4}\zeta_3-\frac{2611}{4320}\pi^4-\frac{149}{6}\pi^2\zeta_3+143\zeta_5\right)\nonumber &\\
&+ \epsilon^2\Bigg(-\frac{49385}{16}+\frac{19681}{72}\pi^2+\frac{11411}{12}\zeta_3-\frac{1949}{576}\pi^4\nonumber &\\
&\phantom{+ \epsilon^2 \quad\,,}-\frac{1945}{27}\pi^2\zeta_3+\frac{7621}{15}\zeta_5+\frac{5357}{60480}\pi^6-\frac{1345}{9}\zeta_3^{2}\Bigg)+ \mathcal{O}(\epsilon^3), &
\end{flalign}
\begin{flalign}\label{eq:HbbX40}
{\cal T}^{q\bar{q},(2)}_{q\bar{q}gg}\Big|_{N^{-1}} = 
&+ \frac{1}{\epsilon^4}\left(-\frac{1}{2}\right)
+ \frac{1}{\epsilon^3}\left(-\frac{3}{2}\right)
+ \frac{1}{\epsilon^2}\left(-7+\frac{3}{4}\pi^2\right)
+ \frac{1}{\epsilon}\left(-\frac{965}{32}+\frac{9}{4}\pi^2+\frac{40}{3}\zeta_3\right)\nonumber &\\
&+ \left(-\frac{8201}{64}+\frac{253}{24}\pi^2+40\zeta_3-\frac{17}{144}\pi^4\right)\nonumber &\\
&+ \epsilon\left(-\frac{68803}{128}+\frac{8717}{192}\pi^2+\frac{2267}{12}\zeta_3-\frac{43}{120}\pi^4-\frac{121}{6}\pi^2\zeta_3+\frac{636}{5}\zeta_5\right)\nonumber &\\
&
+ \epsilon^2\Bigg(-\frac{571465}{256}+\frac{24683}{128}\pi^2+\frac{4879}{6}\zeta_3-\frac{383}{240}\pi^4\nonumber &\\
&\phantom{+ \epsilon^2 \quad\,,}-\frac{241}{4}\pi^2\zeta_3+\frac{1878}{5}\zeta_5+\frac{4763}{90720}\pi^6-\frac{3281}{18}\zeta_3^{2}\Bigg)+ \mathcal{O}(\epsilon^3), &
\end{flalign}
\begin{flalign}
{\cal T}^{q\bar{q},(2)}_{q\bar{q}gg}\Big|_{N} = 
&+ \frac{1}{\epsilon^4}\left(\frac{3}{4}\right)
+ \frac{1}{\epsilon^3}\left(\frac{65}{24}\right)
+ \frac{1}{\epsilon^2}\left(\frac{235}{18}-\frac{13}{12}\pi^2\right)
+ \frac{1}{\epsilon}\left(\frac{49847}{864}-\frac{589}{144}\pi^2-\frac{71}{4}\zeta_3\right)\nonumber &\\
&+ \left(\frac{1290485}{5184}-\frac{4243}{216}\pi^2-\frac{1327}{18}\zeta_3+\frac{373}{1440}\pi^4\right)\nonumber &\\
&
+ \epsilon\left(\frac{32779133}{31104}-\frac{449551}{5184}\pi^2-\frac{37951}{108}\zeta_3+\frac{5207}{8640}\pi^4+\frac{1891}{72}\pi^2\zeta_3-\frac{2661}{20}\zeta_5\right)\nonumber &\\
&+ \epsilon^2\Bigg(+\frac{821060357}{186624}-\frac{11633629}{31104}\pi^2-\frac{250801}{162}\zeta_3+\frac{19717}{6480}\pi^4\nonumber &\\
&\phantom{+ \epsilon^2 \quad\,,}+\frac{11969}{108}\pi^2\zeta_3-\frac{4211}{6}\zeta_5-\frac{139}{30240}\pi^6+\frac{2723}{12}\zeta_3^{2}\Bigg)+ \mathcal{O}(\epsilon^3), &
\end{flalign}
\begin{flalign}
{\cal T}^{q\bar{q},(2)}_{q\bar{q}q\bar{q}}\Big|_{N^{-1}} = 
&+ \frac{1}{\epsilon}\left(\frac{13}{16}-\frac{1}{8}\pi^2+\frac{1}{2}\zeta_3\right)
+ \left(\frac{253}{32}-\frac{3}{8}\pi^2-6\zeta_3+\frac{2}{45}\pi^4\right)\nonumber &\\
&
+ \epsilon\left(\frac{3331}{64}-\frac{221}{96}\pi^2-17\zeta_3-\frac{1}{6}\pi^4-\frac{11}{12}\pi^2\zeta_3+22\zeta_5\right)\nonumber &\\
&+ \epsilon^2\Bigg(+\frac{37001}{128}-\frac{2909}{192}\pi^2-\frac{803}{12}\zeta_3-\frac{37}{90}\pi^4\nonumber &\\
&\phantom{+ \epsilon^2 \quad\,,}+\frac{59}{6}\pi^2\zeta_3-\frac{489}{4}\zeta_5+\frac{277}{11340}\pi^6-\frac{95}{6}\zeta_3^{2}\Bigg)+ \mathcal{O}(\epsilon^3), &
\end{flalign}
\begin{flalign}
{\cal T}^{q\bar{q},(2)}_{q\bar{q}Q\bar{Q}}\Big|_{(N_{F}-1)} = 
&+ \frac{1}{\epsilon^3}\left(-\frac{1}{12}\right)
+ \frac{1}{\epsilon^2}\left(-\frac{7}{18}\right)
+ \frac{1}{\epsilon}\left(-\frac{443}{216}+\frac{11}{72}\pi^2\right)\nonumber &\\
&
+ \left(-\frac{12923}{1296}+\frac{17}{27}\pi^2+\frac{67}{18}\zeta_3\right)\nonumber &\\
&+ \epsilon\left(-\frac{358115}{7776}+\frac{4171}{1296}\pi^2+\frac{695}{54}\zeta_3+\frac{137}{4320}\pi^4\right)\nonumber &\\
&+ \epsilon^2\Bigg(-\frac{9579035}{46656}+\frac{119635}{7776}\pi^2+\frac{5051}{81}\zeta_3-\frac{49}{6480}\pi^4\nonumber &\\
&\phantom{+ \epsilon^2 \quad\,,}-\frac{629}{108}\pi^2\zeta_3+\frac{1651}{30}\zeta_5\Bigg)+ \mathcal{O}(\epsilon^3) &
\end{flalign}

\subsection{Higgs to gluons}

\subsubsection{NLO}

\begin{flalign}
{\cal T}^{gg,(1)}_{gg}\Big|_{N_{F}} = 
&+ \frac{1}{\epsilon}\left(\frac{2}{3}\right), &
\end{flalign}
\begin{flalign}
{\cal T}^{gg,(1)}_{gg}\Big|_{N} = 
&+ \frac{1}{\epsilon^2}\left(-2\right)
+ \frac{1}{\epsilon}\left(-\frac{11}{3}\right)
+ \left(+\frac{7}{6}\pi^2\right)
+ \epsilon\left(-2+\frac{14}{3}\zeta_3\right)\nonumber &\\
&+ \epsilon^2\left(-6-\frac{73}{720}\pi^4\right)
+ \epsilon^3\left(-14+\frac{7}{6}\pi^2-\frac{49}{18}\pi^2\zeta_3+\frac{62}{5}\zeta_5\right)\nonumber &\\
&+ \epsilon^4\left(-30+\frac{7}{2}\pi^2+\frac{14}{3}\zeta_3-\frac{437}{60480}\pi^6-\frac{49}{9}\zeta_3^{2}\right)+ \mathcal{O}(\epsilon^5), &
\end{flalign}
\begin{flalign}
{\cal T}^{gg,(1)}_{ggg}\Big|_{N} = 
&+ \frac{1}{\epsilon^2}\left(2\right)
+ \frac{1}{\epsilon}\left(\frac{11}{3}\right)
+ \left(\frac{73}{6}-\frac{7}{6}\pi^2\right)
+ \epsilon\left(\frac{451}{12}-\frac{77}{36}\pi^2-\frac{50}{3}\zeta_3\right)\nonumber &\\
&+ \epsilon^2\left(\frac{2729}{24}-\frac{511}{72}\pi^2-\frac{275}{9}\zeta_3-\frac{71}{720}\pi^4\right)\nonumber &\\
&+ \epsilon^3\left(\frac{16411}{48}-\frac{3157}{144}\pi^2-\frac{1825}{18}\zeta_3-\frac{781}{4320}\pi^4+\frac{175}{18}\pi^2\zeta_3-\frac{482}{5}\zeta_5\right)\nonumber &\\
&+ \epsilon^4\Bigg(+\frac{98513}{96}-\frac{19103}{288}\pi^2-\frac{11275}{36}\zeta_3-\frac{5183}{8640}\pi^4\nonumber &\\
&\phantom{+ \epsilon^2 \quad\,,}+\frac{1925}{108}\pi^2\zeta_3-\frac{2651}{15}\zeta_5-\frac{4027}{60480}\pi^6+\frac{625}{9}\zeta_3^{2}\Bigg)+ \mathcal{O}(\epsilon^5), &
\end{flalign}
\begin{flalign}
{\cal T}^{gg,(1)}_{q\bar{q}g}\Big|_{N_{F}} = 
&+ \frac{1}{\epsilon}\left(-\frac{2}{3}\right)
+ \left(-\frac{7}{3}\right)
+ \epsilon\left(-\frac{15}{2}+\frac{7}{18}\pi^2\right)\nonumber &\\
&+ \epsilon^2\left(-\frac{93}{4}+\frac{49}{36}\pi^2+\frac{50}{9}\zeta_3\right)
+ \epsilon^3\left(-\frac{567}{8}+\frac{35}{8}\pi^2+\frac{175}{9}\zeta_3+\frac{71}{2160}\pi^4\right)\nonumber &\\
&+ \epsilon^4\left(-\frac{3429}{16}+\frac{217}{16}\pi^2+\frac{125}{2}\zeta_3+\frac{497}{4320}\pi^4-\frac{175}{54}\pi^2\zeta_3+\frac{482}{15}\zeta_5\right)\nonumber &\\&+ \mathcal{O}(\epsilon^5), &
\end{flalign}

\subsubsection{NNLO}

\begin{flalign}
{\cal T}^{gg,(2,\left[1\times 1\right])}_{gg}\Big|_{N_{F}^2} = 
&+ \frac{1}{\epsilon^2}\left(\frac{1}{9}\right), &
\end{flalign}
\begin{flalign}
{\cal T}^{gg,(2,\left[1\times 1\right])}_{gg}\Big|_{N_{F} N} = 
&+ \frac{1}{\epsilon^3}\left(-\frac{2}{3}\right)
+ \frac{1}{\epsilon^2}\left(-\frac{11}{9}\right)
+ \frac{1}{\epsilon}\left(+\frac{7}{18}\pi^2\right)
+ \left(-\frac{2}{3}+\frac{14}{9}\zeta_3\right)\nonumber &\\
&+ \epsilon\left(-2-\frac{73}{2160}\pi^4\right)
+ \epsilon^2\left(-\frac{14}{3}+\frac{7}{18}\pi^2-\frac{49}{54}\pi^2\zeta_3+\frac{62}{15}\zeta_5\right)\nonumber &\\&+ \mathcal{O}(\epsilon^3), &
\end{flalign}
\begin{flalign}
{\cal T}^{gg,(2,\left[1\times 1\right])}_{gg}\Big|_{N^2} = 
&+ \frac{1}{\epsilon^4}\left(1\right)
+ \frac{1}{\epsilon^3}\left(\frac{11}{3}\right)
+ \frac{1}{\epsilon^2}\left(\frac{121}{36}-\frac{1}{6}\pi^2\right)\nonumber &\\
&+ \frac{1}{\epsilon}\left(2-\frac{77}{36}\pi^2-\frac{14}{3}\zeta_3\right)
+ \left(\frac{29}{3}-\frac{77}{9}\zeta_3-\frac{7}{120}\pi^4\right)\nonumber &\\
&+ \epsilon\left(25-\frac{1}{3}\pi^2+\frac{803}{4320}\pi^4+\frac{7}{9}\pi^2\zeta_3-\frac{62}{5}\zeta_5\right)\nonumber &\\
&+ \epsilon^2\Bigg(+\frac{170}{3}-\frac{113}{36}\pi^2-\frac{28}{3}\zeta_3\nonumber &\\
&\phantom{+ \epsilon^2 \quad\,,}+\frac{539}{108}\pi^2\zeta_3-\frac{341}{15}\zeta_5-\frac{31}{3024}\pi^6+\frac{98}{9}\zeta_3^{2}\Bigg)+ \mathcal{O}(\epsilon^3), &
\end{flalign}
\begin{flalign}
{\cal T}^{gg,(2,\left[2\times 0\right])}_{gg }\Big|_{N_{F}N^{-1}} = 
&+ \frac{1}{\epsilon}\left(-\frac{1}{4}\right)
+ \left(\frac{67}{24}-2\zeta_3\right)
+ \epsilon\left(\frac{2027}{144}-\frac{43}{72}\pi^2-\frac{23}{3}\zeta_3-\frac{1}{27}\pi^4\right)\nonumber &\\
&+ \epsilon^2\left(\frac{47491}{864}-\frac{2621}{432}\pi^2-\frac{281}{9}\zeta_3-\frac{23}{162}\pi^4+\frac{41}{9}\pi^2\zeta_3-8\zeta_5\right)
\nonumber &\\& + \mathcal{O}(\epsilon^3), &
\end{flalign}
\begin{flalign}
{\cal T}^{gg,(2,\left[2\times 0\right])}_{gg }\Big|_{N_{F}^2} = 
&+ \frac{1}{\epsilon^2}\left(\frac{2}{9}\right), &
\end{flalign}
\begin{flalign}
{\cal T}^{gg,(2,\left[2\times 0\right])}_{gg }\Big|_{N_{F}N} = 
&+ \frac{1}{\epsilon^3}\left(-\frac{7}{6}\right)
+ \frac{1}{\epsilon^2}\left(-\frac{13}{6}\right)
+ \frac{1}{\epsilon}\left(\frac{155}{108}+\frac{13}{36}\pi^2\right)\nonumber &\\
&+ \left(-\frac{5905}{648}-\frac{25}{36}\pi^2+1\zeta_3\right)
+ \epsilon\left(-\frac{162805}{3888}+\frac{305}{216}\pi^2-\frac{95}{27}\zeta_3+\frac{127}{720}\pi^4\right)\nonumber &\\
&+ \epsilon^2\Bigg(-\frac{3663205}{23328}+\frac{21911}{1296}\pi^2+\frac{548}{81}\zeta_3+\frac{527}{1296}\pi^4\nonumber &\\
&\phantom{+ \epsilon^2 \quad\,,}+\frac{175}{54}\pi^2\zeta_3-\frac{9}{5}\zeta_5\Bigg)+ \mathcal{O}(\epsilon^3), &
\end{flalign}
\begin{flalign}
{\cal T}^{gg,(2,\left[2\times 0\right])}_{gg}\Big|_{N^2} = 
&+ \frac{1}{\epsilon^4}\left(1\right)
+ \frac{1}{\epsilon^3}\left(\frac{77}{12}\right)
+ \frac{1}{\epsilon^2}\left(\frac{175}{36}-\frac{25}{12}\pi^2\right)\nonumber &\\
&+ \frac{1}{\epsilon}\left(-\frac{119}{27}-\frac{143}{72}\pi^2-\frac{25}{6}\zeta_3\right)
+ \left(\frac{8237}{324}+\frac{335}{72}\pi^2-\frac{33}{2}\zeta_3+\frac{31}{40}\pi^4\right)\nonumber &\\
&+ \epsilon\left(\frac{200969}{1944}-\frac{83}{54}\pi^2-\frac{1139}{54}\zeta_3-\frac{5071}{4320}\pi^4+\frac{323}{36}\pi^2\zeta_3+\frac{71}{10}\zeta_5\right)\nonumber &\\
&+ \epsilon^2\Bigg(+\frac{4082945}{11664}-\frac{25735}{648}\pi^2-\frac{13109}{81}\zeta_3-\frac{15343}{4320}\pi^4\nonumber &\\
&\phantom{+ \epsilon^2 \quad\,,}+\frac{781}{108}\pi^2\zeta_3-\frac{341}{10}\zeta_5+\frac{491}{10080}\pi^6+\frac{901}{18}\zeta_3^{2}\Bigg)+ \mathcal{O}(\epsilon^3), &
\end{flalign}
\begin{flalign}
{\cal T}^{gg,(2)}_{ggg}\Big|_{N_{F}N} = 
&+ \frac{1}{\epsilon^3}\left(2\right)
+ \frac{1}{\epsilon^2}\left(\frac{11}{3}\right)
+ \frac{1}{\epsilon}\left(\frac{37}{3}-\frac{7}{6}\pi^2\right)
+ \left(\frac{467}{12}-\frac{77}{36}\pi^2-\frac{50}{3}\zeta_3\right)\nonumber &\\
&+ \epsilon\left(\frac{26149}{216}-\frac{529}{72}\pi^2-\frac{275}{9}\zeta_3-\frac{71}{720}\pi^4\right)\nonumber &\\
&+ \epsilon^2\left(\frac{162827}{432}-\frac{3445}{144}\pi^2-\frac{629}{6}\zeta_3-\frac{781}{4320}\pi^4+\frac{175}{18}\pi^2\zeta_3-\frac{482}{5}\zeta_5\right)\nonumber &\\&+ \mathcal{O}(\epsilon^3), &
\end{flalign}
\begin{flalign}
{\cal T}^{gg,(2)}_{ggg}\Big|_{N^2} = 
&+ \frac{1}{\epsilon^4}\left(-\frac{9}{2}\right)
+ \frac{1}{\epsilon^3}\left(-\frac{121}{6}\right)
+ \frac{1}{\epsilon^2}\left(-\frac{170}{3}+\frac{71}{12}\pi^2\right)\nonumber &\\
&+ \frac{1}{\epsilon}\left(-\frac{23195}{108}+\frac{341}{18}\pi^2+72\zeta_3\right)\nonumber &\\
&+ \left(-\frac{173249}{216}+\frac{13831}{216}\pi^2+\frac{2266}{9}\zeta_3-\frac{199}{144}\pi^4\right)\nonumber &\\
&+ \epsilon\left(-\frac{11793239}{3888}+\frac{332557}{1296}\pi^2+\frac{15799}{18}\zeta_3-\frac{1991}{864}\pi^4-\frac{940}{9}\pi^2\zeta_3+\frac{3224}{5}\zeta_5\right)\nonumber &\\
&+ \epsilon^2\Bigg(-\frac{90432685}{7776}+\frac{7855165}{7776}\pi^2+\frac{1168555}{324}\zeta_3-\frac{53491}{5184}\pi^4\nonumber &\\
&\phantom{+ \epsilon^2 \quad\,,}-\frac{31361}{108}\pi^2\zeta_3+\frac{10186}{5}\zeta_5+\frac{29303}{90720}\pi^6-728\zeta_3^{2}\Bigg)+ \mathcal{O}(\epsilon^3), &
\end{flalign}
\begin{flalign}
{\cal T}^{gg,(2)}_{q\bar{q}g}\Big|_{N_{F}N^{-1}} = 
&+ \frac{1}{\epsilon^3}\left(-\frac{1}{3}\right)
+ \frac{1}{\epsilon^2}\left(-\frac{41}{18}\right)
+ \frac{1}{\epsilon}\left(-\frac{325}{27}+\frac{1}{2}\pi^2\right)\nonumber &\\
&+ \left(-\frac{18457}{324}+\frac{41}{12}\pi^2+\frac{74}{9}\zeta_3\right)\nonumber &\\
&+ \epsilon\left(-\frac{493931}{1944}+\frac{55}{3}\pi^2+\frac{1409}{27}\zeta_3-\frac{67}{1080}\pi^4\right)\nonumber &\\
&+ \epsilon^2\Bigg(-\frac{12729823}{11664}+\frac{6289}{72}\pi^2+\frac{22349}{81}\zeta_3-\frac{751}{1296}\pi^4\nonumber &\\
&\phantom{+ \epsilon^2 \quad\,,}-\frac{127}{9}\pi^2\zeta_3+\frac{1522}{15}\zeta_5\Bigg)+ \mathcal{O}(\epsilon^3), &
\end{flalign}
\begin{flalign}
{\cal T}^{gg,(2)}_{q\bar{q}g}\Big|_{N_{F}^2} = 
&+ \frac{1}{\epsilon^2}\left(-\frac{4}{9}\right)
+ \frac{1}{\epsilon}\left(-\frac{7}{9}\right)
+ \left(\frac{85}{162}+\frac{1}{18}\pi^2\right)
+ \epsilon\left(\frac{161}{12}-\frac{35}{36}\pi^2+\frac{26}{27}\zeta_3\right)\nonumber &\\
&+ \epsilon^2\left(\frac{506689}{5832}-\frac{1655}{216}\pi^2-\frac{343}{27}\zeta_3+\frac{47}{432}\pi^4\right)+ \mathcal{O}(\epsilon^3), &
\end{flalign}
\begin{flalign}
{\cal T}^{gg,(2)}_{q\bar{q}g}\Big|_{N_{F} N} = 
&+ \frac{1}{\epsilon^3}\left(\frac{4}{3}\right)
+ \frac{1}{\epsilon^2}\left(\frac{25}{3}\right)
+ \frac{1}{\epsilon}\left(\frac{805}{27}-\frac{16}{9}\pi^2\right)\nonumber &\\
&+ \left(\frac{2926}{27}-\frac{947}{108}\pi^2-\frac{188}{9}\zeta_3\right)
+ \epsilon\left(\frac{98302}{243}-\frac{21449}{648}\pi^2-\frac{1043}{9}\zeta_3+\frac{41}{90}\pi^4\right)\nonumber &\\
&+ \epsilon^2\Bigg(+\frac{83557}{54}-\frac{499517}{3888}\pi^2-\frac{74869}{162}\zeta_3+\frac{6107}{4320}\pi^4\nonumber &\\
&\phantom{+ \epsilon^2 \quad\,,}+\frac{782}{27}\pi^2\zeta_3-\frac{2588}{15}\zeta_5\Bigg)+ \mathcal{O}(\epsilon^3), &
\end{flalign}
\begin{flalign}
{\cal T}^{gg,(2)}_{gggg}\Big|_{N^2} = 
&+ \frac{1}{\epsilon^4}\left(\frac{5}{2}\right)
+ \frac{1}{\epsilon^3}\left(\frac{121}{12}\right)
+ \frac{1}{\epsilon^2}\left(\frac{436}{9}-\frac{11}{3}\pi^2\right)\nonumber &\\&
+ \frac{1}{\epsilon}\left(\frac{23455}{108}-\frac{1067}{72}\pi^2-\frac{379}{6}\zeta_3\right)\nonumber &\\
&+ \left(\frac{304951}{324}-\frac{7781}{108}\pi^2-\frac{2288}{9}\zeta_3+\frac{479}{720}\pi^4\right)\nonumber &\\
&+ \epsilon\left(\frac{7761353}{1944}-\frac{210323}{648}\pi^2-\frac{68353}{54}\zeta_3+\frac{4081}{1440}\pi^4+\frac{3409}{36}\pi^2\zeta_3-\frac{1129}{2}\zeta_5\right)\nonumber &\\
&+ \epsilon^2\Bigg(+\frac{194419813}{11664}-\frac{2737255}{1944}\pi^2-\frac{466313}{81}\zeta_3+\frac{19703}{1620}\pi^4\nonumber &\\
&\phantom{+ \epsilon^2 \quad\,,}+\frac{20515}{54}\pi^2\zeta_3-\frac{66011}{30}\zeta_5-\frac{1849}{11340}\pi^6+\frac{15301}{18}\zeta_3^{2}\Bigg)+ \mathcal{O}(\epsilon^3), &
\end{flalign}
\begin{flalign}
{\cal T}^{gg,(2)}_{q\bar{q}gg}\Big|_{N_{F}N^{-1}} = 
&+ \frac{1}{\epsilon^3}\left(\frac{1}{3}\right)
+ \frac{1}{\epsilon^2}\left(\frac{41}{18}\right)
+ \frac{1}{\epsilon}\left(\frac{1327}{108}-\frac{1}{2}\pi^2\right)
+ \left(\frac{4864}{81}-\frac{41}{12}\pi^2-\frac{86}{9}\zeta_3\right)\nonumber &\\
&+ \epsilon\left(\frac{134897}{486}-\frac{1331}{72}\pi^2-\frac{1709}{27}\zeta_3+\frac{23}{1080}\pi^4\right)\nonumber &\\
&+ \epsilon^2\Bigg(+\frac{898435}{729}-\frac{4865}{54}\pi^2-\frac{27545}{81}\zeta_3+\frac{419}{1296}\pi^4\nonumber &\\
&\phantom{+ \epsilon^2 \quad\,,}+\frac{131}{9}\pi^2\zeta_3-\frac{1672}{15}\zeta_5\Bigg)+ \mathcal{O}(\epsilon^3), &
\end{flalign}
\begin{flalign}
{\cal T}^{gg,(2)}_{q\bar{q}gg}\Big|_{N_{F} N} = 
&+ \frac{1}{\epsilon^3}\left(-\frac{3}{2}\right)
+ \frac{1}{\epsilon^2}\left(-\frac{155}{18}\right)
+ \frac{1}{\epsilon}\left(-\frac{523}{12}+\frac{79}{36}\pi^2\right)\nonumber &\\
&+ \left(-\frac{16579}{81}+\frac{1385}{108}\pi^2+37\zeta_3\right)\nonumber &\\
&+ \epsilon\left(-\frac{74282}{81}+\frac{42251}{648}\pi^2+\frac{6065}{27}\zeta_3-\frac{1007}{2160}\pi^4\right)\nonumber &\\
&+ \epsilon^2\Bigg(-\frac{5799143}{1458}+\frac{74429}{243}\pi^2+\frac{10376}{9}\zeta_3-\frac{2921}{1296}\pi^4\nonumber &\\
&\phantom{+ \epsilon^2 \quad\,,}-\frac{2971}{54}\pi^2\zeta_3+\frac{1503}{5}\zeta_5\Bigg)+ \mathcal{O}(\epsilon^3), &
\end{flalign}
\begin{flalign}
{\cal T}^{gg,(2)}_{q\bar{q}q\bar{q}}\Big|_{N_{F}N^{-1}} = 
&+ \left(-\frac{5}{12}+\frac{1}{3}\zeta_3\right)
+ \epsilon\left(-\frac{313}{72}+\frac{23}{18}\zeta_3+\frac{1}{36}\pi^4\right)\nonumber &\\
&+ \epsilon^2\left(-\frac{12521}{432}+\frac{5}{8}\pi^2+\frac{127}{27}\zeta_3+\frac{23}{216}\pi^4-\frac{1}{2}\pi^2\zeta_3+12\zeta_5\right)\nonumber &\\&+ \mathcal{O}(\epsilon^3), &
\end{flalign}
\begin{flalign}
{\cal T}^{gg,(2)}_{q\bar{q}Q\bar{Q}}\Big|_{N_{F}(N_{F}-1)} = 
&+ \frac{1}{\epsilon^2}\left(\frac{1}{9}\right)
+ \frac{1}{\epsilon}\left(\frac{7}{9}\right)
+ \left(\frac{677}{162}-\frac{1}{6}\pi^2\right)
+ \epsilon\left(\frac{241}{12}-\frac{7}{6}\pi^2-\frac{80}{27}\zeta_3\right)\nonumber &\\
&+ \epsilon^2\left(\frac{529217}{5832}-\frac{677}{108}\pi^2-\frac{560}{27}\zeta_3+\frac{29}{1080}\pi^4\right)+ \mathcal{O}(\epsilon^3) &
\end{flalign}

\endgroup

\newpage

\section{Colour factors up to N$^3$LO}\label{app:coltables}

\begin{table}[h]
  \centering
  \begin{tabular}{c c c}
    \cmidrule{1-3}\morecmidrules\cmidrule{1-3}  
    Final-state $\mathcal{I}$ & N$^k$LO ($\ell_1 \times \ell_2$) & Colour factors \\
    \cmidrule{1-3}\morecmidrules\cmidrule{1-3}  
    \multirow{6}{*}{$q\bar{q}$}
    & 1 & $N^0$ \\
    \cmidrule{2-3}  
    & 2, $1 \times 1$ & $N$, $N^{-1}$ \\
    & 2, $2 \times 0$ & $N$, $N^{-1}$, $\NF$ \\
    \cmidrule{2-3}  
    & 3, $2 \times 1$ & $N^2$, $N^0$, $N^{-2}$, $\NF N$, $\NF N^{-1}$ \\
    & 3, $3 \times 0$ & $N^2$, $N^0$, $N^{-2}$, $\NF N$, $\NF N^{-1}$, $\NF^2$ \\
    \specialrule{1pt}{0.2em}{0.2em}
    \multirow{4}{*}{$q\bar{q}g$}
    & 1 & $N^0$ \\
    \cmidrule{2-3}  
    & 2 & $N$, $N^{-1}$, $\NF$ \\
    \cmidrule{2-3}  
    & 3, $1 \times 1$ & $N^2$, $N^0$, $N^{-2}$, $\NF N$, $\NF N^{-1}$, $\NF^2$ \\
    & 3, $2 \times 0$ & $N^2$, $N^0$, $N^{-2}$, $\NF N$, $\NF N^{-1}$, $\NF^2$ \\
    \specialrule{1pt}{0.2em}{0.2em}      
    \multirow{2}{*}{$q\bar{q}gg$}
    & 2 & $N$, $N^{-1}$ \\
    \cmidrule{2-3}  
    & 3 & $N^2$, $N^0$, $N^{-2}$, $\NF N$, $\NF N^{-1}$ \\
    \specialrule{1pt}{0.2em}{0.2em}
    \multirow{2}{*}{$q\bar{q}q'\bar{q}'$}
    & 2 & $(\NF-1)$ \\
    \cmidrule{2-3}  
    & 3 & $(\NF-1)N$, $(\NF-1)N^{-1}$, $(\NF-1)\NF$ \\
    \specialrule{1pt}{0.2em}{0.2em}      
    \multirow{2}{*}{$q\bar{q}q\bar{q}$}
    & 2 & $N^{0}$, $N^{-1}$ \\
    \cmidrule{2-3}  
    & 3 & $N$, $N^{0}$, $N^{-1}$, $N^{-2}$, $\NF$, $\NF N^{-1}$ \\
    \specialrule{1pt}{0.2em}{0.2em}  
    \multirow{1}{*}{$q\bar{q}ggg$}
    & 3 & $N^2$, $N^0$, $N^{-2}$ \\
    \specialrule{1pt}{0.2em}{0.2em}
    \multirow{1}{*}{$q\bar{q}q'\bar{q}'g$}
    & 3 & $(\NF-1)N$, $(\NF-1)N^{-1}$ \\
    \specialrule{1pt}{0.2em}{0.2em}      
    \multirow{1}{*}{$q\bar{q}q\bar{q}g$}
    & 3 & $N$, $N^{0}$, $N^{-1}$, $N^{-2}$ \\
    \cmidrule{1-3}\morecmidrules\cmidrule{1-3}    
  \end{tabular}
  \caption{Colour factors appearing in the Higgs decay to bottom quarks,
    organised by final-state particles $\mathcal{I}$, perturbative order $k$ and
    loop configuration $\ell_1 \times \ell_2$, in case of ambiguity. Note that
    an overall factor of $2\CF$ is factored out as indicated
    in~\eqref{eq:Hbb0}.}
  \label{tab:Hbb}
\end{table}

\begin{table}[h]
  \centering
  \begin{tabular}{c c c}
    \cmidrule{1-3}\morecmidrules\cmidrule{1-3}  
    Final-state $\mathcal{I}$ & N$^k$LO ($\ell_1 \times \ell_2$) & Colour factors \\
    \cmidrule{1-3}\morecmidrules\cmidrule{1-3}  
    \multirow{5}{*}{$gg$}
    & 1 & $N$, $\NF$ \\
    \cmidrule{2-3}  
    & 2, $1 \times 1$ & $N^2$, $\NF N$, $\NF^2$ \\
    & 2, $2 \times 0$ & $N^2$, $\NF N$, $\NF N^{-1}$, $\NF^2$ \\
    \cmidrule{2-3}  
    & 3, $2 \times 1$ & $N^3$, $\NF N^2$, $\NF$, $\NF^2 N$, $\NF^2 N^{-1}$, $\NF^3$ \\
    & 3, $3 \times 0$ & $N^3$, $\NF N^2$, $\NF$, $\NF N^{-2}$, $\NF^2 N$, $\NF^2 N^{-1}$, $\NF^3$ \\
    \specialrule{1pt}{0.2em}{0.2em}
    \multirow{4}{*}{$ggg$}
    & 1 & $N$ \\
    \cmidrule{2-3}  
    & 2 & $N^2$, $\NF N$ \\
    \cmidrule{2-3}  
    & 3, $1 \times 1$ & $N^3$, $\NF N^2$, $\NF^2 N$ \\
    & 3, $2 \times 0$ & $N^3$, $\NF N^2$, $\NF$, $\NF^2 N$ \\
    \specialrule{1pt}{0.2em}{0.2em}  
    \multirow{4}{*}{$q\bar{q}g$}
    & 1 & $\NF$ \\
    \cmidrule{2-3}  
    & 2 & $\NF N$, $\NF N^{-1}$, $\NF^2$ \\
    \cmidrule{2-3}  
    & 3, $1 \times 1$ & $\NF N^2$, $\NF$, $\NF N^{-2}$, $\NF^2 N$, $\NF^2 N^{-1}$, $\NF^3$ \\
    & 3, $2 \times 0$ & $\NF N^2$, $\NF$, $\NF N^{-2}$, $\NF^2 N$, $\NF^2 N^{-1}$, $\NF^3$ \\
    \specialrule{1pt}{0.2em}{0.2em}  
    \multirow{2}{*}{$gggg$}
    & 2 & $N^2$ \\
    \cmidrule{2-3}  
    & 3 & $N^3$, $\NF N^2$ \\
    \specialrule{1pt}{0.2em}{0.2em}  
    \multirow{2}{*}{$q\bar{q}gg$}
    & 2 & $\NF N$, $\NF N^{-1}$ \\
    \cmidrule{2-3}  
    & 3 & $\NF N^2$, $\NF$, $\NF N^{-2}$, $\NF^2 N$, $\NF^2 N^{-1}$ \\
    \specialrule{1pt}{0.2em}{0.2em}  
    \multirow{2}{*}{$q\bar{q}q'\bar{q}'$}
    & 2 & $\NF(\NF-1)$ \\
    \cmidrule{2-3}  
    & 3 & $\NF(\NF-1)N$, $\NF(\NF-1)N^{-1}$, $\NF^2(\NF-1)$ \\
    \specialrule{1pt}{0.2em}{0.2em}  
    \multirow{2}{*}{$q\bar{q}q\bar{q}$}
    & 2 & $\NF$, $\NF N^{-1}$ \\
    \cmidrule{2-3}  
    & 3 & $\NF N$, $\NF$, $\NF N^{-1}$, $\NF N^{-2}$, $\NF^2$, $\NF^2 N^{-1}$  \\
    \specialrule{1pt}{0.2em}{0.2em}  
    \multirow{1}{*}{$ggggg$}
    & 3 & $N^3$ \\
    \specialrule{1pt}{0.2em}{0.2em}  
    \multirow{1}{*}{$q\bar{q}ggg$}
    & 3 & $\NF N^2$, $\NF$, $\NF N^{-2}$ \\
    \specialrule{1pt}{0.2em}{0.2em}  
    \multirow{1}{*}{$q\bar{q}q'\bar{q}'g$}
    & 3 & $\NF (\NF - 1) N$, $\NF (\NF - 1) N^{-1}$ \\
    \specialrule{1pt}{0.2em}{0.2em}  
    \multirow{1}{*}{$q\bar{q}q\bar{q}g$}
    & 3 & $\NF N$, $\NF$, $\NF N^{-1}$, $\NF N^{-2}$ \\
    \cmidrule{1-3}\morecmidrules\cmidrule{1-3}  
  \end{tabular}
  \caption{ Colour factors appearing in the Higgs decay to gluons, organised by
    final-state particles $\mathcal{I}$, perturbative order $k$ and loop
    configuration $\ell_1 \times \ell_2$, in case of ambiguity.  }
  \label{tab:Hgg}
\end{table}

\clearpage

\bibliographystyle{JHEP}
\bibliography{main}

\end{document}